\documentclass[11pt]{article}
\usepackage[margin=1in]{geometry}
\usepackage{amsmath,amssymb,amsthm}
\usepackage{graphicx}
\usepackage{booktabs}
\usepackage{longtable}
\usepackage{array}
\usepackage{caption}
\usepackage{authblk}
\usepackage{xcolor}
\usepackage{enumitem}
\usepackage[hidelinks]{hyperref}
\usepackage{url}
\usepackage{float}
\usepackage{multirow}
\usepackage{tabularx}
\usepackage{placeins}
\usepackage[most]{tcolorbox}

\makeatletter
\@namedef{cite@einstein1905}{1}
\@namedef{cite@rubner1908}{2}
\@namedef{cite@lindstedt1981}{3}
\@namedef{cite@calder1984}{4}
\@namedef{cite@livingstone1979}{5}
\@namedef{cite@levine1997}{6}
\@namedef{cite@stahl1967}{7}
\@namedef{cite@escala2022}{8}
\@namedef{cite@pearl1928}{9}
\@namedef{cite@speakman2005}{10}
\@namedef{cite@glazier2022}{11}
\@namedef{cite@schrodinger1944}{12}
\@namedef{cite@prigogine1967}{13}
\@namedef{cite@kleiber1932}{14}
\@namedef{cite@west1997}{15}
\@namedef{cite@mortola2006}{16}
\@namedef{cite@mckechnie2004}{17}
\@namedef{cite@deMagalhaes2009}{18}
\@namedef{cite@hulbert2007}{19}
\@namedef{cite@mcnab2008}{20}
\@namedef{cite@white2003}{21}
\@namedef{cite@genoud2018}{22}
\@namedef{cite@pontzer2014}{23}
\@namedef{cite@fahlman2025}{24}
\@namedef{cite@gompertz1825}{25}
\@namedef{cite@herculano2011}{26}
\@namedef{cite@friston2010}{27}
\@namedef{cite@wilkinson2002}{28}
\@namedef{cite@barja1998}{29}
\@namedef{cite@brand2000}{30}
\@namedef{cite@ogburn2001}{31}
\@namedef{cite@goldbogen2019}{32}
\@namedef{cite@williams2015}{33}
\@namedef{cite@noren2000}{34}
\@namedef{cite@anage2023}{35}
\@namedef{cite@jones2009}{36}
\@namedef{cite@bininda2007}{37}
\@namedef{cite@christian1999}{38}
\@namedef{cite@gillooly2001}{39}
\@namedef{cite@felsenstein1985}{40}
\@namedef{cite@horvath2013}{41}
\@namedef{cite@colman2014}{42}
\@namedef{cite@brown2004}{43}
\@namedef{cite@yegian2024}{44}
\@namedef{cite@lyman1982}{45}
\@namedef{cite@prinzinger1991}{46}
\@namedef{cite@clarke2008}{47}
\@namedef{cite@ponganis2015}{48}
\newcommand{\citenum}[1]{%
  \@ifundefined{cite@#1}{?}{\csname cite@#1\endcsname}}
\newcommand{\@citelist}{}
\DeclareRobustCommand{\cite}[1]{%
  \begingroup
  \let\@citelist\@empty
  \@for\@ci:=#1\do{%
    \edef\@citelist{\@citelist\ifx\@citelist\@empty\else,\,\fi\citenum{\@ci}}%
  }%
  [\@citelist]%
  \endgroup}
\makeatother

\title{Biological proper time and entropy-cost invariance in cardiac and respiratory lifespan scaling}

\author[1]{Mesfin Asfaw Taye}
\affil[1]{West Los Angeles College, Science Division, 9000 Overland Ave, Culver City, CA 90230, USA.\\
Correspondence: \texttt{tayem@wlac.edu}}
\date{\today}

\begin{document}
\maketitle

\begin{abstract}
\noindent
Warm-blooded vertebrates accumulate approximately conserved numbers of physiological cycles over a natural lifetime: of order $10^9$ heartbeats and $10^8$--$3\times10^8$ breaths. These regularities are not exact constants, but their persistence across orders-of-magnitude variation in body mass, metabolic power, physiological frequency, and lifespan suggests that biological time is not measured by chronological duration alone. We develop the Principle of Biological Time Equivalence (PBTE), a thermodynamic framework in which lifetime cycle count is determined by the ratio between total lifetime entropy production and the entropy cost of one physiological cycle. Starting from the open-system entropy balance $\dot S=\dot e_p-\dot h_d$, we define the entropy cost per cycle as $\sigma_0=d\Sigma/dN$, where $d\Sigma$ is the entropy produced as the physiological clock advances by $dN$ cycles. For an adult homeostatic regime, this gives the cycle-count relation $N_\star=\Sigma/\langle\sigma_0\rangle$, with $\Sigma=\int_0^L \dot e_p(t)\,dt$, where $N_\star$ is the lifetime cycle count, $\Sigma$ is total lifetime entropy production, and $\langle\sigma_0\rangle$ is the lifetime-averaged entropy cost per cycle. In the homeostatic limit, $\dot e_p\simeq P/T$, so direct measurement of metabolic power $P$, body temperature $T$, and physiological frequency $f$ gives $\sigma_0\simeq P/(Tf)$.
PBTE converts the empirical lifetime-cycle invariants into entropy-cost invariants. Under Kleiber metabolic scaling and quarter-power physiological-frequency scaling, the mass-specific entropy cost satisfies $\bar\sigma_0=P/(TfM)\propto M^{3/4+1/4-1}=M^0$, providing a thermodynamic interpretation of allometric mass cancellation. The framework also treats clade deviations as structured changes in effective cycle budget rather than residual scatter. We write $N_{\star,C}=N_{\star,0}\Phi_C$, where the clade multiplier $\Phi_C$ decomposes contributions from duty cycle, temperature-dependent kinetics, mitochondrial and antioxidant efficiency, and ecological hazard. Once $\Phi_C$ is determined, the same relation predicts lifespan from physiological frequency through $L_{\rm pred}=N_{\star,0}\Phi_C/(f\mathcal{T})$.
We apply this framework to cardiac and respiratory clocks. The cardiac analysis, on a $230$-species dataset, yields a reference lifetime count of order $10^9$ beats and a nearly mass-independent entropy cost per beat per unit mass. The respiratory analysis, based on the $65$-species subset with reliable resting breath rates, yields a coarser lifetime breath count of order $10^8$--$3\times10^8$ and provides an independent test of the same thermodynamic construction. As the decisive non-circular test, we recompute the respiratory entropy cost from \emph{measured} species-level basal metabolic rates (He et al.\ 2023) rather than imposed Kleiber scaling. On this independent dataset ($n=29$ mammals with both measured BMR and measured resting breath rate) the mass cancellation does \emph{not} survive: the mass-specific respiratory entropy cost rises with body mass (fitted slope $\simeq+0.21$, not statistically resolved, $p\simeq0.11$; coefficient of variation $\sim\!100\%$). The positive, dispersed trend is driven by aquatic mammals, whose low resting breath rates relative to metabolic rate sharply inflate the entropy cost per breath at large body size. The respiratory clock therefore fails the non-circular cancellation test that the cardiac clock passes---an informative asymmetry rather than a confirmation, and the central empirical finding of the respiratory analysis. Building on the validated cardiac invariant, we promote biological proper time to an internal-age coordinate: the entropy-normalized biological age $A_{\mathrm{PBTE}}(t)=\Sigma(t)/\Sigma_{\rm ref}$ measures the fraction of a reference entropy--cycle budget already consumed, and its rate $dA_{\mathrm{PBTE}}/dt=\dot e_p/\Sigma_{\rm ref}$ defines an aging velocity set by entropy production. This coordinate organizes aging and longevity into three thermodynamic mechanism classes---time dilation (caloric restriction, torpor, cetacean bradycardia), entropy-cost reduction and budget expansion (avian and primate maintenance efficiency), and hypertemporal pathologies (inflammation, metabolic syndrome, neurodegeneration, cancer)---all visible in the clade structure of this dataset. The cardiac clock supplies a non-circular reference cost for this coordinate while the respiratory clock does not, so the empirical asymmetry directly governs which rhythm can serve as the clock for aging. The resulting theory identifies biological proper time as the accumulated physiological cycle count and makes PBTE experimentally falsifiable: its decisive test is simultaneous calorimetric, cardiac, respiratory, temperature, and body-mass measurement across species to determine whether $\sigma_0=P/(Tf)$ is truly mass-independent within defined physiological regimes.
\end{abstract}


\section{Introduction}
\label{sec:intro}

Time has always stood at the boundary between science, philosophy, and lived experience. We call it a blessing when it is abundant, a curse when it disappears too quickly, and a sacred gift when it is shared. Civilizations have measured it by the motion of stars, the turning of seasons, the growth of trees, the rise and fall of tides, and the pulse within the body; yet its deepest meaning remains elusive. Modern physics showed that time is not an absolute background: the time measured by an observer depends on motion and gravity~\cite{einstein1905}. A related, but biologically distinct, idea appears in living systems. In inert matter, velocity and gravitation shape physical proper time; in living matter, metabolism, organization, and entropy production shape biological time. A mayfly, a shrew, an elephant, and a human do not merely pass through the same external clock for different durations; each advances through an internal rhythm set by heartbeats, breaths, molecular turnover, repair, and the energetic cost of maintaining order. The analogy with relativity is therefore structural, not literal: biological proper time is not defined by spacetime geometry, but by the dynamics of open dissipative systems. Every physiological tick transforms energy and produces entropy; life may proceed faster or slower in chronological years, but it advances by spending a finite budget of internal cycles.

Among the most striking regularities in comparative physiology is a temporal invariant that is both simple and physically suggestive. A pygmy shrew and an African elephant inhabit radically different chronological worlds: the shrew has a body mass of only a few grams, may sustain a resting heart rate of several hundred beats per minute, and lives only a few years, whereas the elephant has a mass of several thousand kilograms, beats only tens of times per minute, and may live for many decades. Yet this disparity is greatly reduced when physiological rate is multiplied by natural lifespan: both animals accumulate of order $10^9$ cardiac cycles over a lifetime~\cite{rubner1908,lindstedt1981,calder1984,livingstone1979,levine1997}. A parallel regularity occurs in respiration, where resting breath rate multiplied by lifespan gives a lifetime breath count of order $10^8$--$3\times10^8$ in warm-blooded vertebrates~\cite{stahl1967,escala2022}. These two empirical clocks can be summarized compactly as
\begin{equation}
N_H=f_H L\mathcal{T}\simeq10^9,
\qquad
N_R=f_R L\mathcal{T}\simeq3\times10^8,
\label{eq:intro_cycle_invariants}
\end{equation}
where $N_H$ and $N_R$ are the lifetime numbers of heartbeats and breaths, $f_H$ is the resting heart rate, $f_R$ is the resting breath rate, $L$ is lifespan in years, and $\mathcal{T}=525{,}960\,\mathrm{min\,yr^{-1}}$ converts years into minutes when rates are measured per minute. The numerical values in Eq.~\eqref{eq:intro_cycle_invariants} are not exact constants. Their significance is that they remain concentrated despite orders-of-magnitude variation in body mass, metabolic power, physiological rate, and lifespan. Moreover, the residual deviations are structured: primates, bats, birds, and cetaceans show reproducible offsets from the reference mammalian pattern, indicating that lifetime cycle counts are shaped by physiology and clade history rather than by body mass alone.

To explain this near constancy, previous studies have emphasized allometric cancellation. Small animals live at a faster physiological pace, whereas large animals live more slowly: heart rate decreases with body mass, while lifespan increases in the opposite direction. Their product therefore becomes nearly mass independent. This argument explains why the leading body-mass dependence is weak, but it does not identify the physical quantity whose lifetime accumulation determines the number of physiological cycles. Nor does it explain why clades are displaced systematically from the same reference line. Classical rate-of-living arguments face a related limitation~\cite{pearl1928,speakman2005}: they connect lifespan to metabolic expenditure, but they do not specify the entropy cost of an individual cardiac or respiratory cycle. Fixed-exponent scaling arguments are further limited by the dependence of metabolic exponents on taxon, metabolic level, and physiological state~\cite{glazier2022}. Thus allometry explains the cancellation, but not the thermodynamic content of the invariant.

Our proposed Principle of Biological Time Equivalence (PBTE) addresses this gap by identifying irreversible entropy production as the thermodynamic measure of biological progression. A living organism is an open dissipative system: it continuously produces entropy through metabolism, molecular turnover, and physiological activity, while exporting entropy to maintain internal organization. In the homeostatic regime relevant to resting comparative physiology, entropy production is approximately balanced by entropy export. The instantaneous entropy-production rate may therefore be estimated, to leading order, by the metabolic power divided by body temperature. As derived in the next section, PBTE interprets the entropy cost of one physiological cycle as the dissipative price paid when the biological clock advances by one internal tick. This leads to the cycle-count identity
\begin{equation}
N_\star=\frac{\Sigma}{\langle\sigma_0\rangle},
\qquad
\Sigma=\int_0^L \dot e_p(t),dt,
\label{eq:intro_pbte_identity}
\end{equation}
where $N_\star$ denotes the lifetime physiological cycle count, $\Sigma$ is the total entropy produced over the lifespan $L$, and $\langle\sigma_0\rangle$ is the lifetime-averaged entropy cost per cycle. Thus the proposed invariant is neither chronological lifespan alone nor total lifetime energy expenditure alone. It is the ratio between the total dissipative expenditure of the organism and the entropy required to advance its physiological clock by one cycle.

As shown in the next section, this formulation gives allometric cancellation a thermodynamic meaning. Metabolic power, physiological frequency, and body mass scale together. Their scalings partly compensate one another. As a result, the entropy cost per physiological cycle, normalized by body mass, becomes nearly independent of body size. The lifetime-cycle invariant is therefore recast as an entropy-cost invariant. Small and large organisms dissipate entropy at different rates in chronological time. Yet the cost assigned to one unit of biological time remains approximately conserved. PBTE therefore identifies the physical quantity that allometry cancels. The detailed derivation and empirical tests are developed in the sections that follow.

At this point, we emphasize that the framework becomes most informative when the invariant is not exact.  In PBTE, clade deviations are not treated as residual scatter around a single allometric line, but as systematic changes in the effective cycle budget. We write the lifetime cycle count in clade $C$ as
\begin{equation}
N_{\star,C}=N_{\star,0}\Phi_C,
\label{eq:intro_clade_count}
\end{equation}
where $N_{\star,0}$ is the reference cycle budget and $\Phi_C$ is a dimensionless clade multiplier. The multiplier decomposes the departure from the baseline into physiological contributions,
\begin{equation}
\Phi_C=
\Phi_{\rm duty}\,
\Phi_{\rm thermal}\,
\Phi_{\rm mito+oxid}\,
\Phi_{\rm haz}.
\label{eq:intro_clade_multiplier}
\end{equation}
Here $\Phi_{\rm duty}$ accounts for the fraction of life spent in suppressed or accelerated physiological states, $\Phi_{\rm thermal}$ accounts for temperature-dependent changes in metabolic and damage kinetics, $\Phi_{\rm mito+oxid}$ represents mitochondrial efficiency and antioxidant protection, and $\Phi_{\rm haz}$ represents the extent to which ecological mortality permits the intrinsic thermodynamic budget to be realized. Once $\Phi_C$ is determined, Eq.~\eqref{eq:intro_clade_count} can be inverted to predict lifespan from a measured physiological frequency,
\begin{equation}
L_{\rm pred}
=
\frac{N_{\star,0}\Phi_C}{f\mathcal{T}},
\label{eq:intro_lifespan_prediction}
\end{equation}
where $\mathcal{T}=525{,}960\,\mathrm{min\,yr^{-1}}$ when $f$ is measured in cycles per minute and $L_{\rm pred}$ is expressed in years. Thus $\Phi_C$ is not a post hoc correction; it is the predictive component of the theory. After the clade multiplier is derived, lifespan can be estimated from physiological frequency and compared directly with observation.

The aim of this paper is to develop PBTE as a falsifiable thermodynamic parametrisation of biological time and to test its predictive content using two physiological clocks. First, we derive $\sigma_0$ from the entropy balance. We then show how allometric mass cancellation leads to an approximately invariant entropy cost per cycle per unit mass. Second, we derive the clade multiplier $\Phi_C$. This converts the baseline invariant into clade-specific lifespan predictions. Third, we apply the framework to cardiac and respiratory cycle counts. The purpose of using two clocks is not to determine which rhythm is more fundamental. Rather, it is to test whether distinct physiological cycles are governed by the same thermodynamic structure. Fourth, we promote biological proper time to a normalized internal-age coordinate $A_{\rm PBTE}(t)$ and show how the validated cardiac entropy-cost invariant supplies a non-circular clock for aging and longevity, organizing interventions and pathologies into time-dilation, entropy-cost, and hypertemporal mechanism classes, whereas the respiratory clock---which fails the same invariance test---does not. The theory is directly testable. Its decisive test requires simultaneous measurements of metabolic power $P$, body temperature $T$, physiological frequency $f$, and body mass $M$ across species. These measurements allow $\sigma_0=P/(Tf)$ to be determined directly, rather than inferred from allometric scaling. We believe that PBTE provides a thermodynamically motivated scaling framework for biological time. Lifespan is then constrained by the number of dissipative physiological cycles that an organism can realize within its entropy budget.

Furthermore, in this work we show that the approximate lifetime-cycle invariants observed in vertebrates are not merely consequences of allometric scaling. Rather, they reflect an underlying thermodynamic organization of biological time. We further show that systematic clade-dependent departures from the baseline invariant can be interpreted within a unified entropy-budget framework. In this view, primates achieve exceptional longevity through physiological strategies that reduce the effective entropy burden of each biological cycle. This allows a larger number of cycles to be accumulated over a lifetime. Bats extend lifespan through prolonged periods of metabolic suppression, during which biological time advances more slowly. Birds, despite high metabolic rates and elevated body temperatures, achieve extended longevity through enhanced cellular maintenance, mitochondrial performance, and resistance to oxidative damage. Cetaceans follow another strategy. Their lifespan extension is associated with profound cardiovascular slowing during diving and altered physiological pacing over the life course. These mechanisms are biologically distinct. Yet we show that they can be understood as different realizations of the same thermodynamic principle. The resulting framework explains why different clades exhibit systematic differences in lifetime cycle counts. It also provides a quantitative basis for predicting lifespan from physiological and thermodynamic characteristics.

\section{Thermodynamic Foundation}
\label{sec:shared}

\begin{tcolorbox}[colback=gray!4,colframe=gray!55,boxrule=0.5pt,
  arc=2pt,left=8pt,right=8pt,top=6pt,bottom=6pt,
  title={\textbf{The three logical levels of PBTE}},
  fonttitle=\bfseries,coltitle=black,colbacktitle=gray!18]
\small
The framework developed below is best read as three logically distinct
statements of decreasing generality. Conflating them is the most common
source of misunderstanding, so we separate them explicitly here and
maintain the distinction throughout.

\smallskip
\noindent\textbf{Level 1 --- exact identity (no physical assumption).}
Given the definitions of cycle count $N(t)=\int_0^t f\,dt'$ and entropy
cost per cycle $\sigma_0=\dot e_p/f=d\Sigma/dN$, the relation
\[
N_\star=\frac{\Sigma}{\langle\sigma_0\rangle},
\qquad
\Sigma=\int_0^L \dot e_p(t)\,dt,
\]
is an exact algebraic identity. It introduces no biology and cannot be
falsified; it merely names the lifetime cycle count as a ratio of two
defined quantities.

\smallskip
\noindent\textbf{Level 2 --- thermodynamic closure (an approximation).}
In an adult homeostatic regime, entropy export tracks production, so that
$\dot e_p\simeq P/T$ and the cost per cycle is estimated operationally by
\[
\langle\sigma_0\rangle\simeq\frac{P}{Tf}.
\]
This is a constitutive approximation, not an identity. It is testable in
principle by measuring $P$, $T$, and $f$ on the same individuals.

\smallskip
\noindent\textbf{Level 3 --- empirical claim (the falsifiable content).}
The substantive, refutable assertion of PBTE is that the mass-specific
cost
\[
\bar\sigma_0=\frac{\sigma_0}{M}\simeq\frac{P}{TfM}
\]
is approximately invariant \emph{within defined physiological regimes}.
This is an empirical regularity, expected to hold only approximately, and
it is the level at which the framework can fail. The clade multipliers of
Section~\ref{sec:clade_multiplier} are the structured departures from this
approximate invariance.
\end{tcolorbox}

\medskip
A living organism is an open nonequilibrium system~\cite{schrodinger1944,prigogine1967}. It continuously transforms chemical free energy into mechanical work, ionic gradients, biosynthesis, repair, signalling, and heat, while exporting entropy in order to preserve internal organization. Let $S(t)$ denote the coarse-grained internal entropy at chronological time $t$. The open-system entropy balance is
\begin{equation}
\dot S(t)=\dot e_p(t)-\dot h_d(t),
\label{eq:balance}
\end{equation}
where $\dot e_p(t)\geq0$ is the irreversible entropy production rate and $\dot h_d(t)\geq0$ is the entropy export rate. To leading thermodynamic order, entropy export is estimated from metabolic power and absolute temperature as
\begin{equation}
\dot h_d(t)\simeq \frac{P(t)}{T(t)} ,
\label{eq:entropy_export}
\end{equation}
where $P(t)$ is metabolic power and $T(t)$ is body temperature.

Equation~\eqref{eq:balance} contains both the homeostatic regime used for the lifetime cycle-count relation and the non-steady regime relevant to aging. In physiological homeostasis,
\begin{equation}
\dot S(t)\simeq0,
\qquad
\dot e_p(t)\simeq\dot h_d(t)\simeq\frac{P(t)}{T(t)} .
\label{eq:homeostasis}
\end{equation}
If entropy export becomes less efficient during aging, then $\dot h_d(t)<\dot e_p(t)$ and internal entropy accumulates according to
\begin{equation}
S(t)=S(0)+\int_0^t
\left[
\dot e_p(t')-\dot h_d(t')
\right]\,dt' .
\label{eq:Saccum}
\end{equation}
This accumulated entropy is interpreted as a coarse-grained burden associated with loss of physiological order, impaired repair, damaged macromolecules, mitochondrial dysfunction, and weakened regulatory control.

We now introduce the physiological cycle coordinate. Let $f(t)$ denote the frequency of a recurrent physiological process, measured in $\mathrm{s}^{-1}$. Its cumulative count is
\begin{equation}
N(t)=\int_0^t f(t')\,dt',
\qquad
dN=f(t)\,dt .
\label{eq:N_def}
\end{equation}
The entropy produced over the same interval is
\begin{equation}
d\Sigma=\dot e_p(t)\,dt .
\label{eq:dSigma}
\end{equation}
Eliminating $dt$ gives
\begin{equation}
d\Sigma=
\frac{\dot e_p(t)}{f(t)}\,dN .
\label{eq:dSigma_cycle}
\end{equation}
Thus the instantaneous entropy cost of one physiological cycle is
\begin{equation}
\sigma_0(t)
\equiv
\frac{\dot e_p(t)}{f(t)}
=
\frac{d\Sigma}{dN}.
\label{eq:sigma0_def}
\end{equation}
This quantity is not introduced as a dimensional convention; it is the entropy produced when the biological clock advances by one cycle.

Integration over lifespan gives
\begin{equation}
\Sigma
=
\int_0^L \dot e_p(t)\,dt
=
\int_0^{N_\star}
\sigma_0(N)\,dN ,
\label{eq:Sigma_N}
\end{equation}
where $\Sigma$ is the total lifetime entropy production and $N_\star$ is the total lifetime number of cycles. Defining the lifetime-averaged entropy cost per cycle,
\begin{equation}
\langle\sigma_0\rangle
=
\frac{1}{N_\star}
\int_0^{N_\star}
\sigma_0(N)\,dN ,
\label{eq:sigma0_mean}
\end{equation}
one obtains
\begin{equation}
N_\star=
\frac{\Sigma}{\langle\sigma_0\rangle}.
\label{eq:fundamental}
\end{equation}
Equation~\eqref{eq:fundamental} is the basic PBTE cycle-count relation: the lifetime number of physiological cycles equals total lifetime entropy production divided by the mean entropy cost of one cycle.

The Principle of Biological Time Equivalence is the constitutive approximation that, within a defined adult homeostatic regime, the entropy cost per cycle is represented by its average value. In that regime,
\begin{equation}
\dot e_p(t)\simeq \langle\sigma_0\rangle f(t).
\label{eq:closure}
\end{equation}
Combining this closure with Eq.~\eqref{eq:homeostasis} gives the operational estimator
\begin{equation}
\langle\sigma_0\rangle
\simeq
\frac{P}{Tf}.
\label{eq:sigma_estimator}
\end{equation}
The corresponding mass-specific entropy cost is
\begin{equation}
\bar{\sigma}_0
=
\frac{\langle\sigma_0\rangle}{M}
\simeq
\frac{P}{TfM}.
\label{eq:sigma_mass_specific}
\end{equation}
Under Kleiber metabolic scaling~\cite{kleiber1932,west1997}, $P\propto M^{3/4}$, and quarter-power physiological-frequency scaling, $f\propto M^{-1/4}$, this gives
\begin{equation}
\bar{\sigma}_0
\propto
\frac{M^{3/4}}{M^{-1/4}M}
=
M^0 .
\label{eq:mass_cancel}
\end{equation}
Thus the allometric cancellation of lifetime cycle count is recast as an approximate invariance of the entropy cost per cycle per unit mass.

Finally, the same construction defines biological proper time. Chronological time $t$ is the external clock time, whereas biological proper time is the accumulated physiological cycle count,
\begin{equation}
\theta(t)=\int_0^t f(t')\,dt',
\qquad
\hat{\theta}(t)=\frac{\theta(t)}{N_\star}.
\label{eq:theta_def}
\end{equation}
The quantity $\theta(t)$ measures the advancement of the biological clock rather than the passage of chronological time. If the total lifetime cycle budget $N_\star$ is approximately invariant across comparable organisms, then lifespan is determined primarily by the rate at which biological proper time accumulates. For an approximately constant physiological frequency,
\begin{equation}
N_\star\simeq fL,
\label{eq:Nstar_fL}
\end{equation}
where $L$ is the chronological lifespan. Consequently,
\begin{equation}
L\simeq \frac{N_\star}{f}.
\label{eq:L_inverse_f}
\end{equation}
Thus, a lower physiological frequency implies a slower advance of biological proper time and therefore a longer chronological lifespan. In this view, longevity is associated not with slowing chronological time itself, but with slowing the rate at which the organism consumes its finite budget of physiological cycles.

Under the PBTE closure,
\begin{equation}
d\Sigma=\langle\sigma_0\rangle\,d\theta,
\qquad
\frac{d\Sigma}{d\theta}=\langle\sigma_0\rangle .
\label{eq:uniform_entropy_theta}
\end{equation}
Biological proper time is therefore the coordinate in which entropy accumulates uniformly. The cardiac and respiratory clocks are two realizations of the same construction:
\begin{equation}
N_H=
\frac{\Sigma}{\langle\sigma_H\rangle},
\qquad
N_R=
\frac{\Sigma}{\langle\sigma_R\rangle}.
\label{eq:cardiac_respiratory_master}
\end{equation}
The following sections apply this identity to heartbeat and breathing.
In the next section, we develop the clade modulation of the lifetime cycle budget and show how physiological differences enter the PBTE framework.

\section{Clade Modulation of the Lifetime Cycle Budget}
\label{sec:clade_multiplier}

As discussed above, the PBTE relation $N_\star=\Sigma/\langle\sigma_0\rangle$ defines a reference lifetime cycle budget. Empirical observations, however, show that lifetime cycle counts are only approximately invariant. Systematic departures occur among major vertebrate clades. This indicates that the effective biological-time budget is modulated by physiological and ecological factors beyond the leading allometric scaling.

Within PBTE, these departures are not interpreted as residual scatter around a universal scaling law. Rather, they reflect systematic renormalizations of the effective lifetime cycle budget. We therefore write
\begin{equation}
N_{\star,C}
=
N_{\star,0}\Phi_C ,
\label{eq:clade_budget}
\end{equation}
where $N_{\star,0}$ is the reference cycle budget and $\Phi_C$ is a dimensionless clade multiplier. Values $\Phi_C>1$ correspond to clades that realize a larger effective cycle budget than the reference state, whereas $\Phi_C<1$ corresponds to a reduced budget.

The multiplier summarizes the dominant mechanisms that modify either the lifetime entropy budget $\Sigma$ or the entropy cost per cycle $\langle\sigma_0\rangle$. Therefore, to leading order, we write
\begin{equation}
\Phi_C
=
\Phi_{\rm duty}\,
\Phi_{\rm thermal}\,
\Phi_{\mathrm{mito+oxid}}\,
\Phi_{\rm haz}.
\label{eq:Phi_decomp}
\end{equation}
Here $\Phi_{\rm duty}$ captures changes in biological-clock accumulation caused by intermittent physiological states such as torpor, hibernation, dormancy, or diving bradycardia. The factor $\Phi_{\rm thermal}$ represents temperature-dependent changes in metabolic and damage kinetics. The factor $\Phi_{\mathrm{mito+oxid}}$ accounts for mitochondrial efficiency, oxidative-stress resistance, and cellular maintenance mechanisms that alter the entropy generated per physiological cycle. Finally, $\Phi_{\rm haz}$ represents extrinsic ecological mortality, which determines how fully the intrinsic thermodynamic budget can be realized in practice.

The resulting lifetime cycle count becomes
\begin{equation}
N_{\star,C}
=
N_{\star,0}\,
\Phi_{\rm duty}\,
\Phi_{\rm thermal}\,
\Phi_{\mathrm{mito+oxid}}\,
\Phi_{\rm haz}.
\label{eq:general_budget}
\end{equation}
Equation~\eqref{eq:general_budget} extends the PBTE invariant from a single reference class to a hierarchy of biological strategies. Clade-specific variation is therefore interpreted not as statistical scatter, but as evidence of distinct thermodynamic mechanisms that modify biological-time accumulation, the entropy cost of physiological cycles, or the accessible lifetime entropy budget. In the following sections, we examine two independent realizations of this framework: the cardiac clock, in which biological proper time is measured by accumulated heartbeats, and the respiratory clock, in which it is measured by accumulated breaths. The presence of approximate lifetime invariants in both systems provides a direct test of whether distinct physiological rhythms are governed by a common thermodynamic structure. In Supplementary Sec.~\ref{app:clade_multiplier}, we derive the clade multipliers in detail for major vertebrate lineages, including nonplacental mammals, primates, chiropterans, avian species, and cetaceans. We then use these multipliers to generate quantitative lifespan predictions and compare the predicted values directly with observed lifespans.

To make the predictive content of $\Phi_C$ transparent and to guard against the impression that the multipliers are tuned post hoc, we state explicitly how each factor is obtained. The four channels differ sharply in their epistemic status: two are fixed by measured inputs through closed-form expressions, while two are supplied from independent literature and enter phenomenologically. Table~\ref{tab:phi_provenance} records this classification. Throughout, $\Phi_{\rm duty}$ and $\Phi_{\rm thermal}$ are computed from quantities ($q$, $f_{H,k}$, $T_b$) measured per species and inserted into the exact duty-cycle identity and the Arrhenius expression, so they introduce no free parameters once those quantities are known. By contrast, $\Phi_{\mathrm{mito+oxid}}$ and $\Phi_{\rm haz}$ are not derived from the closure; they are taken from comparative biochemistry and demography and carry the larger share of the residual uncertainty in the predictions.

\begin{table}[ht]
\centering
\caption{Provenance of the four clade-multiplier channels. ``Derived'' means
computed from a closed-form expression once its inputs are fixed;
``measured input'' identifies the per-species quantities those expressions
consume; ``literature/phenomenological'' means taken from independent
comparative studies rather than from the PBTE closure.}
\label{tab:phi_provenance}
\small
\renewcommand{\arraystretch}{1.25}
\begin{tabularx}{\textwidth}{@{}l l l X@{}}
\toprule
Factor & Status & Measured inputs & How obtained \\
\midrule
$\Phi_{\rm duty}$ & Derived (exact identity) &
$q,\ f_{H,k},\ f_{H,\rm ref}$ &
Closed-form $\kappa^{-1}$ from the state-occupancy average; no free parameter. \\
$\Phi_{\rm thermal}$ & Derived (Arrhenius) &
$T_b,\ T_{\rm ref}$ &
Closed-form exponential with a single literature activation energy $E_a$; no per-species fitting. \\
$\Phi_{\mathrm{mito+oxid}}$ & Literature / phenomenological &
--- (clade-level) &
Estimated from comparative biochemistry (coupling efficiency, ROS production, repair capacity); not derived from the closure. \\
$\Phi_{\rm haz}$ & Literature / phenomenological &
--- (clade-level) &
Inferred from comparative demography (extrinsic mortality, ecological shielding); not derived from the closure. \\
$\Phi_{\rm neuro}$ (primates) & Derived form, calibrated exponent &
$\varphi=P_{\rm brain}/P_{\rm body}$ &
Power-law form derived from additive logarithmic sensitivities; the exponent $\alpha$ is calibrated, with $0<\alpha<1$ imposed by the second law. \\
\bottomrule
\end{tabularx}
\end{table}

\begin{figure}[t]
\centering
\includegraphics[width=\textwidth]{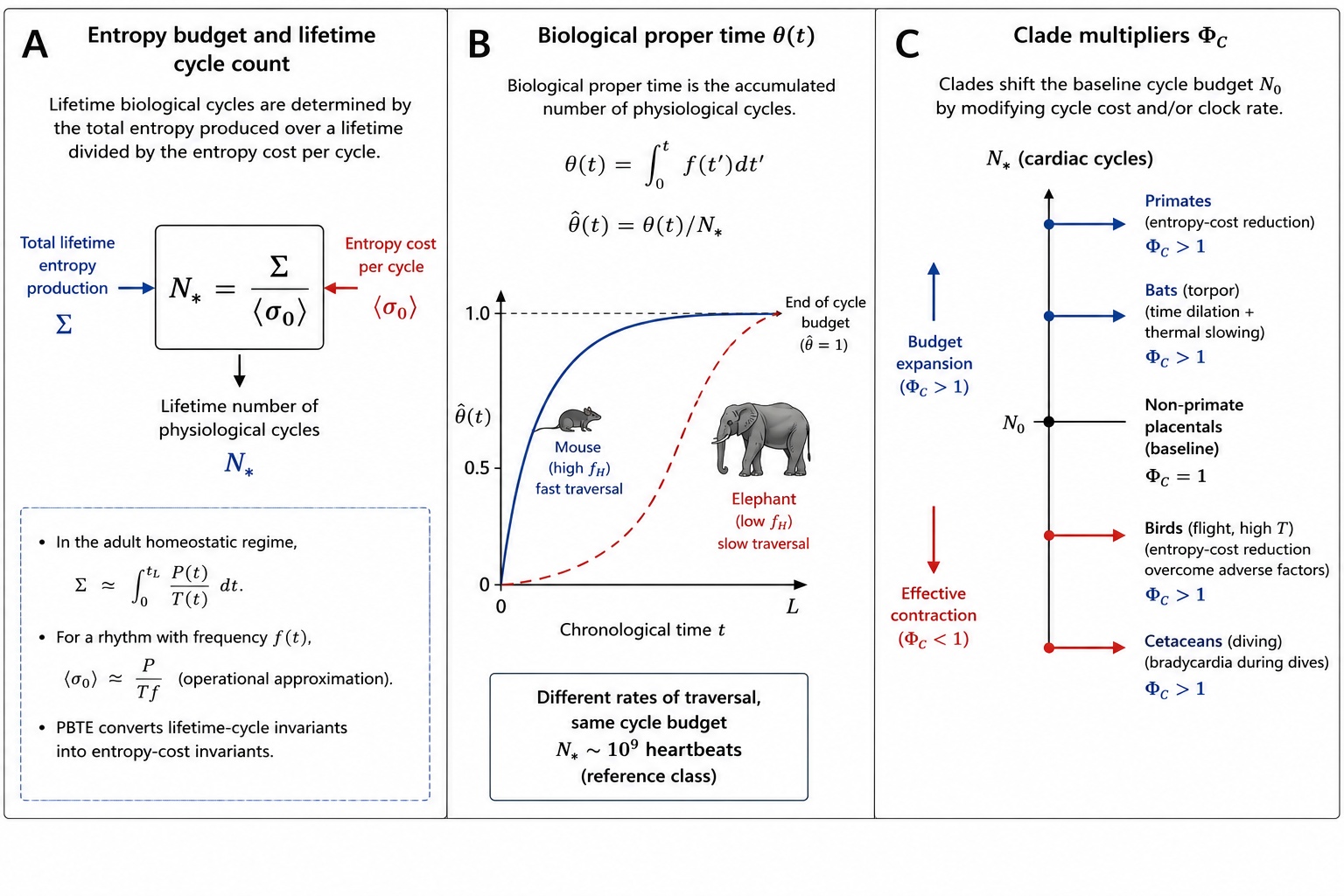}
\caption{Schematic summary of the PBTE framework. 
(A) The lifetime number of physiological cycles is determined by the ratio of total lifetime entropy production $\Sigma$ to the mean entropy cost per cycle $\langle\sigma_0\rangle$. 
(B) Biological proper time $\theta(t)$ is the accumulated number of physiological cycles. Small mammals, with high cardiac frequency, traverse the reference cycle budget rapidly in chronological time, whereas large mammals traverse the same budget more slowly. 
(C) Clade multipliers $\Phi_C$ shift the baseline cardiac budget $N_0$ by modifying the entropy cost per cycle and/or the rate at which biological time advances. Non-primate placentals define the reference class, while primates, bats, birds, and cetaceans represent distinct thermodynamic strategies for modifying the effective lifetime cycle budget.}
\label{fig:PBTE_framework}
\end{figure}

We next examine the entropy cost, mass cancellation, and biological proper time associated with the cardiac and respiratory clocks. In PBTE, a physiological cycle is not merely a countable biological event; it is a thermodynamic event with a definite dissipative cost. For the cardiac clock, this cost is associated with each heartbeat. For the respiratory clock, it is associated with each breath. In both cases, metabolic power, body temperature, physiological frequency, and body mass determine the entropy cost per cycle. Allometric scaling then provides the mechanism by which body-size dependence is largely canceled at the level of the mass-normalized cycle cost. Thus heartbeats and breaths define two distinct, but structurally related, measures of biological proper time. Each records the progression of life through accumulated dissipative cycles rather than through chronological duration alone. Figure~\ref{fig:PBTE_framework} summarizes the three central elements of the framework: the entropy-budget cycle-count relation, biological proper time as accumulated cycle count, and the clade multipliers that displace the baseline budget.

\section{Entropy cost, mass cancellation, and biological proper time}
\label{sec:entropy_cost_mass_cancellation}

\subsection{The cardiac clock}
\label{subsec:cardiac_clock}

We first apply PBTE to the cardiac rhythm, which provides the most direct physiological realization of biological proper time. For this clock, the relevant frequency is the resting heart rate, $f=f_H$, and the entropy cost per cycle is the entropy cost per heartbeat, denoted by $\sigma_H$. In the adult homeostatic regime, the entropy-production rate is estimated by $\dot e_p\simeq P/T$, where $P$ is metabolic power and $T$ is body temperature. The entropy cost per heartbeat is therefore
\begin{equation}
\sigma_H
=
\frac{P}{T f_H}.
\label{eq:sigmaH}
\end{equation}
The corresponding mass-specific entropy cost is
\begin{equation}
\bar{\sigma}_H^{(M)}
=
\frac{\sigma_H}{M}
=
\frac{P}{T f_H M}.
\label{eq:sigmaH_mass}
\end{equation}
This is the central quantity for the cardiac form of PBTE. The total entropy produced by an organism increases strongly with body size. By contrast, $\bar{\sigma}_H^{(M)}$ measures the entropy cost of one heartbeat per unit body mass. It is this normalized cycle cost, not the total dissipated entropy, that is expected to become approximately invariant after allometric cancellation.

The cancellation follows directly from the leading allometric scalings. If metabolic power scales as $P\propto M^{3/4}$ and resting heart rate scales as $f_H\propto M^{-1/4}$, then Eq.~\eqref{eq:sigmaH_mass} gives
\begin{equation}
\bar{\sigma}_H^{(M)}
\propto
\frac{M^{3/4}}{M^{-1/4}M}
=
M^0 .
\label{eq:cardiac_cancel}
\end{equation}
Thus PBTE predicts that the entropy cost of one heartbeat per unit body mass is approximately independent of body size. This does not mean that every organism dissipates the same entropy per heartbeat. Rather, it means that after normalization by body mass, the dissipative cost assigned to one cardiac tick becomes nearly conserved across mammals.

\begin{table}[t]
\centering
\caption{Mass-specific entropy cost per heartbeat for representative mammals. Metabolic power is estimated from $P=3.4M^{0.75}\,\mathrm{W}$, body temperature is taken as $T\simeq310\,\mathrm{K}$, and resting heart rate follows $f_H\simeq241M^{-0.25}\,\mathrm{min^{-1}}$.}
\label{tab:cardiac_sigma_table}
\resizebox{\columnwidth}{!}{%
\begin{tabular}{lccccc}
\hline
Species & $M$ (kg) & $P$ (W) & $f_H$ (bpm) & $\sigma_H$ ($10^{-3}$ J K$^{-1}$ beat$^{-1}$) & $\bar{\sigma}_H^{(M)}$ ($10^{-3}$ J K$^{-1}$ beat$^{-1}$ kg$^{-1}$)\\
\hline
House mouse & 0.020 & 0.18 & 600 & 0.058 & 2.9\\
Rat & 0.300 & 1.38 & 420 & 0.634 & 2.1\\
Rabbit & 2.0 & 5.72 & 205 & 2.7 & 2.7\\
Dog & 23 & 35.7 & 90 & 19.2 & 3.3\\
Human & 70 & 82.3 & 70 & 56.9 & 3.3\\
Horse & 500 & 360 & 40 & 273 & 3.5\\
Elephant & 4000 & 1710 & 28 & 1180 & 3.0\\
\hline
Mean $\pm$ s.d. &  &  &  &  & $3.0\pm0.5$\\
Coefficient of variation &  &  &  &  & $16\%$\\
\hline
\end{tabular}%
}
\end{table}

\begin{figure}[t]
\centering
\includegraphics[width=\columnwidth]{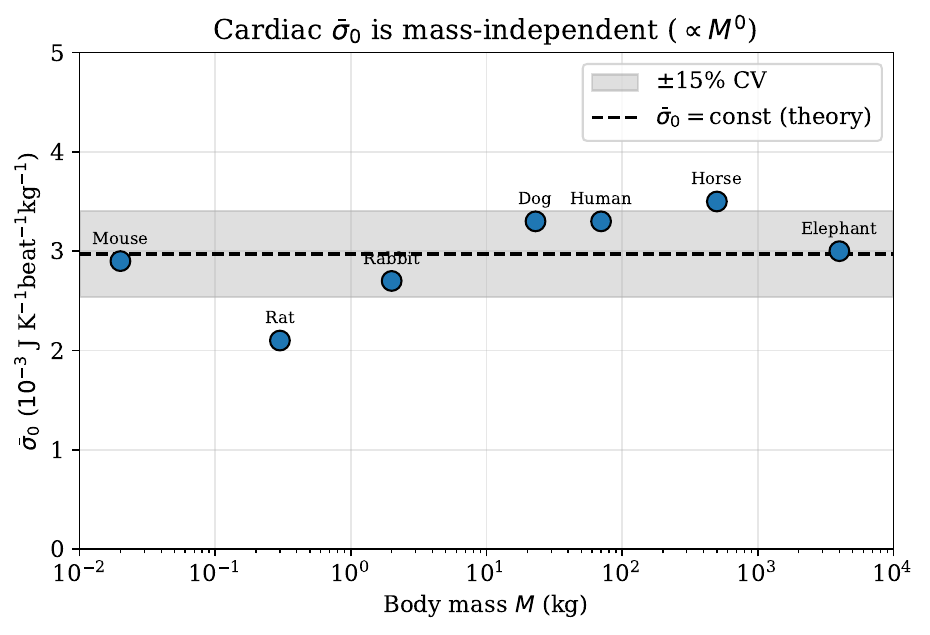}
\caption{Mass-specific cardiac entropy cost $\bar{\sigma}_H^{(M)}$ versus body mass. The dashed line shows the PBTE prediction $\bar{\sigma}_H^{(M)}\propto M^0$.}
\label{fig:C1}
\end{figure}

Table~\ref{tab:cardiac_sigma_table} illustrates this point. The absolute entropy cost per heartbeat, $\sigma_H$, varies by orders of magnitude from small to large mammals. This is expected, because a heartbeat in a large animal supports a much larger body mass. However, the mass-specific cost $\bar{\sigma}_H^{(M)}$ remains narrowly distributed around
\begin{equation}
\bar{\sigma}_H^{(M)}
\simeq
3.0\times10^{-3}\,
\mathrm{J\,K^{-1}\,beat^{-1}\,kg^{-1}} .
\label{eq:sigmaH_value}
\end{equation}
The approximate constancy of $\bar{\sigma}_H^{(M)}$ is therefore the empirical signature of cardiac entropy-cost cancellation. In physical terms, the cardiac invariant means that one unit of cardiac biological time carries nearly the same entropy cost per unit mass across a broad range of mammalian body sizes. Figure~\ref{fig:C1} shows that the mass-specific cardiac entropy cost remains essentially flat across nearly six orders of magnitude in body mass, consistent with the predicted $\bar{\sigma}_H^{(M)}\propto M^0$ scaling.

The same cancellation gives the thermodynamic interpretation of the lifetime heartbeat count. Using the allometric estimates above yields the \emph{a priori} reference cardiac budget
\begin{equation}
N_{H,0}^{\rm (allo)}
\simeq
1.5\times10^9
\quad
\mathrm{beats}.
\label{eq:NH_value}
\end{equation}
We stress that two distinct reference budgets appear in this paper, and we keep them notationally separate throughout. Equation~\eqref{eq:NH_value} is the \emph{a priori} estimate obtained by inserting the canonical resting allometries ($a\simeq3.4\,\mathrm{W\,kg^{-3/4}}$, $b\simeq241\,M^{-1/4}\,\mathrm{min^{-1}}$, $T\simeq310\,$K) into the closure; it carries no fitting and serves only to show that the lifetime-heartbeat invariant emerges naturally from mass cancellation. For all clade-level lifespan \emph{predictions} (Section~\ref{sec:clade_applications} and Supplementary Sec.~\ref{app:clade_multiplier}) we instead use the \emph{fitted empirical anchor}
\begin{equation}
N_{H,0}^{\rm (emp)}
=
10^{\,8.995}
\simeq
9.9\times10^{8}
\quad
\mathrm{beats},
\label{eq:NH_emp_anchor}
\end{equation}
the measured mean of $\log_{10}N_H^\star$ over the $n=46$ non-primate placental species (Eq.~Supplementary Eq.~\eqref{eq:appC_baseline}, Supplementary Table~\ref{tab:dataset_summary}). The two differ by about $50\%$ ($0.18\,$dex, where one dex is a factor of ten in $\log_{10}$) because the \emph{a priori} estimate uses idealized exponents while the empirical anchor absorbs the residual scatter of real allometries; using the fitted value for prediction avoids importing that idealization into the comparisons with observed lifespans. Where the round figure ``$N_0=10^9$'' appears in worked examples it is shorthand for $N_{H,0}^{\rm (emp)}$, to which all predicted lifespans are referred.

The lifetime cardiac invariant is therefore not introduced as a numerical coincidence. It follows from the compensation among metabolic power, cardiac frequency, and body mass. Small mammals dissipate energy rapidly and beat their hearts rapidly. Large mammals dissipate energy more slowly per unit mass and beat their hearts more slowly. These trends cancel in such a way that the mass-normalized entropy cost per heartbeat remains approximately conserved.

The cardiac rhythm also defines a biological proper time. Let
\begin{equation}
\theta_H(t)
=
\int_0^t f_H(t')\,dt'
\label{eq:thetaH}
\end{equation}
denote the accumulated number of cardiac cycles up to chronological time $t$. The corresponding normalized cardiac age is
\begin{equation}
\hat{\theta}_H(t)
=
\frac{\theta_H(t)}{N_H^\star}.
\label{eq:thetaH_hat}
\end{equation}
Species with high heart rates advance rapidly in cardiac biological time, whereas species with low heart rates advance more slowly. Normalization by $N_H^\star$ places organisms with very different chronological lifespans on a common cardiac-time coordinate.

\begin{figure}[t]
\centering
\includegraphics[width=\columnwidth]{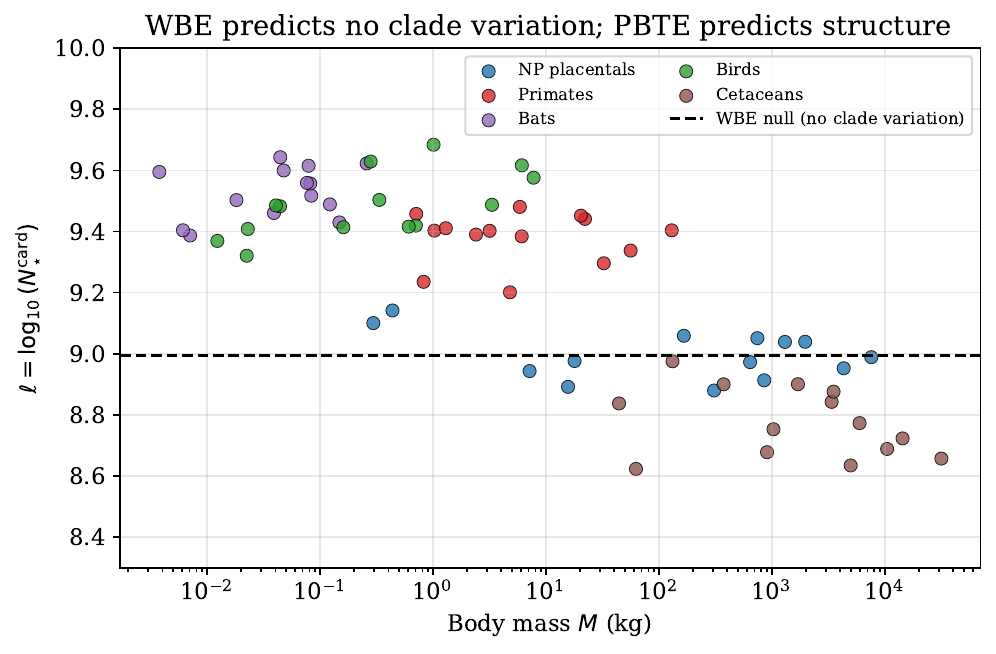}
\caption{Lifetime cardiac cycle count $\ell=\log_{10}(N_H^\star)$ versus body mass. PBTE predicts flat within-clade profiles with clade-dependent vertical offsets $\Delta\ell=\log_{10}\Phi_C$.}
\label{fig:C3}
\end{figure}

Finally, the cardiac clock exposes structured clade-level variation around the baseline invariant. We write
\begin{equation}
N_H^\star
=
N_{H,0}\Phi_C ,
\label{eq:cardiac_phi}
\end{equation}
where $\Phi_C$ is the clade multiplier. In PBTE, vertical offsets in $\log_{10}N_H^\star$ are not treated as residual scatter around a universal allometric line. They are interpreted as changes in the effective entropy cost per heartbeat, the accessible lifetime entropy budget, or the physiological pacing of biological time. Figure~\ref{fig:C3} displays the lifetime cardiac cycle count against body mass, showing the flat within-clade profiles and clade-dependent vertical offsets $\Delta\ell=\log_{10}\Phi_C$ predicted by the framework.

The main conclusion of this subsection is that the cardiac entropy cost becomes invariant in its mass-specific form. The absolute entropy cost per heartbeat, $\sigma_H$, increases with body size because a heartbeat in a large organism supports a larger physiological mass. The PBTE invariant is therefore the mass-normalized quantity $\bar{\sigma}_H^{(M)}=\sigma_H/M$. After allometric cancellation, $\bar{\sigma}_H^{(M)}$ is approximately invariant across body size. This distinction is essential. PBTE does not claim that all mammals spend the same total entropy per heartbeat. It claims that one heartbeat carries an invariant entropy cost per unit body mass. The cardiac clock therefore provides a clean realization of biological proper time: heartbeats count physiological progression, while $\bar{\sigma}_H^{(M)}$ defines the invariant thermodynamic cost of each cardiac tick on a mass-normalized scale. The respiratory clock examined next provides an independent, but more protocol-sensitive, test of the same principle.

\subsection{The respiratory clock}
\label{sec:resp_clock}

We next apply PBTE to the respiratory rhythm, which provides an independent physiological clock against which the cardiac construction can be tested. For this clock, the relevant frequency is the resting breathing frequency, $f=f_R$, and the entropy cost per cycle is the entropy cost per breath, denoted by $\sigma_R$. In the homeostatic regime, the entropy-production rate is estimated by $\dot e_p\simeq P/T$, where $P$ is metabolic power and $T$ is body temperature. The entropy cost per breath is therefore
\begin{equation}
\sigma_R
=
\frac{P}{T f_R}.
\label{eq:sigmaR}
\end{equation}
When the respiratory frequency is reported in breaths per minute, this becomes
\begin{equation}
\sigma_R
=
\frac{60P}{T f_{R,\mathrm{bpm}}},
\qquad
\bar{\sigma}_R^{(M)}
=
\frac{\sigma_R}{M}
=
\frac{60P}{T f_{R,\mathrm{bpm}}M}.
\label{eq:sigmaR_mass}
\end{equation}
The relevant PBTE quantity is again the mass-specific entropy cost, $\bar{\sigma}_R^{(M)}$. The absolute cost $\sigma_R$ depends on body size because a breath in a large organism supports a larger physiological mass. By contrast, $\bar{\sigma}_R^{(M)}$ measures the entropy cost of one respiratory cycle per unit body mass. This is the quantity expected to become approximately invariant if metabolic power and respiratory frequency scale in compensating ways.

The leading cancellation has the same structure as in the cardiac case. Under Kleiber metabolic scaling, $P\propto M^{3/4}$, and the approximate respiratory allometry $f_R\propto M^{-1/4}$~\cite{mortola2006}, consistent with the broader metabolic theory of ecology~\cite{brown2004}, Eq.~\eqref{eq:sigmaR_mass} gives
\begin{equation}
\bar{\sigma}_R^{(M)}
\propto
\frac{M^{3/4}}{M^{-1/4}M}
=
M^0 .
\label{eq:resp_cancel}
\end{equation}
Thus PBTE predicts that the entropy cost of one breath per unit body mass is approximately independent of body size. This does not imply that all species dissipate the same total entropy per breath. Rather, it means that the respiratory tick carries an approximately invariant entropy cost after normalization by body mass.

\begin{figure}[t]
\centering
\includegraphics[width=\columnwidth]{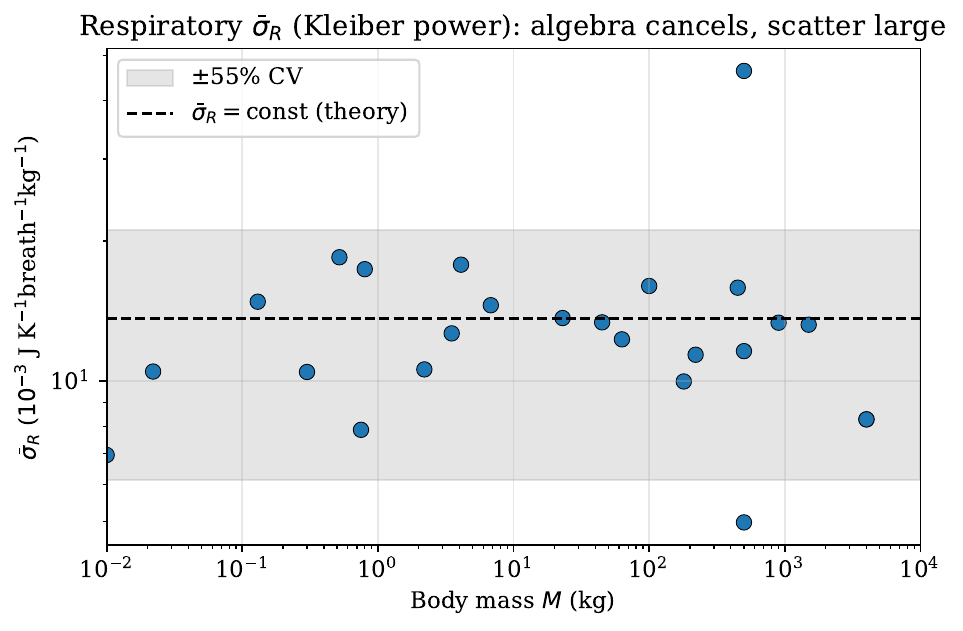}
\caption{Mass-specific respiratory entropy cost $\bar{\sigma}_R^{(M)}$ estimated with Kleiber metabolic power. The leading allometric exponents cancel, yielding an approximately scale-invariant respiratory entropy cost per breath per unit mass, although the scatter is larger than in the cardiac clock.}
\label{fig:R1}
\end{figure}

For the respiratory analysis we use the $65$-species subset of the full
cardiac dataset (Supplementary Sec.~\ref{app:resp_data}) for which a reliable
resting breath rate $f_R$ is available in addition to $f_H$; these
species and their $f_R$ values are flagged in Supplementary Data
File~1. For this subset the non-primate placental baseline is
\begin{equation}
\ell_0
=
\log_{10}N_R^\star
=
8.417\pm0.177,
\qquad
N_{R,0}^\star
\simeq
2.6\times10^8
\quad
\mathrm{breaths}.
\label{eq:resp_baseline}
\end{equation}
This value defines the reference respiratory cycle budget. It is smaller than the cardiac cycle budget because breaths occur less frequently than heartbeats, but it plays the same conceptual role in PBTE: it measures the number of respiratory ticks through which biological proper time is accumulated. Figure~\ref{fig:R1} shows the mass-specific respiratory entropy cost estimated with Kleiber metabolic power, in which the leading allometric exponents cancel to give an approximately scale-invariant cost per breath per unit mass, albeit with larger scatter than the cardiac clock.

The respiratory clock is a more stringent and more protocol-sensitive test of PBTE than the cardiac clock. This is because the cancellation in Eq.~\eqref{eq:resp_cancel} can be partly imposed by using Kleiber-derived metabolic power. To reduce this circularity, we also compute $\sigma_R$ using measured species-level basal metabolic rates wherever available~\cite{white2003,genoud2018}, recognizing that field and total energy throughput can depart substantially from basal values~\cite{pontzer2014}. In that analysis, $P$, $f_R$, $T$, and $M$ enter as independent empirical quantities.

Two conclusions follow, and they differ from the Kleiber-power case. First, replacing Kleiber power with measured metabolic power does \emph{not} reduce the scatter; it increases it markedly, to a coefficient of variation of order $100\%$. Second, the measured analysis reveals a clear positive mass dependence rather than near-cancellation. The slope of $\log_{10}\bar{\sigma}_R^{(M)}$ versus $\log_{10}M$ changes from approximately zero under Kleiber scaling to about $+0.21$ when measured basal metabolic rates are used, although on $n=29$ species this slope is not statistically resolved ($p\simeq0.11$). The mass-specific respiratory entropy cost therefore \emph{rises} with body size on the measured data, contrary to the cardiac case. This positive trend is dominated by aquatic mammals: marine species breathe far more slowly for their metabolic rate than terrestrial species of similar size (the ``aquatic breathing strategy''), so the entropy cost per breath, $\sigma_R=P/(Tf_R)$, is strongly inflated at the large-mass, aquatic end of the sample.

\begin{figure}[t]
\centering
\includegraphics[width=\columnwidth]{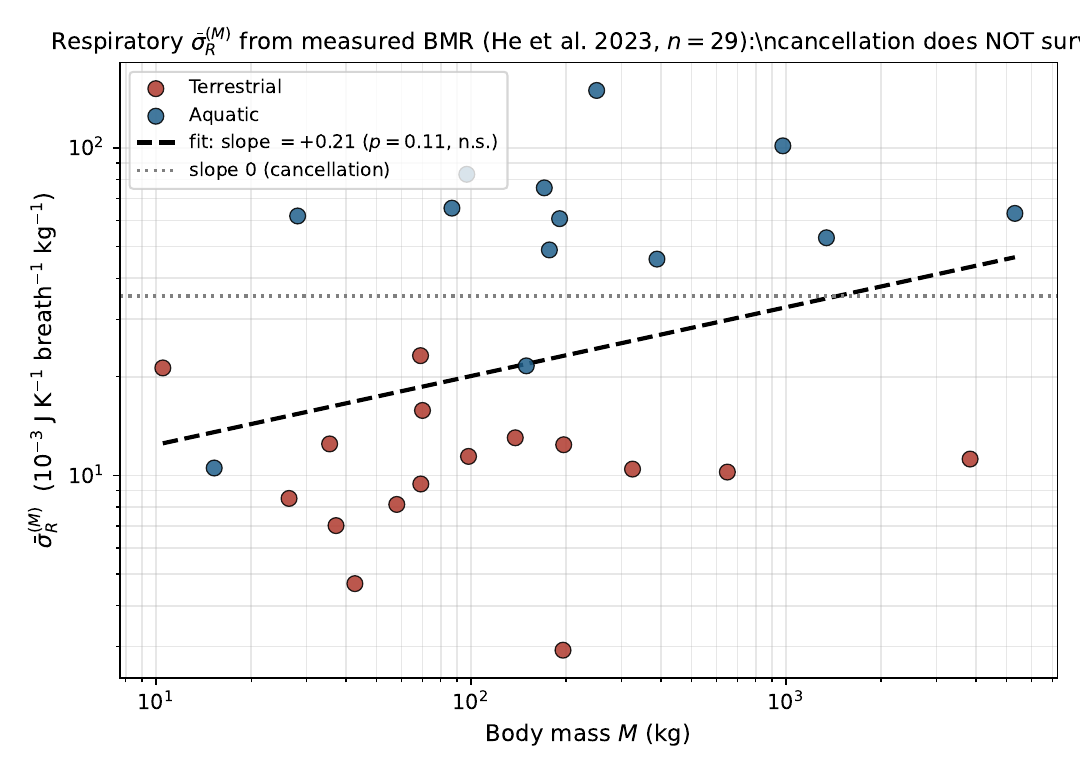}
\caption{\textbf{The non-circular test of the respiratory clock: mass cancellation does not survive measured BMR.}
Mass-specific respiratory entropy cost $\bar{\sigma}_R^{(M)}$ computed from
\emph{measured} species-level basal metabolic rates (He et al.\ 2023), in which $P$, $f_R$,
$T$, and $M$ enter as independent empirical quantities rather than through
imposed Kleiber scaling. Unlike Figure~\ref{fig:R1}, where flatness is
partly guaranteed by construction ($P\propto M^{3/4}$), here no canonical
exponent is assumed. On the $n=29$ mammals with both measured BMR and measured
resting breath rate, the fitted slope is $+0.21$ (cf.\ $+0.005$ under
Kleiber power); it is not statistically resolved ($p\simeq0.11$), and the
scatter is large (coefficient of variation $\sim\!100\%$). The cancellation seen
under Kleiber power therefore does not persist under independent inputs. The
positive trend is dominated by aquatic mammals (blue), whose low resting breath
rate relative to metabolic rate inflates $\sigma_R$ at large body mass. This is
the central empirical result of the respiratory analysis: the respiratory clock
fails the non-circular cancellation test that the cardiac clock passes. Dataset
and regression details are given in Supplementary Sec.~\ref{app:respiratory_measured}.}
\label{fig:R_measured}
\end{figure}

We stress that this comparison is the central empirical test of the respiratory clock, not a minor methodological aside. Mass cancellation computed from Kleiber-derived power is partly circular: because $P\propto M^{3/4}$ is imposed, a flat $\bar{\sigma}_R^{(M)}(M)$ is guaranteed by construction and carries no independent evidential weight. The decisive question is whether the cancellation survives when $P$ is taken from measured basal metabolic rates that do not assume the canonical exponent. On the He et al.\ (2023) dataset it does \emph{not}, as Figure~\ref{fig:R_measured} shows directly: the fitted slope is $+0.21$ (cf.\ $+0.005$ under Kleiber power), the scatter \emph{increases} to a coefficient of variation of order $100\%$, and the slope is not statistically resolved on the $n=29$ matched species ($p\simeq0.11$). The respiratory mass cancellation is therefore an artefact of imposing Kleiber scaling; once $P$ is supplied by independent measurement, it disappears. We regard this as an informative negative result rather than a failure of the framework as a whole. The cardiac clock, where measured resting allometries do yield mass cancellation (Section~\ref{subsec:cardiac_clock}, Figure~\ref{fig:C1}), provides the clean realization of the entropy-cost invariant; the respiratory clock does not, because breathing is a strongly regulated control variable---governed by tidal volume, ventilation strategy, thermoregulation, and especially the aquatic versus terrestrial breathing dichotomy---rather than a passive metabolic tick. The asymmetry between the two clocks is itself the substantive finding: the Level-3 invariance claim is supported for the cardiac coordinate and refuted for the respiratory coordinate on currently available measured data. The matched 29-species dataset, its provenance, and the regression are documented in Supplementary Sec.~\ref{app:respiratory_measured}.

The respiratory invariant also exhibits systematic clade structure. We write
\begin{equation}
N_R^\star
=
N_{R,0}^\star\Phi_C^{(R)},
\label{eq:resp_phi}
\end{equation}
where $\Phi_C^{(R)}$ is the respiratory clade multiplier. This multiplier measures how far a given clade lies from the non-primate placental baseline. Primates, bats, birds, and cetaceans occupy distinct bands in $\log_{10}N_R^\star$. The observed variation is therefore not random scatter around a single universal respiratory count. It is structured variation associated with differences in physiology, ecological risk, thermal regulation, ventilation strategy, and life-history organization.

Once the clade multiplier is specified, the respiratory invariant can be inverted to estimate lifespan from breathing frequency:
\begin{equation}
L_{\mathrm{pred}}
=
\frac{N_{R,0}^\star\Phi_C^{(R)}}
{f_{R,\mathrm{bpm}}\,525960}.
\label{eq:Lpred_resp}
\end{equation}
Here $f_{R,\mathrm{bpm}}$ is measured in breaths per minute and $L_{\mathrm{pred}}$ is expressed in years. This equation makes the respiratory clock predictive. The measured breathing frequency supplies the physiological pace, whereas $\Phi_C^{(R)}$ supplies the clade-level correction to the accessible biological-time budget.

\begin{figure}[t]
\centering
\includegraphics[width=\columnwidth]{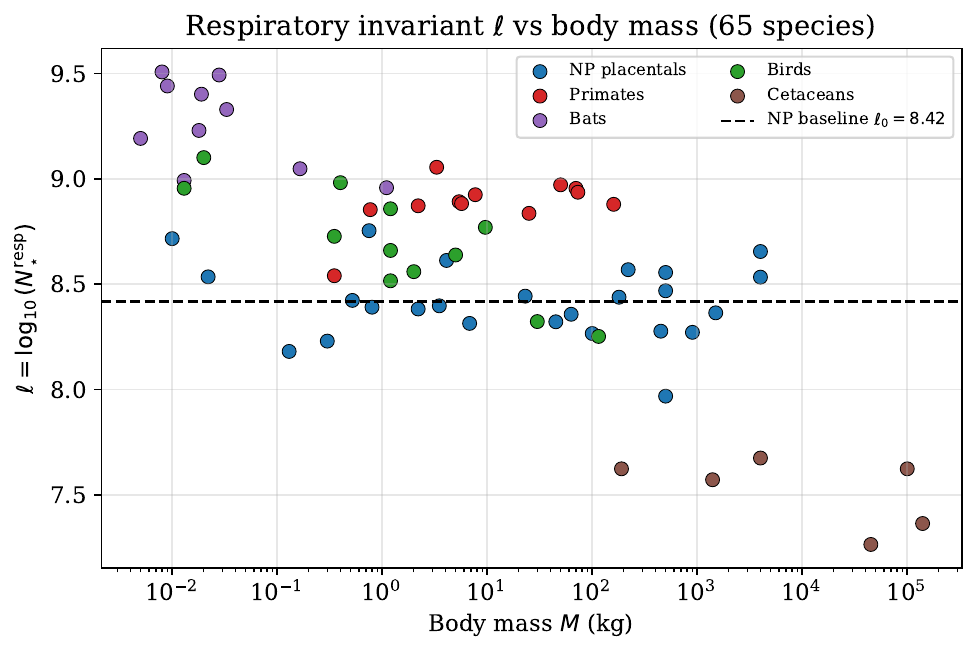}
\caption{Respiratory lifetime cycle count $\ell=\log_{10}(N_R^\star)$ versus body mass. The dominant pattern is clade-dependent vertical displacement rather than continuous mass dependence.}
\label{fig:R5}
\end{figure}

The main conclusion of this subsection is that the respiratory entropy cost becomes approximately invariant in its mass-specific form. The absolute entropy cost per breath, $\sigma_R$, increases with body size because each breath supports a larger physiological mass in larger organisms. The PBTE-relevant quantity is therefore $\bar{\sigma}_R^{(M)}=\sigma_R/M$, the entropy cost per breath per unit body mass. After allometric cancellation, this quantity is approximately scale-invariant, although less tightly than the corresponding cardiac quantity. As Figure~\ref{fig:R5} shows, the dominant pattern in the respiratory lifetime cycle count is clade-dependent vertical displacement rather than continuous mass dependence. The difference is biologically expected. Respiration is shaped not only by metabolic throughput, but also by tidal volume, ventilation strategy, thermoregulation, diving behavior, ecological state, and measurement protocol. The respiratory clock therefore supports PBTE as an independent biological-time coordinate, while also revealing where physiological complexity enters beyond the leading allometric cancellation.

\section{Clade-specific application of the PBTE multiplier}
\label{sec:clade_applications}

The detailed derivation of the clade multipliers, together with the full data analysis, is presented in Supplementary Sec.~\ref{app:clade_derivations}. In this section, we give the main results and emphasize how the PBTE multiplier is used to convert the reference cardiac cycle budget into quantitative lifespan predictions. Once the relevant cardiac frequency and physiological correction factors are specified, the clade multiplier maps the baseline invariant onto a clade-specific effective cycle budget. For consistency with the empirical cardiac dataset, we use the non-primate placental reference
\begin{equation}
N_{H,0}^{(\mathrm{emp})}
=
10^{8.995}
\simeq
9.9\times10^8
\quad
\mathrm{beats},
\label{eq:NH0_emp}
\end{equation}
where $8.995$ is the mean value of $\log_{10}N_H^\star$ for the non-primate placental baseline. For a cardiac frequency reported in beats per minute, the corresponding clade-level lifespan prediction is
\begin{equation}
L_{\mathrm{pred},C}
=
\frac{
N_{H,0}^{(\mathrm{emp})}\Phi_C
}{
f_{H,\mathrm{ref}}\,\mathcal{T}_{\mathrm{yr}}
},
\qquad
\mathcal{T}_{\mathrm{yr}}
=
525960\ \mathrm{min\,yr^{-1}} .
\label{eq:Lpred_clade}
\end{equation}
Here $f_{H,\mathrm{ref}}$ is the cardiac frequency appropriate to the physiological state used as the reference. For ordinary single-state organisms, it is the resting heart rate. For duty-cycled organisms, such as hibernating bats or diving cetaceans, the active or surface heart rate is used, while suppression of the effective biological clock is included through $\Phi_{\rm duty}$. This convention avoids double counting cardiac slowing and keeps the frequency measurement separate from the physiological multiplier.

\subsection{Non-primate placental reference: the mouse-to-elephant baseline}
\label{subsec:baseline_mammals}

The non-primate placental mammals define the reference PBTE class. In this
group no clade-level multiplier is introduced,
\begin{equation}
\Phi_{\rm NP}=1 .
\label{eq:Phi_NP}
\end{equation}
The predicted lifespan is therefore determined only by the resting cardiac
frequency,
\begin{equation}
L_{\mathrm{pred,NP}}
=
\frac{N_{H,0}^{(\mathrm{emp})}}
{f_H\,\mathcal{T}_{\mathrm{yr}}}.
\label{eq:Lpred_NP}
\end{equation}
This is the unrenormalized PBTE clock. Small mammals have large $f_H$ and
therefore consume the reference cardiac budget rapidly in chronological
time; large mammals have small $f_H$ and reach the same biological endpoint
over a longer chronological interval.

\begin{table}[t]
\centering
\small
\caption{Reference-class PBTE prediction for representative non-primate
placental mammals. All entries use
$N_{H,0}^{(\mathrm{emp})}=10^{8.995}$ beats and $\Phi_{\rm NP}=1$.}
\label{tab:baseline_mammal_predictions}
\begin{tabular}{lcccc}
\hline
Species &
$f_H$ (bpm) &
$\Phi_C$ &
$L_{\mathrm{pred}}$ (yr) &
$L_{\mathrm{obs}}$ (yr) \\
\hline
\emph{Mus musculus} & 632 & 1.00 & 3.0 & 3.5 \\
\emph{Rattus norvegicus} & 420 & 1.00 & 4.5 & 3.8 \\
\emph{Oryctolagus cuniculus} & 205 & 1.00 & 9.2 & 9.0 \\
\emph{Felis catus} & 150 & 1.00 & 12.5 & 15.0 \\
\emph{Canis lupus familiaris} & 90 & 1.00 & 20.9 & 20.0 \\
\emph{Equus caballus} & 38 & 1.00 & 49.5 & 46.0 \\
\emph{Loxodonta africana} & 28 & 1.00 & 67.1 & 65.0 \\
\emph{Elephas maximus} & 27 & 1.00 & 69.6 & 86.0 \\
\hline
\end{tabular}
\end{table}

Table~\ref{tab:baseline_mammal_predictions} shows that the same reference
cycle budget captures the correct lifespan scale across several orders of
magnitude in body mass. We emphasize that this table is an internal
\emph{consistency check}, not an independent prediction: because the
anchor $N_{H,0}^{(\mathrm{emp})}$ is itself the mean of
$\log_{10}N_H^\star$ over precisely these non-primate placentals
(Eq.~Supplementary Eq.~\eqref{eq:appC_baseline}), the agreement between $L_{\rm pred}$ and
$L_{\rm obs}$ tests only that individual placental species scatter
tightly about their own clade mean once $\Phi_{\rm NP}=1$ is imposed. The
genuinely predictive content of the framework lies in the
\emph{other} clades, whose multipliers $\Phi_C\neq1$ are fixed by
independent physiological inputs (Sections~\ref{subsec:clade_primates}--\ref{subsec:clade_cetaceans})
and then compared with observed lifespans that played no part in setting
the anchor. With that caveat, the mouse and elephant are not special
exceptions requiring separate mechanisms; they represent the canonical
PBTE cancellation. Their different chronological lifespans arise primarily
because the mouse advances through cardiac biological proper time rapidly,
whereas the elephant advances through the same coordinate slowly.

\subsection{Primates}
\label{subsec:clade_primates}

Primates are treated as single-state organisms in the cardiac-clock
approximation. Thus,
\begin{equation}
\Phi_{\rm duty}^{(\mathrm{prim})}=1 .
\label{eq:Phi_duty_prim}
\end{equation}
Their elevated lifetime cycle count is therefore interpreted primarily as a
reduction in the entropy cost per beat, rather than as a slowing of the
cardiac clock. We write
\begin{equation}
\Phi_{\rm prim}
=
\Phi_{\rm neuro}\Phi_{\rm thermal}\Phi_{\rm haz}.
\label{eq:Phi_prim}
\end{equation}
The dominant contribution is $\Phi_{\rm neuro}$, which represents the
reduction in entropy cost per beat associated with neural metabolic
investment, predictive regulation, and enhanced physiological maintenance.
The thermal correction is comparatively small because primate body
temperatures remain close to the mammalian reference.

\begin{table}[t]
\centering
\small
\caption{Representative primate lifespan predictions. All entries are
generated from the single power law
$\Phi_{\rm neuro}=(\varphi/\varphi_0)^{\alpha}$ with the canonical
calibration $\alpha=0.40$, $\beta=3$, $\varphi_0=0.02$, and the empirical
anchor $N_{H,0}^{(\mathrm{emp})}=10^{8.995}$ beats, via
Eq.~\eqref{eq:Lpred_clade} with
$\Phi_C=\Phi_{\rm neuro}\Phi_{\rm thermal}$ and $\Phi_{\rm haz}=1$. The
same calibration is used in Supplementary Sec.~\ref{app:phiC_primates},
Supplementary Table~\ref{tab:appC_primates}(a); every value here is reproducible by
inspection from $f_H$, $\varphi$, and $T_b$.}
\label{tab:primate_predictions}
\resizebox{\columnwidth}{!}{%
\begin{tabular}{lcccccc}
\hline
Species & $f_H$ (bpm) & $\Phi_{\rm neuro}$ & $\Phi_{\rm thermal}$ &
$\Phi_C$ & $L_{\mathrm{pred}}$ (yr) & $L_{\mathrm{obs}}$ (yr) \\
\hline
\emph{Callithrix jacchus} & 220 & 1.55 & 1.01 & 1.56 & 13.3 & 10--17 \\
\emph{Macaca mulatta} & 120 & 1.65 & 1.01 & 1.67 & 26.1 & 25--40 \\
\emph{Pan troglodytes} & 75 & 2.05 & 1.03 & 2.11 & 52.8 & 45--60 \\
\emph{Gorilla gorilla} & 65 & 1.83 & 1.03 & 1.88 & 54.3 & 40--55 \\
\emph{Homo sapiens} & 70 & 2.51 & 1.04 & 2.60 & 69.8 & 70--85 \\
\hline
\end{tabular}%
}
\end{table}

Table~\ref{tab:primate_predictions} shows that primate lifespan extension
is captured by increasing the effective cardiac cycle budget rather than by
reducing heart rate. In PBTE terminology, primate longevity is therefore a
budget-expansion mechanism: more cardiac cycles can be realized because the
entropy burden per cycle is reduced.

\subsection{Bats}
\label{subsec:clade_bats}

Bats realize an extended cardiac budget through intermittent physiological
suppression. In temperate hibernating bats, a substantial fraction of the
year is spent in torpor, where both heart rate and body temperature are
strongly reduced. The effective multiplier is
\begin{equation}
\Phi_{\rm bat}
=
\Phi_{\rm duty}\Phi_{\rm thermal}\Phi_{\rm haz}.
\label{eq:Phi_bat}
\end{equation}
The duty-cycle factor accounts for the slowing of cardiac biological time,
the thermal factor accounts for Arrhenius suppression of damage kinetics,
and the hazard factor accounts for ecological truncation or protection.

For a torpor fraction $q$, active-phase heart rate $f_{H,\mathrm{act}}$,
and torpid heart rate $f_{H,\mathrm{tor}}$, the duty factor is
\begin{equation}
\Phi_{\rm duty}^{(\mathrm{bat})}
=
\left[
(1-q)
+
q\frac{f_{H,\mathrm{tor}}}{f_{H,\mathrm{act}}}
\right]^{-1}.
\label{eq:Phi_bat_duty}
\end{equation}
The active-phase rate $f_{H,\mathrm{act}}$ is used as
$f_{H,\mathrm{ref}}$ in Eq.~\eqref{eq:Lpred_clade}, because torpor-induced
slowing is already included through $\Phi_{\rm duty}$.

\begin{table}[t]
\centering
\small
\caption{Representative bat lifespan predictions. For bats,
$f_{H,\mathrm{ref}}$ is the active-phase heart rate; torpor-induced slowing
is included through $\Phi_{\rm duty}$.}
\label{tab:bat_predictions}
\resizebox{\columnwidth}{!}{%
\begin{tabular}{lccccccc}
\hline
Species & $f_{H,\mathrm{ref}}$ (bpm) & $\Phi_{\rm duty}$ &
$\Phi_{\rm thermal}$ & $\Phi_{\rm haz}$ & $\Phi_C$ &
$L_{\mathrm{pred}}$ (yr) & $L_{\mathrm{obs}}$ (yr) \\
\hline
\emph{Myotis lucifugus} & 300 & 1.94 & 4.10 & 0.68 & 5.40 & 33.8 & 34 \\
\emph{Eptesicus fuscus} & 280 & 1.79 & 4.54 & 0.35 & 2.84 & 19.1 & 19 \\
\emph{Pteropus vampyrus} & 250 & 1.07 & 1.22 & 1.60 & 2.09 & 15.7 & 15--23 \\
\hline
\end{tabular}%
}
\end{table}

Table~\ref{tab:bat_predictions} illustrates the multiplicative character of
the bat strategy. Duty-cycle slowing alone is insufficient, and thermal
suppression alone is insufficient; the observed lifespan scale emerges from
their product, modified by extrinsic hazard.

\subsection{Birds}
\label{subsec:clade_birds}

Birds provide a contrasting case because two leading factors are adverse:
elevated body temperature accelerates damage kinetics, and flight increases
cardiac workload. The avian multiplier is written as
\begin{equation}
\Phi_{\rm bird}
=
\Phi_{\rm duty}\Phi_{\rm thermal}\Phi_{\rm mito\mbox{-}oxid}\Phi_{\rm haz}.
\label{eq:Phi_bird}
\end{equation}
Because $\Phi_{\rm duty}$ and $\Phi_{\rm thermal}$ are typically less than
unity, avian longevity requires compensating reductions in entropy cost per
cycle through mitochondrial efficiency, oxidative resistance, cellular
repair, and reduced ecological hazard.

\begin{table}[t]
\centering
\small
\caption{Representative bird lifespan predictions. The first row gives the
mechanistic small-passerine calculation. The remaining rows use the
representative avian multiplier to illustrate the cardiac lifespan scale for
longer-lived birds.}
\label{tab:bird_predictions}
\begin{tabular}{lcccc}
\hline
Species or case & $f_H$ (bpm) & $\Phi_C$ &
$L_{\mathrm{pred}}$ (yr) & $L_{\mathrm{obs}}$ (yr) \\
\hline
Small passerine & 320 & 2.97 & 17.4 & 10--20 \\
\emph{Columba livia} & 190 & 3.41 & 33.7 & 35 \\
\emph{Diomedea exulans} & 100 & 3.41 & 64.1 & 70 \\
\emph{Aquila chrysaetos} & 130 & 3.41 & 49.3 & 46 \\
\hline
\end{tabular}
\end{table}

Table~\ref{tab:bird_predictions} shows that birds do not extend lifespan by
simply slowing biological time. Instead, adverse thermal and duty-cycle
effects are overcome by a larger biochemical and ecological multiplier. In
PBTE terminology, the avian strategy is primarily an entropy-cost reduction
mechanism rather than a pure clock-slowing mechanism.

\subsection{Cetaceans}
\label{subsec:clade_cetaceans}

For cetaceans, the dominant correction is sustained diving bradycardia. The
surface heart rate is used as the reference frequency, while the reduced
cardiac rate during dives is included through the duty-cycle factor. The
cetacean multiplier is written as
\begin{equation}
\Phi_{\rm cet}
=
\Phi_{\rm duty}\Phi_{\rm thermal}\Phi_{\mathrm{O}_2}\Phi_{\rm haz},
\label{eq:Phi_cet}
\end{equation}
where $\Phi_{\mathrm{O}_2}$ represents oxygen-storage and dive-associated
protection.

For dive fraction $p_d$, surface heart rate $f_{H,\mathrm{surf}}$, and dive
heart rate $f_{H,\mathrm{dive}}$, the duty factor is
\begin{equation}
\Phi_{\rm duty}^{(\mathrm{cet})}
=
\left[
(1-p_d)
+
p_d\frac{f_{H,\mathrm{dive}}}{f_{H,\mathrm{surf}}}
\right]^{-1}.
\label{eq:Phi_cet_duty}
\end{equation}
As in the bat case, using a time-averaged dive-corrected heart rate and
also applying $\Phi_{\rm duty}$ would count bradycardia twice.

\begin{table}[t]
\centering
\small
\caption{Representative cetacean lifespan predictions. For cetaceans,
$f_{H,\mathrm{ref}}$ is the surface cardiac rate; dive bradycardia is
included through $\Phi_{\rm duty}$.}
\label{tab:cetacean_predictions}
\resizebox{\columnwidth}{!}{%
\begin{tabular}{lcccccccc}
\hline
Species & $f_{H,\mathrm{surf}}$ (bpm) & $\Phi_{\rm duty}$ &
$\Phi_{\rm thermal}$ & $\Phi_{\mathrm{O}_2}$ & $\Phi_{\rm haz}$ &
$\Phi_C$ & $L_{\mathrm{pred}}$ (yr) & $L_{\mathrm{obs}}$ (yr) \\
\hline
\emph{Balaenoptera musculus} & 37 & 2.70 & 1.17 & 1.40 & 0.50 & 2.21 & 112 & 80--90 \\
\emph{Balaena mysticetus} & 30 & 3.08 & 1.17 & 1.50 & 0.60 & 3.24 & 203 & 150--200 \\
\emph{Tursiops truncatus} & 80 & 1.50 & 1.09 & 1.20 & 0.65 & 1.28 & 30 & 40--50 \\
\hline
\end{tabular}%
}
\end{table}

Table~\ref{tab:cetacean_predictions} illustrates why raw heartbeat counts
in cetaceans can be misleading. A large fraction of life is spent in
bradycardic dive states, so the surface heart rate alone does not represent
the lifetime cardiac clock. After duty-cycle correction, cetacean longevity
is interpreted primarily as a time-dilation mechanism: biological proper
time advances more slowly during prolonged diving physiology.

The tables above are intended as representative applications of the PBTE
multiplier, not as species-level fitted models. Their purpose is to show how
a single reference cardiac budget can be renormalized by clade-specific
physiological mechanisms. Detailed derivations of the duty-cycle identities,
thermal factors, biochemical factors, and hazard corrections are given in
Supplementary Sec.~\ref{app:clade_multiplier} and in the Supplementary Material.

% =====================================================================
\section{Biological proper time and the thermodynamics of aging}
\label{sec:aging}

The cardiac and respiratory analyses above establish a lifetime cycle
budget $N_\star$ and an entropy cost per cycle $\langle\sigma_0\rangle$.
This section promotes those lifetime totals to an internal coordinate for
the \emph{progression} of an organism through its life, and develops the
thermodynamic theory of aging that follows. The construction is the same
entropy accounting used above, now read as a function of age rather than
as a lifetime total. It connects the present cardiac/respiratory results
to the PBTE treatment of aging and longevity.

\paragraph{Thermodynamic foundation.}
A living organism is an open nonequilibrium system that maintains
macroscopic order by continuously transforming chemical free energy into
mechanical work, ion gradients, biosynthesis, molecular repair,
signaling, and heat~\cite{schrodinger1944,prigogine1967}. It is therefore
not an isolated body that passively degrades in calendar time, but a
regulated dissipative system whose internal state is sustained by
continuous thermodynamic throughput. With $S(t)$ the coarse-grained
internal entropy, $\dot e_p(t)\ge0$ the irreversible entropy-production
rate, and $\dot h_d(t)\ge0$ the entropy exported to the environment, the
open-system balance is
\begin{equation}
\frac{dS}{dt}=\dot e_p(t)-\dot h_d(t).
\label{eq:aging_balance}
\end{equation}
In adult homeostasis the organism is held near a nonequilibrium steady
state in which internal entropy is regulated within a bounded
physiological range. On timescales long compared with metabolic
fluctuations but short compared with lifespan-scale deterioration,
Eq.~\eqref{eq:aging_balance} gives
\begin{equation}
\frac{dS}{dt}\simeq0,
\qquad
\dot e_p(t)\simeq\dot h_d(t),
\label{eq:aging_homeostasis}
\end{equation}
and the entropy exported as heat is estimated to leading order by the
dissipated metabolic power divided by absolute body temperature,
\begin{equation}
\dot h_d(t)\simeq\frac{P(t)}{T(t)} .
\label{eq:aging_heat}
\end{equation}
Combining Eqs.~\eqref{eq:aging_homeostasis} and \eqref{eq:aging_heat}
gives the homeostatic closure
\begin{equation}
\dot e_p(t)\simeq\dot h_d(t)\simeq\frac{P(t)}{T(t)} .
\label{eq:aging_closure}
\end{equation}
Aging becomes visible when this balance is no longer exact. If entropy
export, molecular repair, proteostasis, immune regulation, and cellular
renewal fail to fully compensate irreversible entropy production, the
coarse-grained internal burden associated with disorder, damage, and loss
of regulation accumulates. Integrating Eq.~\eqref{eq:aging_balance} from
the initial life-history time to age $t$,
\begin{equation}
S(t)=S(0)+\int_0^t\bigl[\dot e_p(s)-\dot h_d(s)\bigr]\,ds .
\label{eq:aging_Saccum}
\end{equation}
PBTE does not identify this coarse-grained entropy with a single
biomarker; instead, entropy production is used to assign a thermodynamic
cost to the advancement of an internal physiological clock.

\paragraph{Internal-time coordinate and entropy cost per cycle.}
Let $f(t)$ be a physiological frequency---heart rate, respiratory rate, or
another biologically meaningful rhythm. The accumulated biological proper
time is the total number of physiological cycles completed up to age $t$,
already introduced in Eq.~\eqref{eq:theta_def},
\begin{equation}
\theta(t)=\int_0^t f(s)\,ds,
\qquad
d\theta=f(t)\,dt .
\label{eq:aging_theta}
\end{equation}
Thus $\theta$ is an internal cycle coordinate, not an external time
coordinate. The entropy produced in the same infinitesimal interval is
$d\Sigma=\dot e_p(t)\,dt$; using $dt=d\theta/f(t)$ this becomes
\begin{equation}
d\Sigma=\frac{\dot e_p(t)}{f(t)}\,d\theta ,
\label{eq:aging_dSigma}
\end{equation}
which identifies the instantaneous entropy cost per biological tick as
\begin{equation}
\sigma_0(t)=\frac{\dot e_p(t)}{f(t)}=\frac{d\Sigma}{d\theta},
\label{eq:aging_sigma0}
\end{equation}
the entropy produced per physiological cycle, i.e.\ the thermodynamic
price of one increment of biological proper time. A cycle is more
expensive when entropy production is high relative to the physiological
frequency, and less expensive when the same cycle is executed with lower
irreversible dissipation. Over the lifespan $L$, with total entropy
production $\Sigma=\int_0^L\dot e_p\,dt$ and total cycle count
$N_\star=\theta(L)=\int_0^L f\,dt$, the relation $d\Sigma=\sigma_0\,d\theta$
gives $\Sigma=\int_0^{N_\star}\sigma_0(\theta)\,d\theta$, so the
lifetime-average cost per tick is
\begin{equation}
\langle\sigma_0\rangle
=\frac{1}{N_\star}\int_0^{N_\star}\sigma_0(\theta)\,d\theta
=\frac{\Sigma}{N_\star},
\label{eq:aging_avgsigma}
\end{equation}
and rearranging recovers the PBTE cycle-count relation
\begin{equation}
\boxed{\,N_\star=\frac{\Sigma}{\langle\sigma_0\rangle}\,}.
\label{eq:aging_cyclecount}
\end{equation}
This is an accounting identity with biological content: the number of
cycles completed over life equals the total entropy produced divided by
the average entropy cost of one cycle. It does not assert death at an
exact universal cycle number; it states that biological time advances by
consuming an entropy-normalized cycle budget. In the constant limit
$f(t)=f$, $\sigma_0(t)=\sigma_0$ one has $N_\star=fL$ and $L=N_\star/f$, so
for a fixed budget a higher physiological frequency implies a shorter
chronological lifespan. The metabolic closure
Eq.~\eqref{eq:aging_closure} gives the complementary thermodynamic form
\begin{equation}
\sigma_0(t)\simeq\frac{P(t)}{T(t)\,f(t)} ,
\label{eq:aging_sigma_metabolic}
\end{equation}
so the cost of one cycle is set by metabolic power per unit temperature
per unit frequency.

\paragraph{Biological age as a normalized internal-time coordinate.}
The simplest PBTE biological age is the normalized internal time of
Eq.~\eqref{eq:theta_def},
\begin{equation}
A_\theta(t)=\frac{\theta(t)}{N_\star},
\label{eq:aging_Atheta}
\end{equation}
running from $A_\theta=0$ at birth to $A_\theta=1$ at the terminal PBTE
boundary; for constant $f$ it reduces to $A_\theta(t)=ft/N_\star=t/L$.
Cycle count alone, however, is not sufficient. The same heartbeat or
breath can carry a different thermodynamic cost depending on mitochondrial
efficiency, oxidative load, inflammation, repair capacity, temperature,
and disease. To carry this information, PBTE weights each cycle by its
entropy cost through the entropy-normalized biological time
\begin{equation}
\Theta_\sigma(t)
=\int_0^t\frac{\sigma_0(s)}{\sigma_{0,\mathrm{ref}}}\,f(s)\,ds ,
\label{eq:aging_Thetasigma}
\end{equation}
where $\sigma_{0,\mathrm{ref}}$ is a reference entropy cost per cycle. A
cycle with $\sigma_0=\sigma_{0,\mathrm{ref}}$ contributes one reference
tick; a more expensive cycle contributes more than one, a cheaper cycle
less. Using $\sigma_0(t)=\dot e_p(t)/f(t)$ the frequency cancels and
\begin{equation}
\Theta_\sigma(t)
=\int_0^t\frac{\dot e_p(s)}{\sigma_{0,\mathrm{ref}}}\,ds
=\frac{\Sigma(t)}{\sigma_{0,\mathrm{ref}}},
\qquad
\Sigma(t)=\int_0^t\dot e_p(s)\,ds ,
\label{eq:aging_Thetasigma_entropy}
\end{equation}
so $\Theta_\sigma$ is both the number of entropy-weighted cycles and the
accumulated entropy production in units of the reference cost. The PBTE
biological age is the normalized coordinate
\begin{equation}
A_{\mathrm{PBTE}}(t)
=\frac{\Theta_\sigma(t)}{N_{\star,\mathrm{ref}}}
=\frac{1}{N_{\star,\mathrm{ref}}}
\int_0^t\frac{\sigma_0(s)}{\sigma_{0,\mathrm{ref}}}\,f(s)\,ds ,
\label{eq:aging_APBTE}
\end{equation}
with $N_{\star,\mathrm{ref}}$ the reference entropy--cycle budget.
Defining $\Sigma_{\mathrm{ref}}=\sigma_{0,\mathrm{ref}}\,
N_{\star,\mathrm{ref}}$ and using
Eq.~\eqref{eq:aging_Thetasigma_entropy} gives the compact form
\begin{equation}
\boxed{\,A_{\mathrm{PBTE}}(t)=\frac{\Sigma(t)}{\Sigma_{\mathrm{ref}}}\,},
\label{eq:aging_APBTE_budget}
\end{equation}
so PBTE biological age is the fraction of a reference lifetime
entropy--cycle budget already consumed. Differentiating
Eq.~\eqref{eq:aging_APBTE} gives the instantaneous aging rate,
\begin{equation}
\frac{dA_{\mathrm{PBTE}}}{dt}
=\frac{1}{N_{\star,\mathrm{ref}}}
\frac{\sigma_0(t)}{\sigma_{0,\mathrm{ref}}}\,f(t)
=\frac{\dot e_p(t)}{\sigma_{0,\mathrm{ref}}N_{\star,\mathrm{ref}}}
=\frac{\dot e_p(t)}{\Sigma_{\mathrm{ref}}},
\label{eq:aging_rate}
\end{equation}
where $\sigma_0 f=\dot e_p$, and with the closure
Eq.~\eqref{eq:aging_closure},
\begin{equation}
\frac{dA_{\mathrm{PBTE}}}{dt}\simeq\frac{P(t)}{T(t)\,\Sigma_{\mathrm{ref}}} .
\label{eq:aging_rate_metabolic}
\end{equation}

\begin{center}\setlength{\fboxsep}{10pt}
\colorbox{black!4}{\begin{minipage}{0.92\linewidth}
\centering
\textbf{Central definition.}\\[4pt]
$\displaystyle
A_{\mathrm{PBTE}}(t)
=\frac{1}{N_{\star,\mathrm{ref}}}
\int_0^t\frac{\sigma_0(s)}{\sigma_{0,\mathrm{ref}}}\,f(s)\,ds
=\frac{\Sigma(t)}{\Sigma_{\mathrm{ref}}}$\\[4pt]
{\small\itshape biological age $=$ fraction of the reference
entropy--cycle budget consumed}
\end{minipage}}
\end{center}

Equations~\eqref{eq:aging_APBTE} and \eqref{eq:aging_rate} make the
thermodynamic content of aging explicit. An organism ages faster when its
physiological clock runs faster (large $f$), when each biological tick is
thermodynamically more expensive (large $\sigma_0$), or when both occur
together. It ages more slowly when physiological rate is reduced without
functional collapse, when entropy production per cycle is lowered by more
efficient maintenance, or when the accessible budget
$N_{\star,\mathrm{ref}}$ is enlarged. This is why chronological age is
sometimes predictive and sometimes misleading: calendar time is a faithful
proxy only when $f(t)$, $\sigma_0(t)$, and the effective budget are similar
across the organisms compared. When they differ---precisely the clade
structure documented in Sec.~\ref{sec:clade_applications}---two organisms
of identical chronological age occupy different positions on the
entropy-normalized internal-time axis. The clade multiplier $\Phi_C$ of
Eq.~\eqref{eq:intro_clade_count}, which rescales
$N_{\star,\mathrm{ref}}$, is therefore not only a lifespan-prediction
factor but the rescaling that maps chronological age onto biological age.

This coordinate organizes aging and longevity into three thermodynamic
mechanism classes, all visible in the clade results of this paper. The
first is \emph{time dilation}: reducing the rate $dA_{\mathrm{PBTE}}/dt$
at which biological proper time accumulates, without lowering the cost or
enlarging the budget. Caloric restriction, torpor, and the bradycardic
diving physiology of cetaceans act here, slowing the internal clock so
that fewer entropy--cycles are spent per calendar year and the same budget
is stretched over a longer chronological lifespan. In PBTE terms these
interventions reduce $f(t)$ or the effective duty cycle, and longevity is
bought by living more slowly in internal time rather than by changing the
total budget.

The second class is \emph{entropy-cost reduction and budget expansion}:
lowering $\sigma_0$ per cycle or raising $N_{\star,\mathrm{ref}}$ so that
more biological time can be traversed before the terminal boundary. Birds,
despite high body temperatures and metabolic rates, and primates, with
their enhanced cellular maintenance, realize this strategy through
superior mitochondrial efficiency, antioxidant protection, and repair
capacity. Each heartbeat is made thermodynamically cheaper, or more of
them are made affordable within the budget, so longevity is bought per
physiological cycle rather than per calendar year. The cardiac
entropy-cost invariant established in this paper is what makes
$\sigma_{0,\mathrm{ref}}$ a well-defined reference against which these
clade-specific reductions can be measured.

The third class is the \emph{hypertemporal} pathologies of aging:
chronic inflammation, metabolic syndrome, neurodegeneration, and cancer,
which raise $f(t)$, $\sigma_0(t)$, or both, and so accelerate
$dA_{\mathrm{PBTE}}/dt$. In this picture such states are not merely damage
endpoints but increases in the velocity of biological time, pushing the
organism through its entropy--cycle budget faster than chronological age
would suggest. The framework yields a falsifiable consequence:
biological-age biomarkers, especially DNA-methylation clocks, should
correlate more tightly with accumulated internal time
$A_{\mathrm{PBTE}}$ than with chronological age, and damage and mortality
trajectories should collapse more cleanly when plotted against the cardiac
$A_{\mathrm{PBTE}}$---whose reference cost survives the non-circular
test---than when plotted against chronological time or against the
respiratory clock that does not.

% =====================================================================

% =====================================================================
\section{Summary and conclusions}
\label{sec:summary_conclusions}

In this work we developed the Principle of Biological Time Equivalence (PBTE)
as a thermodynamic framework for interpreting lifetime physiological
cycle invariants in vertebrates. The central idea is that biological
time is not measured most naturally by chronological duration alone, but
by the accumulated number of recurrent physiological cycles weighted by
their entropy cost. In this formulation, lifespan is governed by how
rapidly an organism consumes an entropy-weighted cycle budget.

The cardiac clock provides the clearest realization of this principle.
Across representative mammals, the absolute entropy cost per heartbeat
increases strongly with body size, but the mass-specific entropy cost
per heartbeat remains approximately constant. This cancellation follows
from the compensation between metabolic power, cardiac frequency, and
body mass. As a result, small mammals and large mammals can have very
different chronological lifespans while still approaching a comparable
endpoint in cardiac biological proper time. The mouse-to-elephant
baseline is therefore not a collection of special cases, but the
canonical expression of the PBTE invariant.

The respiratory clock provides an independent, but coarser, test of the
same construction. The lifetime breath count is also approximately
invariant, and the mass-specific entropy cost per breath is much flatter
than the absolute entropy cost per breath. However, respiration shows
larger scatter than the cardiac clock. This is expected because
breathing is not only a clock, but also a regulated control variable for
gas exchange, tidal volume, thermoregulation, diving, and activity
state. The respiratory results therefore support PBTE while also showing
that not all physiological rhythms provide equally clean measures of
biological time.

The comparison between the cardiac and respiratory clocks gives an
important internal consistency check. A breath is a slower and coarser
cycle than a heartbeat, and therefore carries the entropy cost of
several cardiac cycles. In typical mammals this ratio is of order four
to five, consistent with the observed relation between respiratory and
cardiac frequencies. The same comparison clarifies why surface breath
counts can be misleading in diving mammals: in cetaceans, respiratory
and cardiac pacing are strongly modified by dive physiology, so the
cardiac clock remains the cleaner thermodynamic coordinate.

The clade-multiplier analysis extends the theory beyond the baseline
invariant. Non-primate placental mammals define the reference class.
Primates extend the effective cycle budget primarily by reducing the
entropy burden per cardiac cycle. Bats extend lifespan through the
combined effects of torpor, duty-cycle suppression, and thermal slowing.
Birds overcome adverse thermal and flight-related costs through
mitochondrial efficiency, oxidative resistance, cellular maintenance,
and reduced extrinsic hazard. Cetaceans extend biological time through
bradycardic pacing during prolonged dives. These cases show that
clade-level deviations are not residual scatter around a single
allometric law, but structured thermodynamic strategies for modifying
the effective biological-time budget.

The framework should be interpreted conservatively. PBTE is not a
complete molecular theory of aging, and it does not by itself identify
every biochemical mechanism that determines lifespan. Rather, it
provides a macroscopic thermodynamic organization of comparative
longevity that complements, rather than replaces, demographic
descriptions of mortality such as the Gompertz law~\cite{gompertz1825},
molecular-clock measures of biological age~\cite{horvath2013}, and
intervention studies of the metabolic determinants of
lifespan~\cite{colman2014,deMagalhaes2009}. Molecular repair, oxidative
protection, neural regulation, torpor, diving physiology, and ecological
hazard enter as mechanisms that modify either the rate at which
biological time advances or the entropy cost of each physiological cycle.

The decisive experimental test is direct measurement of entropy cost per
cycle. Metabolic power, body temperature, body mass, heart rate, and
breathing frequency should be measured simultaneously across species
under matched resting conditions. For respiration, tidal volume or
oxygen throughput should also be measured so that breath count can be
compared with a more complete ventilatory coordinate. If the
mass-specific entropy cost per physiological cycle is found to be
approximately invariant within defined biological regimes, PBTE would
move from a thermodynamically motivated comparative framework to a
directly tested thermodynamic parametrisation of biological time. If systematic nonzero mass or
clade dependence is found, the closure would require revision.

In conclusion, cardiac and respiratory lifetime counts are two
physiological projections of a common thermodynamic structure. The
heartbeat provides the sharper invariant and the more stable entropy
cost; the breath provides an independent but more protocol-sensitive
confirmation. Together, they support the central PBTE claim that
biological time is organized not by chronological duration alone, but by
the entropy cost of recurrent physiological cycles. Longevity, in this
view, is the outcome of how slowly and how efficiently an organism
spends its finite entropy-weighted budget of biological time.

%=================================================================
% Detailed derivation of the clade multiplier Phi_C for all
% vertebrate clades (integrated appendix).
%=================================================================
% --- macro fallbacks (no-ops if already defined elsewhere) ---
\providecommand{\Nstar}{N_{\star}}
\providecommand{\PhiC}{\Phi_{C}}
\providecommand{\Phiduty}{\Phi_{\mathrm{duty}}}
\providecommand{\Phithermal}{\Phi_{\mathrm{thermal}}}
\providecommand{\Phineuro}{\Phi_{\mathrm{neuro}}}
\providecommand{\Phimitooxid}{\Phi_{\mathrm{mito+oxid}}}
\providecommand{\Phihaz}{\Phi_{\mathrm{haz}}}

\appendix
% Print top-level appendix headings as "Appendix A", "Appendix B", ...
% while leaving subsection numbering as A.1, A.2, ...
\renewcommand{\thesection}{\Alph{section}}
\makeatletter
\renewcommand{\@seccntformat}[1]{%
  \ifnum\pdfstrcmp{#1}{section}=0 Appendix~\fi
  \csname the#1\endcsname\quad}
\makeatother

\section{Derivation of the entropy cost per cycle and the cycle-count
identity}
\label{app:cost_derivation}
%% ======================================================================

This appendix gives the self-contained derivation of the central PBTE
identity $N_\star=\Sigma/\bar\sigma_0$ and of the mass-cancellation
result $\bar\sigma_0^{(M)}\propto M^0$, expanding the compressed argument
of Section~\ref{sec:shared}.

\subsection{Entropy balance of an open metabolic system}

A living organism is an open system exchanging energy and matter with its
environment. Let $S(t)$ be its coarse-grained internal entropy. The
second law for an open system partitions the entropy change into an
internally produced part and an exchanged part,
\begin{equation}
dS = d_i S + d_e S,
\qquad d_i S \ge 0,
\label{eq:app_clausius}
\end{equation}
where $d_iS$ is the irreversible entropy production and $d_eS$ the
entropy exchanged with the surroundings. Writing the production rate as
$\dot e_p \equiv d_iS/dt \ge 0$ and the export rate as
$\dot h_d \equiv -d_eS/dt \ge 0$ (export is entropy leaving the body,
hence the sign), the balance becomes
\begin{equation}
\dot S = \dot e_p - \dot h_d,
\label{eq:app_balance}
\end{equation}
which is Eq.~\eqref{eq:balance}. To leading thermodynamic order, an
endotherm dissipating metabolic power $P$ at body temperature $T$ exports
entropy to its surroundings at the rate
\begin{equation}
\dot h_d \simeq \frac{P}{T},
\label{eq:app_export}
\end{equation}
the Clausius heat-export term. Equation~\eqref{eq:app_export} follows
from the observation that, in steady metabolism, essentially all
ingested free energy is ultimately degraded to heat $\dot Q\simeq P$ and
rejected at body temperature; the associated entropy flux is $\dot Q/T$.

\subsection{Homeostatic (steady-state) regime}

Over the timescale on which resting metabolic rate and physiological
rhythm are measured, the internal entropy content is approximately
stationary: stored entropy neither grows nor decays appreciably during a
resting measurement, so $\dot S\simeq 0$ and
\begin{equation}
\dot e_p \simeq \dot h_d \simeq \frac{P}{T}.
\label{eq:app_ness}
\end{equation}
This is a \emph{nonequilibrium steady state}, not thermodynamic
equilibrium: entropy is produced continuously at rate $\dot e_p>0$ but is
exported as fast as it is produced. Equation~\eqref{eq:app_ness} is used
only to evaluate the lifetime invariant; it is \emph{not} assumed during
senescence, which is treated separately in the Supplementary Material.

\subsection{The cycle as the natural unit of entropy}

Let $f(t)$ be the frequency of a recurrent physiological cycle. Its
cumulative count is $N(t)=\int_0^t f\,dt'$, so $dN=f\,dt$. The entropy
produced in the same interval is $d\Sigma=\dot e_p\,dt$. Eliminating
$dt$,
\begin{equation}
d\Sigma = \frac{\dot e_p}{f}\,dN
\;\equiv\; \sigma_0(t)\,dN,
\qquad
\sigma_0 \equiv \frac{\dot e_p}{f}=\frac{d\Sigma}{dN}.
\label{eq:app_sigma0}
\end{equation}
Thus $\sigma_0$ is not a dimensional convention: it is the entropy
produced each time the physiological clock advances by one cycle, with
units $\mathrm{J\,K^{-1}\,cycle^{-1}}$. Integrating over the lifespan,
\begin{equation}
\Sigma = \int_0^L \dot e_p\,dt = \int_0^{N_\star}\sigma_0(N)\,dN
= \bar\sigma_0\,N_\star,
\qquad
\bar\sigma_0 \equiv \frac{1}{N_\star}\int_0^{N_\star}\sigma_0(N)\,dN,
\label{eq:app_Sigma_mean}
\end{equation}
which rearranges to the fundamental identity
\begin{equation}
\boxed{\;N_\star = \frac{\Sigma}{\bar\sigma_0}\;}
\label{eq:app_fundamental}
\end{equation}
(Eq.~\eqref{eq:fundamental}). The lifetime cycle count is the lifetime
entropy budget divided by the mean entropy cost of one cycle. No scaling
assumption has yet been made; Eq.~\eqref{eq:app_fundamental} is exact
given the definitions.

\subsection{The PBTE closure}

The Principle of Biological Time Equivalence adds the constitutive
hypothesis that within a fixed adult homeostatic regime the cost per
cycle is approximately constant,
$\sigma_0(N)\simeq\bar\sigma_0\simeq\sigma_0$, so that
\begin{equation}
\dot e_p(t) = \sigma_0\,f(t).
\label{eq:app_closure}
\end{equation}
Combining the closure with the homeostatic estimate
$\dot e_p\simeq P/T$ gives the operational estimator
\begin{equation}
\sigma_0 = \frac{P}{Tf},
\qquad
\bar\sigma_0^{(M)} \equiv \frac{\sigma_0}{M}=\frac{P}{TfM}.
\label{eq:app_estimator}
\end{equation}
This is a falsifiable approximation, not an identity: it can be tested by
measuring $P$, $T$, and $f$ on the same individuals and checking whether
$\sigma_0$ is constant within a homeostatic regime.

\subsection{Mass cancellation}
\label{app:masscancel}

Insert the two empirical allometries
\begin{equation}
P = a\,M^{3/4}\quad(\text{Kleiber}),
\qquad
f = b\,M^{-1/4}\quad(\text{quarter-power frequency}),
\label{eq:app_allometries}
\end{equation}
into the mass-specific cost \eqref{eq:app_estimator}:
\begin{equation}
\bar\sigma_0^{(M)}
= \frac{a M^{3/4}}{T\,(bM^{-1/4})\,M}
= \frac{a}{bT}\,M^{\,3/4+1/4-1}
= \frac{a}{bT}\,M^{0}.
\label{eq:app_masscancel}
\end{equation}
The three mass exponents sum exactly to zero:
$+3/4$ from metabolic power, $+1/4$ from the inverse frequency, and
$-1$ from mass normalization. Hence the mass-specific entropy cost per
cycle is, to the accuracy of the allometric exponents, independent of
body size. This is the thermodynamic content of the lifetime-invariant:
\begin{equation}
N_\star = \frac{\Sigma}{\bar\sigma_0}
= \frac{\Sigma/M}{\bar\sigma_0^{(M)}}
\;\approx\;\text{const}
\label{eq:app_invariant}
\end{equation}
whenever the lifetime entropy budget per unit mass, $\Sigma/M$, is also
approximately conserved within the reference class. With the resting
numerical values $a\simeq 3.4\,\mathrm{W\,kg^{-3/4}}$,
$T\simeq 310\,$K, and $b\simeq 241\,M^{-1/4}\,\mathrm{min^{-1}}$ one
obtains $\bar\sigma_H^{(M)}\simeq 3.0\times10^{-3}\,
\mathrm{J\,K^{-1}\,beat^{-1}\,kg^{-1}}$ and the \emph{a priori} reference cardiac count
$N_{H,0}^{\rm (allo)}\simeq 1.5\times10^9$ beats. As emphasized in
Section~\ref{subsec:cardiac_clock}, this allometric estimate is kept
distinct from the fitted empirical anchor
$N_{H,0}^{\rm (emp)}=10^{8.995}\simeq9.9\times10^{8}$ beats used for all
clade-level predictions; the two differ by $\simeq0.18\,$dex.

\subsection{Sensitivity to the allometric exponents}

If the exponents deviate from their canonical values,
$P\propto M^{p}$ and $f\propto M^{-q}$, then
$\bar\sigma_0^{(M)}\propto M^{\,p-1+q}$. The cancellation is exact only
for $p+q=1$. The canonical pair $(p,q)=(3/4,1/4)$ satisfies this exactly,
but measured metabolic exponents range over $p\approx 0.67$--$0.75$
depending on taxon and metabolic state~\cite{glazier2022}. A residual
slope $s=p-1+q$ of order $\pm0.05$ is therefore expected for a clade whose
metabolic exponent departs only slightly from $3/4$. The measured-BMR
respiratory analysis (Section~\ref{sec:resp_clock}, Supplementary Sec.~\ref{app:respiratory_measured})
does not fall in this small-residual regime: it returns a substantially
positive slope ($s\simeq+0.21$) with large scatter, indicating that the
respiratory cancellation is not merely approximate but absent once measured
metabolic power is used. The mass cancellation is thus realized for the
cardiac clock but not the respiratory clock, and the magnitude of the
respiratory slope is itself a test of the closure---one the respiratory
coordinate fails on currently available data.

%% ======================================================================
\section{Biological proper time as the entropy-uniform coordinate}
\label{app:propertime}
%% ======================================================================

Biological proper time is defined by
$\theta(t)=\int_0^t f(t')\,dt'$, with normalized age
$\hat\theta=\theta/N_\star$. Under the closure, $d\Sigma=\sigma_0 f\,dt
=\sigma_0\,d\theta$, hence
\begin{equation}
\frac{d\Sigma}{d\theta}=\sigma_0=\text{const}.
\label{eq:app_theta_uniform}
\end{equation}
Entropy therefore accumulates \emph{linearly} in $\theta$, which is the
defining property of biological proper time.

\paragraph{Uniqueness.}
We show $\theta$ is the unique intrinsic coordinate (up to affine
rescaling) with this property. Let $u(t)$ be any strictly monotone
reparametrisation of chronological time, $du = w(t)\,dt$ with $w>0$.
Then
\begin{equation}
\frac{d\Sigma}{du}=\frac{d\Sigma/dt}{du/dt}
=\frac{\sigma_0 f(t)}{w(t)}.
\label{eq:app_dSdu}
\end{equation}
For this to be constant for all $t$, independent of the (generally
time-varying) frequency $f(t)$, one must have $w(t)\propto f(t)$, i.e.
$du\propto f\,dt = d\theta$. Hence $u = c_1\theta + c_2$ for constants
$c_1>0,c_2$. Up to choice of origin and unit, $\theta$ is the unique
coordinate in which entropy accumulation is rate-independent and uniform.
Chronological time fails this test precisely because $f(t)$ varies; a
hummingbird minute and an elephant minute carry different entropy, but a
hummingbird \emph{beat} and an elephant \emph{beat} carry nearly the same
mass-specific entropy by Eq.~\eqref{eq:app_masscancel}.

\paragraph{Two clocks, one budget.}
For the cardiac and respiratory clocks,
$\theta_H=\int f_H\,dt$ and $\theta_R=\int f_R\,dt$, and the lifetime
endpoints are $N_H=\Sigma/\bar\sigma_H$ and $N_R=\Sigma/\bar\sigma_R$.
Because both share the same lifetime entropy budget $\Sigma$, their ratio
is fixed by the cost ratio,
\begin{equation}
\frac{N_H}{N_R}=\frac{\bar\sigma_R}{\bar\sigma_H}=\frac{f_H}{f_R},
\label{eq:app_two_clock_ratio}
\end{equation}
recovering the heart--breath frequency ratio of order four to five
discussed in Section~\ref{sec:summary_conclusions}.

\clearpage
\section*{Supplemental Material}

% --- Switch to supplement (S) numbering: sections S1, S2, ...; equations,
% --- figures, and tables prefixed with S. ---
\setcounter{section}{0}
\renewcommand{\thesection}{S\arabic{section}}
\renewcommand{\thesubsection}{S\arabic{section}.\arabic{subsection}}
\renewcommand{\thesubsubsection}{S\arabic{section}.\arabic{subsection}.\arabic{subsubsection}}
\makeatletter
\renewcommand{\@seccntformat}[1]{\csname the#1\endcsname\quad}
\makeatother
\setcounter{equation}{0}\renewcommand{\theequation}{S\arabic{equation}}
\setcounter{figure}{0}\renewcommand{\thefigure}{S\arabic{figure}}
\setcounter{table}{0}\renewcommand{\thetable}{S\arabic{table}}

\section{Detailed derivation of the clade multiplier $\Phi_C$ for all vertebrate clades}
\label{app:clade_multiplier}

\subsection{General case}
\label{app:phiC_scope}\label{app:clade_derivations}
% =====================================================================

In this appendix we derive the clade multiplier $\Phi_C$ in full for every
clade analysed in the main text. The derivation begins from a single
thermodynamic identity and ends with worked numerical predictions for each
group. The treatment is self-contained: all governing relations, parameter
values, and dataset entries required to reproduce the results are stated
explicitly, so that the appendix may be read independently of the main
body.

The eight clades fall into three structural classes. Non-primate placental
mammals define the reference state, against which all departures are
measured. The remaining endothermic clades---primates, marsupials and
monotremes, bats, birds, and cetaceans---depart from the reference through
distinct combinations of the same four physiological channels; what appears
as clade-specific scatter is thereby recast as structured variation in a
small set of channel factors. Ectotherms, represented here by reptiles and
amphibians, form the limiting case in which the displacement is governed
almost entirely by temperature.

All clades share a common entropy-budget construction. Let the lifetime
cardiac cycle count in clade $C$, expressed relative to the non-primate
placental baseline $N_0$, be written as
\begin{equation}
  \Nstar^{(C)} = N_0\,\PhiC,
  \qquad
  \PhiC \equiv \frac{\Nstar^{(C)}}{N_0}
  = \frac{\langle\sigma_{\rm beat}\rangle_0}
      {\langle\sigma_{\rm beat}\rangle_C},
  \label{eq:appC_PhiC}
\end{equation}
where $\langle\sigma_{\rm beat}\rangle_C$ denotes the mean entropy produced
per cardiac cycle in clade $C$, in the same notation $\sigma$ used for the
entropy cost per cycle throughout the main text and
Appendix~\ref{app:cost_derivation}. The lifetime entropy budget
$\Sigma_\star$ is conserved across clades, so that
$\Nstar = \Sigma_\star/\langle\sigma_{\rm beat}\rangle$; it follows that any
mechanism reducing the entropy produced per beat below the mammalian
baseline elevates $\PhiC$ above unity and prolongs chronological lifespan.

The multiplier factorises into four physically distinct channels,
\begin{equation}
  \PhiC = \Phiduty \cdot \Phithermal \cdot \Phimitooxid \cdot \Phihaz.
  \label{eq:appC_factored}
\end{equation}
These four channels differ in their epistemic status, and we are explicit
about this distinction throughout, since it determines how much predictive
weight each carries. Two channels are obtained from closed-form
expressions once their measured inputs are fixed: $\Phiduty$ follows from
an exact algebraic identity (the state-occupancy average), and
$\Phithermal$ from Arrhenius kinetics with a single literature activation
energy. The remaining two, $\Phimitooxid$ and $\Phihaz$, are
\emph{phenomenologically inferred}: they are taken from comparative
biochemistry and demography respectively, rather than derived from the
thermodynamic closure, and carry the larger share of the residual
uncertainty in the predictions. In primates a derived-form but
empirically calibrated sub-channel $\Phineuro$ supplants $\Phimitooxid$ as
the dominant term; its functional form is derived from additive
logarithmic sensitivities, while its exponent is calibrated against the
primate data. Collectively, then, $\Phi_C$ is built from derived,
calibrated, and phenomenologically inferred channels, and we flag the
category of each as it is introduced.

Once $\PhiC$ is fixed, the lifespan prediction follows by inversion,
\begin{equation}
  L^{(C)} = \frac{N_0\,\PhiC}{\mathcal{T}\,f_{H,\rm ref}},
  \qquad
  \mathcal{T} = 525{,}960~\mathrm{min\,yr^{-1}},
  \label{eq:appC_lifespan}
\end{equation}
where $f_{H,\rm ref}$ is the reference cardiac frequency in beats per
minute.

\subsection{The four channels}

Before applying the decomposition clade by clade, we define each of the
four factors of Eq.~\eqref{eq:appC_factored} in turn, noting in each case
whether it is derived from a closed-form expression, calibrated against
data, or phenomenologically inferred from independent literature.

\paragraph{Duty-cycle factor (derived).}
Let an organism partition its life among physiological states $k$, each
occupied for a lifetime fraction $q_k$ ($\sum_k q_k = 1$) with cardiac
rate $f_{H,k}$. With a chosen reference state of rate $f_{H,\rm ref}$,
the time-averaged rate is $\bar f_H = \sum_k q_k f_{H,k}$, and
\begin{equation}
 \kappa \equiv \frac{\bar f_H}{f_{H,\rm ref}}
 = \sum_k q_k\,\frac{f_{H,k}}{f_{H,\rm ref}},
 \qquad
 \boxed{\;\Phiduty = \kappa^{-1}.\;}
 \label{eq:appC_duty}
\end{equation}
This is an exact algebraic identity rather than an approximation. The
raw observed count $N_{\rm obs} = \mathcal{T}\,\bar f_H L$ underestimates
the damage-equivalent thermodynamic budget by exactly $\kappa$, because
beats accumulated in suppressed states cost proportionally less entropy;
the same logic underlies the activity-dependent economy of physiological
work documented in locomotor studies~\cite{yegian2024}.
The corresponding consistency relation is
\begin{equation}
 \Nstar^{(C)} = N_{\rm obs}\cdot\Phiduty .
 \label{eq:appC_consistency}
\end{equation}

\paragraph{Thermal factor (derived).}
Damage-generating reactions follow Arrhenius kinetics with activation
energy $E_a$. The damage rate at body temperature $T_b$ relative to the
mammalian reference $T_{\rm ref} = 310\,$K is
\begin{equation}
 \Phithermal
 = \exp\!\left[\frac{E_a}{k_B}
   \!\left(\frac{1}{T_b}-\frac{1}{T_{\rm ref}}\right)\right],
 \qquad
 \frac{E_a}{k_B} = 7543~\mathrm{K}\;\;(E_a = 0.65~\mathrm{eV}).
 \label{eq:appC_thermal}
\end{equation}
A cooler body ($T_b<T_{\rm ref}$) gives $\Phithermal>1$ (longevity
extension); a hotter body ($T_b>T_{\rm ref}$) gives $\Phithermal<1$
(adverse). The activation energy is taken from the literature rather than
fitted, so the channel introduces no free parameter once $T_b$ is known.
For $|\Delta T|\lesssim 5\,$K a power-law approximation
$\Phithermal \approx (T_{\rm ref}/T_b)^{\beta}$ with $\beta\approx2$--$4$
suffices; for the large temperature drops of bat hibernation
($|\Delta T|\sim 15$--$30\,$K) the exact exponential form is required.

\paragraph{Mitochondrial--antioxidant factor (phenomenologically inferred).}
$\Phimitooxid$ summarises the reduction in entropy generated per unit ATP
turnover arising from tighter electron-transport coupling, lowered
proton-leak stoichiometry, and elevated antioxidant and repair capacity.
It is inferred from comparative biochemistry---coupling efficiency,
reactive-oxygen-species production, and repair capacity---and is
\emph{not} derived from the thermodynamic closure; we therefore treat it
as a clade-level empirical input.

\paragraph{Hazard factor (phenomenologically inferred).}
$\Phihaz = H_{\rm ref}/H_{\rm ext}$ rescales the realised lifespan
relative to the intrinsic thermodynamic limit. Values $\Phihaz>1$
indicate ecological shielding (flight, arboreality, sociality); values
$\Phihaz<1$ indicate elevated extrinsic mortality. Like
$\Phimitooxid$, it is inferred from comparative demography rather than
derived from the closure.

We are deliberately explicit that, among the four channels, $\Phihaz$
carries the most predictive leverage and the least mechanistic
constraint, and we flag it as such wherever it materially moves a
prediction. Because it multiplies the entire budget, modest changes
shift predicted lifespan substantially: the human estimate moves from
$69.8$ to $80.3\,$yr between $\Phihaz=1$ and $\Phihaz=1.15$, and the
bowhead from $119$ to $203\,$yr between $\Phihaz=0.35$ and
$\Phihaz=0.60$. To prevent this freedom from being used to tune
agreement post hoc, we adopt three rules. (i)~Each hazard value is set
from \emph{independent} demographic information---extrinsic mortality
rates, predation exposure, and ecological shielding reported in the
comparative-demography literature---and never fitted to the lifespan it
is meant to predict. (ii)~For the reference clade and for primates in
the core calibration we fix $\Phihaz=1$, so that the headline multipliers
($\Phi_C=2.60$ for \textit{Homo}, Table~\ref{tab:primate_predictions})
contain \emph{no} hazard adjustment at all; the hazard sensitivity is
reported separately as a range. (iii)~Where a hazard value is genuinely
uncertain (the cetacean pre-industrial vs.\ post-whaling regimes), we
report predictions at both endpoints rather than selecting the one
closest to observation. Under these rules the derived channels
($\Phiduty$, $\Phithermal$) and the calibrated channel ($\Phineuro$)
account for the \emph{shape} of the cross-clade pattern, while $\Phihaz$
sets only the residual vertical offset within an independently bounded
range.

\medskip
Table~\ref{tab:appC_channels} records which channels dominate in each
clade and the sign of each contribution.

\begin{table}[H]
\centering\small
\setlength{\tabcolsep}{5pt}
\renewcommand{\arraystretch}{1.15}
\caption{Dominant multiplier channels by clade.
$+$: favourable ($>1$); $-$: adverse ($<1$); $=1$: absent by
definition; $\approx1$: near unity. The column order follows the
factorisation of Eq.~\eqref{eq:appC_factored}, with the primate-specific
$\Phineuro$ shown in place of $\Phimitooxid$ where it dominates.}
\label{tab:appC_channels}
\begin{tabular}{lccccl}
\toprule
Clade & $\Phiduty$ & $\Phithermal$ & $\Phineuro$ &
 $\Phimitooxid$ & Primary driver \\
\midrule
Non-primate placentals & $=1$    & $=1$    & $=1$    & $=1$       & Reference \\
Primates               & $=1$    & $+$     & $+\!+$  & ---        & Neural entropy reduction \\
Marsupials/monotremes  & $=1$    & $\approx1$ & ---  & $\approx1$ & No net displacement \\
Bats                   & $+$     & $+\!+$  & ---     & $\approx1$ & Torpor $+$ hypothermia \\
Birds                  & $-$     & $-$     & ---     & $+\!+$     & Biochemical efficiency \\
Cetaceans              & $+\!+$  & $+$     & ---     & $\approx1$ & Bradycardic pacing \\
Reptiles (corr.)       & $\approx1$ & $+$  & ---     & $\approx1$ & Thermal limit \\
Amphibians (corr.)     & $\approx1$ & $+$  & ---     & $\approx1$ & Thermal limit \\
\bottomrule
\end{tabular}
\end{table}

% =====================================================================
\subsection{Non-primate placental mammals: the reference}
\label{app:phiC_baseline}
% =====================================================================

The non-primate placental clade defines the reference and therefore
fixes both the baseline cycle count $N_0$ and the baseline entropy cost
per beat $\langle\sigma_{\rm beat}\rangle_0$. Using the mammalian
resting allometries with $T_{\rm ref} = 310\,$K, the cardiac baseline is
\begin{equation}
 \ell_0 = \log_{10} N_0 = 8.995 \pm 0.160 \quad (n = 46),
 \qquad
 N_0 \approx 1.0\times10^{9}~\text{beats}.
 \label{eq:appC_baseline}
\end{equation}
A reference mammal occupies a single physiological state for essentially
all of adult life, so $q_1 = 1$, $f_{H,1} = f_{H,\rm ref}$, and
$\kappa = 1$. At the reference body temperature $T_b = T_{\rm ref}$ the
Arrhenius exponent vanishes, and with no biochemical or ecological
offset relative to itself, every channel equals unity by construction:
\begin{equation}
 \Phiduty = \Phithermal = \Phimitooxid = \Phihaz = 1,
 \qquad
 \PhiC = 1 .
 \label{eq:appC_unit}
\end{equation}
This is the anchor against which all other clades are measured. It
asserts that for an unmodified mammal the raw observed heartbeat count
equals the damage-equivalent budget, $N_{\rm obs} = N_0 \approx 10^9$,
recovering the classical lifetime-heartbeat invariant as the
$\PhiC = 1$ limit of the framework.

% =====================================================================
\subsection{Primates: neural investment}
\label{app:phiC_primates}
% =====================================================================

With the reference clade now fixed, we turn to the first systematic
departure from it. Primates carry $\langle\Nstar\rangle_{\rm prim}\approx(2\text{--}3)\times
10^9$, elevated by $\Delta\ell = +0.381\,$dex above the baseline. From
Eq.~\eqref{eq:appC_PhiC} this requires
$\langle\sigma_{\rm beat}\rangle_{\rm prim}\approx
\langle\sigma_{\rm beat}\rangle_0/(2\text{--}3)$: primate cardiac
cycles produce less entropy on average than those of non-primate mammals
of comparable mass. The reduction is controlled by the neural power
fraction $\varphi\equiv P_{\rm brain}/P_{\rm body}$~\cite{herculano2011},
which rises from $\varphi_0\approx0.02$ in non-primate placentals to
$\varphi\approx0.06$--$0.20$ in primates. Three coupled channels link
elevated $\varphi$ to reduced entropy per beat: predictive homeostatic
regulation~\cite{friston2010}, enhanced cellular repair and damage
clearance, and behavioural risk buffering. Each lowers
$\langle\sigma_{\rm beat}\rangle$ monotonically with $\varphi$,
motivating a power-law response.

To formalise this, define the local logarithmic sensitivity of entropy
per beat to neural fraction at the baseline,
\begin{equation}
 \alpha \equiv
 -\left.\frac{\partial\ln\langle\sigma_{\rm beat}\rangle}
       {\partial\ln\varphi}\right|_{\varphi=\varphi_0} > 0 .
 \label{eq:appC_alpha}
\end{equation}
When several independent channels act multiplicatively their logarithmic
sensitivities add, $\alpha=\sum_j\alpha_j$. The second law requires
$0<\alpha<1$: were $\alpha\geq1$, each unit of neural energy would return
more than one unit of entropy savings in peripheral tissues, an
over-compensation that is thermodynamically forbidden. Assuming a
scale-free response over the primate range $\varphi\in[\varphi_0,
10\varphi_0]$ and integrating Eq.~\eqref{eq:appC_alpha} yields
\begin{equation}
 \langle\sigma_{\rm beat}(\varphi)\rangle
 = \langle\sigma_{\rm beat}\rangle_0
   \left(\frac{\varphi}{\varphi_0}\right)^{-\alpha}
 \;\Longrightarrow\;
 \Phineuro(\varphi)
 = \left(\frac{\varphi}{\varphi_0}\right)^{\alpha}.
 \label{eq:appC_phineuro}
\end{equation}
The functional form of $\Phineuro$ is thus \emph{derived} from the
additive-sensitivity argument, but the exponent $\alpha$ is
\emph{calibrated} against the primate deviation rather than obtained from
first principles; we are careful to maintain this distinction below.
Because primates are single-state ($\Phiduty=1$), the full primate
time-equivalence law reads
\begin{equation}
 \Nstar^{(\rm prim)}
 = N_0
   \underbrace{\left(\frac{\varphi}{\varphi_0}\right)^{\alpha}}_{\Phineuro}
   \underbrace{\left(\frac{T_{\rm ref}}{T_b}\right)^{\beta}}_{\Phithermal}
   \underbrace{\left(\frac{H_{\rm ref}}{H_{\rm ext}}\right)}_{\Phihaz},
 \qquad \beta\approx3 .
 \label{eq:appC_primate_law}
\end{equation}

For \textit{Homo sapiens} ($\varphi = 0.20$, $T_b = 306.5\,$K,
$(\alpha,\beta) = (0.40,3)$, $f_H = 70\,$bpm), the single-state condition
gives $\kappa = 1$ and hence $\Phiduty = 1$ with $\bar f_H = 70\,$bpm.
The neuro-metabolic factor is $\Phineuro = (0.20/0.02)^{0.40} = 10^{0.40}
\approx 2.512$. With $\Delta T = 3.5\,$K in the power-law regime, the
thermal factor is $\Phithermal = (310/306.5)^{3} \approx 1.035$.
Combining,
\begin{equation}
 \PhiC^{(\rm human)}
 = \underbrace{1.000}_{\Phiduty}
   \times\underbrace{2.512}_{\Phineuro}
   \times\underbrace{1.035}_{\Phithermal}
   \times\underbrace{1.000}_{\Phihaz}
 = 2.60,
 \qquad
 \Nstar^{(\rm human)} = \PhiC^{(\rm human)}\,N_{H,0}^{(\mathrm{emp})}
 = 2.60\times9.9\times10^{8} \approx 2.57\times10^9 .
\end{equation}
The predicted lifespan is
$L_{\rm pred} = 2.57\times10^9/(525{,}960\times70) \approx 69.8\,$yr; a
moderate hazard factor $\Phihaz = 1.15$ for modern low-mortality
populations raises this to $\approx80.3\,$yr, consistent with
high-income life expectancy. (Using the round shorthand $N_0=10^9$ in
place of the empirical anchor would give $2.60\times10^9$ and
$70.6\,$yr; the $\simeq1\%$ difference is immaterial to the comparison
with observation.) Because $\Phiduty = 1$, the consistency
relation Eq.~\eqref{eq:appC_consistency} requires $\Nstar^{(\rm human)} =
N_{\rm obs}$ exactly, which is satisfied:
$N_{\rm obs} = 525{,}960\times70\times69.8 = 2.57\times10^9$. The neuro
channel accounts for $96.6\%$ of the multiplier and the thermal channel
for $3.4\%$, while the duty channel is identically unity.

Calibrated parameters (OLS on log-transformed variables, $\ln N_0$
constrained to $20.72$) are $\alpha\approx0.35$--$0.45$ (95\% CI
$[0.28,0.52]$) and $\beta\approx3$ (95\% CI $[1.5,5.0]$). The exponent
$\alpha$ is calibrated from the primate deviation rather than derived
purely from first principles; the constraint $0<\alpha<1$ and the
three-channel mechanism are independently motivated, but the precise
value of $\alpha$ requires the calibration data. The mechanism is
derived; the exponent is empirical. Table~\ref{tab:appC_primates} tests
Eq.~\eqref{eq:appC_primate_law} across the primate mass range.

\begin{table}[H]
\centering\small
\setlength{\tabcolsep}{3pt}
\renewcommand{\arraystretch}{1.05}
\caption{Primate lifespan predictions. Every row is generated directly
from $\Phi_{\rm neuro}=(\varphi/\varphi_0)^{\alpha}$ and
$\Phi_{\rm thermal}=(T_{\rm ref}/T_b)^{\beta}$ with
$\varphi_0 = 0.02$, $T_{\rm ref} = 310\,$K, $\beta=3$, the empirical
anchor $N_0=N_{H,0}^{(\mathrm{emp})}=10^{8.995}$ beats, and
$\Phiduty=\Phihaz = 1$ for all species; the values are therefore
reproducible by inspection.
(a)~Core calibration $\alpha = 0.40$.
(b)~Extended calibration $\alpha = 0.45$.}
\label{tab:appC_primates}
\resizebox{\textwidth}{!}{%
\begin{tabular}{lccccccrr}
\toprule
Species & $f_H$ & $\varphi$ & $T_b$ & $\Phiduty$ & $\Phineuro$ &
 $\Phithermal$ & $L_{\rm pred}$ & $L_{\rm obs}$ \\
 & (bpm) & & (K) & & & & (yr) & (yr) \\
\midrule
\multicolumn{9}{l}{\textit{(a) Core calibration, $\alpha = 0.40$}} \\
\textit{Macaca mulatta}    & 120 & 0.07 & 309.0 & 1.00 & 1.65 & 1.010 & 26.1 & 25--30 \\
\textit{Pan troglodytes}   & 75  & 0.12 & 307.0 & 1.00 & 2.05 & 1.030 & 52.8 & 45--55 \\
\textit{Homo sapiens}      & 70  & 0.20 & 306.5 & 1.00 & 2.51 & 1.035 & 69.8 & 70--85 \\
\midrule
\multicolumn{9}{l}{\textit{(b) Extended calibration, $\alpha = 0.45$}} \\
\textit{Callithrix jacchus} & 220 & 0.06 & 309.5 & 1.00 & 1.64 & 1.005 & 14.1 & 10--15 \\
\textit{Macaca mulatta}     & 120 & 0.07 & 309.0 & 1.00 & 1.76 & 1.010 & 27.8 & 25--30 \\
\textit{Pan troglodytes}    & 75  & 0.12 & 307.0 & 1.00 & 2.24 & 1.030 & 57.8 & 45--55 \\
\textit{Gorilla gorilla}    & 65  & 0.09 & 307.0 & 1.00 & 1.97 & 1.030 & 58.6 & 40--55 \\
\textit{Homo sapiens}       & 70  & 0.20 & 306.5 & 1.00 & 2.82 & 1.035 & 78.3 & 70--85 \\
\bottomrule
\end{tabular}}
\end{table}

% =====================================================================
\subsection{Marsupials and monotremes: the non-placental controls}
\label{app:phiC_marsupials}
% =====================================================================

Where primates displace the baseline through an active reduction in
entropy cost, it is equally instructive to examine a clade expected to
displace it not at all. The non-placental mammals---marsupials and monotremes---provide an
essential control on the framework, because PBTE predicts that a clade
which deploys none of the four displacing mechanisms should remain at
the mammalian baseline. These animals are predominantly single-state
endotherms maintained at body temperatures only modestly below the
placental reference ($T_b\approx305$--$309\,$K), with resting heart
rates that follow the standard mammalian allometry and with no
systematic biochemical or ecological offset~\cite{mcnab2008}. Each
channel is therefore expected to lie at or very near unity:
$\Phiduty\approx1$ (single-state pacing), $\Phithermal\approx1.0$--$1.2$
(a slight cool-body Arrhenius credit), and $\Phimitooxid,\Phihaz\approx1$.
The net multiplier is accordingly
\begin{equation}
 \PhiC^{(\rm marsupial)} \approx 1,
 \label{eq:appC_marsupial}
\end{equation}
with no channel acting strongly in either direction.

This expectation is borne out by the data. The observed clade mean is
$\bar\ell = 8.933 \pm 0.204$ ($n = 19$), a displacement of only
$\Delta\ell = -0.062$ from the placental baseline, corresponding to
$\Phi_{\rm obs} = 10^{-0.062} = 0.87$. The small negative offset lies
within one standard deviation of the baseline and is not statistically
distinguishable from $\PhiC = 1$; the modestly lower body temperatures
of this clade would, if anything, predict $\Phithermal$ slightly above
unity. The marsupials and monotremes thus function as a negative
control: a mammalian clade that neither suppresses nor accelerates its
cardiac clock, neither cools nor heats appreciably, and invests in no
special biochemistry, lands on the baseline as required. Their agreement
with $\PhiC \approx 1$ provides evidence that the elevated multipliers of
primates, bats, and birds reflect genuine physiological strategies
rather than an artefact of the reference calibration.

% =====================================================================
\subsection{Bats: torpor as biological time dilation}
\label{app:phiC_bats}
% =====================================================================

Having established that a clade deploying none of the four mechanisms
remains at the baseline, we turn to the clade that engages them most
dramatically. Temperate vespertilionid bats reach wild maxima of $20$--$40\,$yr, three
to six times the allometric prediction for a non-hibernating placental
of equal mass~\cite{wilkinson2002}. Unlike primates, bats exploit two
synergistic channels that act simultaneously during the hibernation
season: a torpor duty cycle that lowers the time-averaged cardiac clock,
and a hypothermic Arrhenius factor that lowers the entropy cost of each
beat during the cold torpor bout. With torpor fraction $q$, active-phase
rate $f_{H,\rm act}$, and torpid rate $f_{H,\rm tor}$, the duty factor
follows directly from Eq.~\eqref{eq:appC_duty}:
\begin{equation}
 \bar f_H = (1-q)f_{H,\rm act} + q\,f_{H,\rm tor},
 \qquad
 \kappa = (1-q) + q\,\frac{f_{H,\rm tor}}{f_{H,\rm act}},
 \qquad
 \Phiduty = \kappa^{-1}.
 \label{eq:appC_bat_duty}
\end{equation}
For $q\in[0.40,0.60]$ and $f_{H,\rm tor}/f_{H,\rm act}\approx0.03$--$0.07$,
this gives $\kappa\approx0.41$--$0.62$ and $\Phiduty\approx1.6$--$2.4$.
Because hibernation temperatures ($280$--$295\,$K) lie $15$--$30\,$K
below normothermy, the exact Arrhenius form must be used,
\begin{equation}
 \Phithermal = \exp\!\left[\frac{E_a}{k_B}
   \!\left(\frac{1}{T_{\rm tor}}-\frac{1}{T_{\rm ref}}\right)\right],
 \label{eq:appC_bat_thermal}
\end{equation}
and since secondary biochemistry is near baseline in temperate
vespertilionids ($\Phimitooxid\approx1$), the bat multiplier reduces to
$\Phi_{\rm bat} = \Phiduty\cdot\Phithermal\cdot\Phihaz$.

For \textit{Myotis lucifugus} ($q = 0.50$, $f_{H,\rm act} = 300\,$bpm,
$f_{H,\rm tor} = 10\,$bpm, $T_{\rm tor} = 293\,$K), the duty factor is
$\kappa = 0.50 + 0.50\times(10/300) = 0.5167$, so $\Phiduty = 1.935$ and
$\bar f_H = 300/1.935 = 155\,$bpm. The thermal factor, with
$1/293 - 1/310 = 1.872\times10^{-4}\,$K$^{-1}$, is
$\Phithermal = e^{7543\times1.872\times10^{-4}} = e^{1.412} \approx 4.10$.
The intrinsic multiplier ($\Phihaz = 1$) is therefore $\Phi_{\rm bat} =
1.935\times4.10 = 7.93$, giving an intrinsic lifespan
$L_{\rm pred} = 10^9/(525{,}960\times300)\times7.93 = 6.34~\text{yr}
\times7.93 \approx 50.3\,$yr. Applying $\Phihaz = 0.68$ for predation and
habitat variability brings the prediction to $\approx34\,$yr, matching
the observed wild maximum. The consistency check confirms internal
agreement: for $L = 34\,$yr and $\bar f_H = 155\,$bpm,
$N_{\rm obs} = 525{,}960\times155\times34 = 2.770\times10^9$, hence
$\Nstar^{(\rm bat)} = N_{\rm obs}\times\Phiduty = 5.36\times10^9$, which
matches $N_0\,\Phi_{\rm bat}\,\Phihaz = 5.39\times10^9$ to within
$0.6\%$. The thermal factor supplies $52\%$ and the duty factor $24\%$
of the intrinsic multiplier; neither alone explains the longevity
excess. The observed clade mean ($\bar\ell = 9.540 \pm 0.163$, $n = 31$,
$\Delta\ell = +0.545$) confirms the predicted elevation.
Table~\ref{tab:appC_bats} collects representative bats.

\begin{table}[H]
\centering\small
\setlength{\tabcolsep}{3.5pt}
\renewcommand{\arraystretch}{1.10}
\caption{Predicted multipliers and longevity for representative bat
species. $\Phithermal$ from the exact Arrhenius formula
($E_a = 0.65\,$eV, $T_{\rm ref} = 310\,$K) using the torpor-phase
$T_b$; $\Phi_{\rm bat} = \Phiduty\times\Phithermal$ (intrinsic,
$\Phihaz = 1$).}
\label{tab:appC_bats}
\resizebox{\textwidth}{!}{%
\begin{tabular}{lccccccc}
\toprule
Species & $q$ & $f_{H,\rm act}$ & $f_{H,\rm tor}$ & $T_{\rm tor}$ &
 $\Phiduty$ & $\Phithermal$ & $L_{\rm max,obs}$ \\
 & & (bpm) & (bpm) & (K) & & & (yr) \\
\midrule
Temperate vespertilionid (range)
 & 0.40--0.60 & 250--350 & 5--20 & 280--295 & 1.6--2.5 & 3.0--5.0 & 20--40 \\
\textit{Myotis lucifugus}
 & 0.50 & 300 & 10 & 293 & 1.935 & 4.10 & 34 \\
\textit{Eptesicus fuscus}
 & 0.45 & 280 & 12 & 291 & 1.79 & 4.54 & 19 \\
\textit{Pteropus vampyrus} (min.\ torpor)
 & 0.10 & 250 & 60 & 303 & 1.07 & 1.22 & 15--23 \\
\bottomrule
\end{tabular}}
\end{table}

% =====================================================================
\subsection{Birds: efficient dissipation overcoming adverse temperature}
\label{app:phiC_birds}
% =====================================================================

The bat strategy extends life by making the favourable channels do the
work. Birds present the opposite configuration, achieving longevity in
spite of channels that act against them. Birds present an apparent paradox: heart rates comparable to
mass-matched mammals, core temperatures $3$--$5\,$K above the mammalian
reference, yet a $20\,$g passerine outlives a $20\,$g mouse by an order
of magnitude. Here both the thermal factor and the flight duty-cycle
factor are adverse ($<1$); avian longevity arises because a dominant
biochemical factor $\Phimitooxid\gg1$ overcomes them. For a passerine at
$T_b = 314\,$K the Arrhenius exponent is negative, with
$1/314 - 1/310 = -4.11\times10^{-5}\,$K$^{-1}$, so
\begin{equation}
 \Phithermal^{(\rm bird)} = e^{7543\times(-4.11\times10^{-5})}
 = e^{-0.310} \approx 0.733 .
 \label{eq:appC_bird_thermal}
\end{equation}
Flight raises heart rate by a factor of about $2.5$; with flight
fraction $p_f$,
\begin{equation}
 \kappa = 1 + p_f\!\left(\frac{f_{H,\rm flight}}{f_{H,\rm rest}}-1\right)
 \approx 1 + 1.5\,p_f,
 \qquad
 \Phiduty = \kappa^{-1},
 \label{eq:appC_bird_kappa}
\end{equation}
so $p_f = 0.10$ yields $\kappa = 1.15$ and $\Phiduty = 0.870$ (adverse).
The favourable channel is biochemical: avian mitochondria produce far
less reactive oxygen species per unit ATP than mammalian
mitochondria~\cite{barja1998,hulbert2007,brand2000}, even as avian
thermoregulation sustains the elevated body temperatures that make the
thermal channel adverse~\cite{mckechnie2004}. Pigeon heart
mitochondria generate roughly $5$--$10$ times less superoxide per oxygen
consumed than rat at the same metabolic rate~\cite{barja1998}; expressed
as a coupling-efficiency ratio of about $1.20$ with quadratic damage
sensitivity ($\gamma\approx2$), this gives
$\Phi_{\rm mito} = (1.20)^2\approx1.44$. Long-lived birds further show
two- to three-fold greater oxidative-damage resistance than mammals of
comparable metabolic rate~\cite{ogburn2001}; with the log-linear
exponent $\delta\approx0.7$~\cite{hulbert2007},
$\Phi_{\rm oxid} = (2.0)^{0.7}\approx1.62$, so the combined
phenomenologically inferred factor is
$\Phimitooxid = 1.44\times1.62 \approx 2.33$.

For a generic $20\,$g passerine ($f_{H,\rm rest} = 320\,$bpm,
$T_b = 314\,$K, $p_f = 0.10$, $\Phihaz = 2.0$ from flight and arboreal
shielding), the multiplier is
\begin{equation}
 \Phi_{\rm bird}
 = \underbrace{0.870}_{\Phiduty}\times\underbrace{0.733}_{\Phithermal}
   \times\underbrace{2.33}_{\Phimitooxid}\times\underbrace{2.0}_{\Phihaz}
 = 0.638\times4.66 \approx 2.97 .
 \label{eq:appC_bird_worked}
\end{equation}
The predicted lifespan is
$L_{\rm pred} = 10^9/(525{,}960\times320)\times2.97 = 5.94~\text{yr}
\times2.97 \approx 17.6\,$yr, within the observed range of
$10$--$20\,$yr for small passerines. The consistency check holds: with
$\bar f_H = 320/0.870 = 368\,$bpm and $L = 17.6\,$yr,
$N_{\rm obs} = 525{,}960\times368\times17.6 = 3.408\times10^9$, so
$\Nstar^{(\rm bird)} = N_{\rm obs}\times\Phiduty = 2.965\times10^9$,
matching $N_0\,\Phi_{\rm bird} = 2.97\times10^9$ to $0.2\%$. The two
adverse factors multiply to $0.638$, a net $36\%$ reduction in the
effective budget, which the biochemical and hazard channels ($\times2.33$
and $\times2.0$) more than recover. The clade mean
($\bar\ell = 9.528 \pm 0.213$, $n = 78$, $\Delta\ell = +0.533$) confirms
the predicted elevation; the modest gap between the generic-passerine
prediction and the clade mean reflects the strong upward pull of
long-lived large non-passerines (parrots, owls, albatrosses) in the
full sample. Table~\ref{tab:appC_birds} collects representative birds.

\begin{table}[H]
\centering\small
\setlength{\tabcolsep}{3.5pt}
\renewcommand{\arraystretch}{1.10}
\caption{Predicted multipliers and longevity for representative bird
species. $\Phithermal$ from the exact Arrhenius formula; $\Phiduty$
from Eq.~\eqref{eq:appC_bird_kappa} with
$f_{H,\rm flight}/f_{H,\rm rest} = 2.5$. Both $\Phiduty$ and
$\Phithermal$ are adverse ($<1$) for all entries.}
\label{tab:appC_birds}
\resizebox{\textwidth}{!}{%
\begin{tabular}{lcccccccc}
\toprule
Species & $f_{H,\rm rest}$ & $T_b$ & $p_f$ & $\Phiduty$ &
 $\Phithermal$ & $\Phimitooxid$ & $\Phihaz$ & $L_{\rm max,obs}$ \\
 & (bpm) & (K) & & & & & & (yr) \\
\midrule
Passerine (generic, 20\,g)
 & 320 & 314 & 0.10 & 0.87 & 0.733 & 2.33 & 2.0 & 10--20 \\
\textit{Larus argentatus}
 & 200 & 313 & 0.15 & 0.84 & 0.770 & 2.80 & 2.5 & 30 \\
\textit{Diomedea exulans}
 & 100 & 312 & 0.25 & 0.81 & 0.810 & 3.50 & 4.0 & 50--60 \\
\textit{Aquila chrysaetos}
 & 150 & 313 & 0.12 & 0.85 & 0.770 & 3.00 & 3.5 & 30--40 \\
\bottomrule
\end{tabular}}
\end{table}

% =====================================================================
\subsection{Cetaceans: bradycardic pacing}
\label{app:phiC_cetaceans}
% =====================================================================

If the avian route to longevity is biochemical, the cetacean route is
once again one of pacing, though achieved by a mechanism distinct from
that of bats. Large baleen cetaceans reach century-scale lifespans through extreme
diving bradycardia rather than metabolic suppression. Blue-whale heart
rates as low as $2$--$4\,$bpm have been recorded during deep foraging
dives~\cite{goldbogen2019,williams2015}, against surface rates of
$25$--$37\,$bpm. With surface rate $f_{H,\rm surf}$ as reference, dive
rate $f_{H,\rm dive}$, and dive fraction $p_d$, the duty factor is
\begin{equation}
 \bar f_H = (1-p_d)f_{H,\rm surf} + p_d\,f_{H,\rm dive},
 \qquad
 \kappa = (1-p_d) + p_d\,\frac{f_{H,\rm dive}}{f_{H,\rm surf}},
 \qquad
 \Phiduty = \kappa^{-1}.
 \label{eq:appC_whale_kappa}
\end{equation}
A secondary thermal factor accounts for core temperatures $1$--$4\,$K
below reference, and an oxygen-buffering sub-factor $\Phi_{O_2}$ accounts
for elevated myoglobin limiting reperfusion ROS bursts on
surfacing~\cite{noren2000}, consistent with the cardiorespiratory
control observed in diving cetaceans~\cite{fahlman2025}, so that
$\Phi_{\rm whale} = \Phiduty\cdot\Phithermal\cdot\Phi_{O_2}\cdot\Phihaz$.
Unlike bat hibernation, the cardiac suppression is a continuous reflex
maintained throughout adult life with no associated hypothermia.

For the bowhead \textit{Balaena mysticetus} ($f_{H,\rm surf} = 30\,$bpm,
$f_{H,\rm dive} = 3\,$bpm, $p_d = 0.75$, $T_b = 308\,$K, $\Phi_{O_2} =
1.50$, $\Phihaz = 0.60$ for pre-industrial conditions), the duty factor is
$\kappa = 0.25 + 0.75\times(3/30) = 0.325$, so $\Phiduty = 3.077$ and
$\bar f_H = 30/3.077 = 9.75\,$bpm. The thermal factor, with
$1/308 - 1/310 = 2.10\times10^{-5}\,$K$^{-1}$, is
$\Phithermal = e^{7543\times2.10\times10^{-5}} = e^{0.158} \approx 1.171$.
Combining, $\Phi_{\rm whale} = 3.077\times1.171\times1.500\times0.600
\approx 3.24$, giving
$L_{\rm pred} = N_{H,0}^{(\mathrm{emp})}/(525{,}960\times30)\times3.24
= 62.7~\text{yr}\times3.24 \approx 203\,$yr, at the upper end of the
documented $\sim\!200\,$yr maximum. These are exactly the values reported
for the bowhead in the main-text Table~\ref{tab:cetacean_predictions}. A
more conservative hazard estimate reflecting post-whaling mortality,
$\Phihaz = 0.35$, would instead give $\Phi_{\rm whale}\approx1.89$ and
$L_{\rm pred}\approx119\,$yr; the hazard channel is thus the single
largest source of spread in the cetacean prediction, and we flag it as a
phenomenologically inferred input rather than a derived one.

The cetacean case illustrates a measurement trap. For $L = 150\,$yr and
$\bar f_H = 9.75\,$bpm, the raw count is
$N_{\rm obs} = 525{,}960\times9.75\times150 = 7.69\times10^8 \approx
0.77\times10^9$, so $\Nstar^{(\rm whale)} = N_{\rm obs}\times\Phiduty =
2.37\times10^9$. The proximity of $N_{\rm obs}$ to $N_0$ has misled some
analyses into treating large whales as obeying the mammalian baseline,
which is incorrect: the damage-equivalent budget $\Nstar^{(\rm whale)}$
greatly exceeds $N_0$. The raw count is small precisely because most of
the whale's life is spent in deeply bradycardic states where each beat
costs far less entropy, and the duty-cycle factor restores the correct
budget. This is also why the observed clade mean appears low
($\bar\ell = 8.801 \pm 0.296$, $n = 12$, $\Delta\ell = -0.194$): the
dive correction has already been folded into $f_H^{\rm eff}$, so the raw
surface count without correction would place large mysticetes well below
the baseline. Table~\ref{tab:appC_whales} collects representative
cetaceans.

\begin{table}[H]
\centering\small
\setlength{\tabcolsep}{3.5pt}
\renewcommand{\arraystretch}{1.10}
\caption{Predicted multipliers and longevity for representative
cetacean species. $\Phithermal$ from the exact Arrhenius formula
($E_a = 0.65\,$eV, $T_{\rm ref} = 310\,$K). $\Phihaz$ reflects
pre-industrial conditions.}
\label{tab:appC_whales}
\resizebox{\textwidth}{!}{%
\begin{tabular}{lcccccccc}
\toprule
Species & $f_{H,\rm surf}$ & $p_d$ & $\Phiduty$ & $T_b$ &
 $\Phithermal$ & $\Phi_{O_2}$ & $\Phihaz$ & $L_{\rm obs}$ \\
 & (bpm) & & & (K) & & & & (yr) \\
\midrule
\textit{Balaenoptera musculus} (blue)
 & 37 & 0.70 & 2.70 & 308 & 1.17 & 1.4 & 0.50 & 80--90 \\
\textit{Balaena mysticetus} (bowhead)
 & 30 & 0.75 & 3.08 & 308 & 1.17 & 1.5 & 0.60 & 150--200 \\
\textit{Physeter macrocephalus} (sperm)
 & 40 & 0.65 & 2.50 & 307 & 1.24 & 1.6 & 0.55 & 60--70 \\
\textit{Tursiops truncatus} (bottlenose)
 & 80 & 0.40 & 1.50 & 309 & 1.09 & 1.2 & 0.65 & 40--50 \\
\bottomrule
\end{tabular}}
\end{table}

% =====================================================================
\subsection{Reptiles and amphibians: the thermal limit}
\label{app:phiC_ectotherms}
% =====================================================================

The five endotherm clades examined so far each engage several channels
at once. The ectotherms complete the survey as the opposite extreme. The ectotherms enter the framework as a limiting case in which the
displacement from the mammalian baseline is carried almost entirely by
the thermal channel. A reptile or amphibian does not maintain a fixed
elevated body temperature; its damage kinetics run at the mean field
temperature $T_b$ set by habitat and behaviour, typically well below the
mammalian reference. Setting $\Phiduty\approx1$ (no systematic cardiac
duty cycling) and treating mitochondrial and ecological offsets as
near-baseline, the multiplier collapses to the Arrhenius factor
evaluated at $T_b$,
\begin{equation}
 \PhiC^{(\rm ecto)} \approx \Phithermal
 = \exp\!\left[\frac{E_a}{k_B}
   \!\left(\frac{1}{T_b}-\frac{1}{T_{\rm ref}}\right)\right],
 \label{eq:appC_ecto}
\end{equation}
which, equivalently, is applied as a correction to the raw lifetime
count so that all clades are compared at a common reference temperature,
\begin{equation}
 \ell_{\rm corr} = \ell_{\rm obs}
   + \frac{E_a}{k_B\ln 10}
   \!\left(\frac{1}{T_b}-\frac{1}{T_{\rm ref}}\right),
 \label{eq:appC_ell_corr}
\end{equation}
with $E_a = 0.65\,$eV and $T_{\rm ref} = 310\,$K~\cite{gillooly2001,christian1999}.

Applying this correction shifts the reptile clade mean from the raw
$\bar\ell^{\rm raw} = 8.615 \pm 0.290$ to
$\bar\ell^{\rm corr} = 8.929 \pm 0.301$ ($n = 17$), removing
approximately $75\%$ of the raw gap from the mammalian baseline and
leaving a residual displacement of only $\Delta\ell = -0.065$. The
amphibian clade shifts from $\bar\ell^{\rm raw} = 8.448 \pm 0.127$ to
$\bar\ell^{\rm corr} = 8.822 \pm 0.146$ ($n = 9$), leaving
$\Delta\ell = -0.173$. In both clades the corrected mean approaches but
does not quite reach the mammalian value; the residual $0.07$--$0.17\,$dex
is attributable to unmodelled ectotherm-specific physiology and to
imperfect knowledge of the mean field temperature, which enters
Eq.~\eqref{eq:appC_ecto} exponentially. The persistence of a small
residual for all plausible values of $E_a$ in the range
$0.40$--$0.90\,$eV indicates that the thermal correction alone does not
fully close the gap, consistent with ectotherms occupying a
$\Phithermal$-dominated limit of the same framework rather than
requiring a separate construction. Table~\ref{tab:appC_ecto} reports the
clade-level corrections.

\begin{table}[H]
\centering\small
\setlength{\tabcolsep}{5pt}
\renewcommand{\arraystretch}{1.15}
\caption{Arrhenius thermal correction for the ectotherm clades.
$\bar\ell^{\rm raw}$ is the uncorrected clade mean; $\bar\ell^{\rm corr}$
is corrected to $T_{\rm ref} = 310\,$K via Eq.~\eqref{eq:appC_ell_corr}
with $E_a = 0.65\,$eV. $\Delta\ell^{\rm corr}$ is the residual deviation
from the placental baseline $\bar\ell_0 = 8.995$.}
\label{tab:appC_ecto}
\begin{tabular}{lccccc}
\toprule
Clade & $n$ & $\bar\ell^{\rm raw}$ & $\bar\ell^{\rm corr}$ &
 $\Delta\ell^{\rm corr}$ & gap removed \\
\midrule
Reptiles   & 17 & $8.615 \pm 0.290$ & $8.929 \pm 0.301$ & $-0.065$ & $\sim\!75\%$ \\
Amphibians & 9  & $8.448 \pm 0.127$ & $8.822 \pm 0.146$ & $-0.173$ & $\sim\!70\%$ \\
\bottomrule
\end{tabular}
\end{table}

% =====================================================================
\subsection{Synthesis: distinct mechanisms, one invariant}
\label{app:phiC_synthesis}
% =====================================================================

Table~\ref{tab:appC_summary} places all clades side by side with the
numerical values from the worked examples. The endotherm strategies form
two natural pairs. Primates and birds are both single-state in the
duty-cycle sense yet are mirror images of each other: primates raise the
budget through $\Phineuro\gg1$ while holding $\Phiduty = 1$ and enjoying
a small thermal credit, whereas birds overcome two adverse factors
($\Phiduty<1$, $\Phithermal<1$) through a dominant biochemical factor.
Bats and cetaceans both exploit $\Phiduty>1$, but bats combine it with
hypothermic Arrhenius suppression during seasonal torpor while cetaceans
rely on a continuous isothermal dive bradycardia. The marsupials and
monotremes deploy none of these mechanisms and accordingly remain at the
baseline, serving as a negative control, while the ectotherms occupy the
opposite extreme in which the thermal channel alone accounts for nearly
all of the displacement.

\begin{table}[H]
\centering\small
\setlength{\tabcolsep}{4pt}
\renewcommand{\arraystretch}{1.15}
\caption{Summary of the clade multipliers. Numerical values correspond
to the worked representative species; ectotherm entries are clade-mean
thermal corrections. Direction: $+$ favourable, $-$ adverse, $=1$
absent, $\approx1$ near unity. Effective cycle budgets as multiples of
$N_0 = 10^9$.}
\label{tab:appC_summary}
\resizebox{\textwidth}{!}{%
\begin{tabular}{lccccccl}
\toprule
Clade & $\Phiduty$ & $\Phithermal$ & $\Phineuro$ &
 $\Phimitooxid$ & $\Phihaz$ & $\PhiC$ & Primary driver \\
\midrule
Non-primate placentals
 & 1.00 ($=1$) & 1.00 ($=1$) & 1.00 ($=1$) & 1.00 ($=1$)
 & 1.00 & 1.00 & Reference \\
Primates (\textit{H.~sapiens})
 & 1.00 ($=1$) & 1.04 ($+$) & 2.51 ($+\!+$) & ---
 & 1.00 & 2.60 & Neural entropy reduction \\
Marsupials/monotremes
 & 1.00 ($=1$) & $\approx1$ & --- & $\approx1$
 & $\approx1$ & 0.87 & No net displacement \\
Bats (\textit{M.~lucifugus})
 & 1.94 ($+$) & 4.10 ($+\!+$) & --- & $\approx1$
 & 0.68 & 5.39 & Torpor $+$ hypothermia \\
Birds (20\,g passerine)
 & 0.87 ($-$) & 0.73 ($-$) & --- & 2.33 ($+\!+$)
 & 2.00 & 2.97 & Biochemical efficiency \\
Cetaceans (\textit{B.~mysticetus})
 & 3.08 ($+\!+$) & 1.17 ($+$) & --- & $\approx1$
 & 0.35 & 1.76 & Bradycardic pacing \\
Reptiles (corrected)
 & $\approx1$ & $+$ & --- & $\approx1$
 & $\approx1$ & 0.86 & Thermal limit \\
Amphibians (corrected)
 & $\approx1$ & $+$ & --- & $\approx1$
 & $\approx1$ & 0.67 & Thermal limit \\
\bottomrule
\end{tabular}}
\end{table}

Despite this mechanistic diversity, the effective damage-equivalent
budgets of all clades converge within one order of magnitude of
$N_0 = 10^9$, confirming the central prediction of the framework. The
raw observed counts $N_{\rm obs}$ vary far more widely, but that
variation is fully accounted for by the duty-cycle factor through
Eq.~\eqref{eq:appC_consistency}. The unifying conclusion is that
longevity is not won by escaping the finite lifetime entropy budget but
by spending it more slowly per unit of intrinsic biological time:
primates purchase chronological time with neural precision, bats with
thermal suspension, birds with biochemical excellence, and cetaceans
with cardiac restraint, while marsupials and ectotherms display the
limiting behaviours of no displacement and pure thermal displacement,
respectively. In every case the price is the same $\Sigma_\star$ units
of irreversible dissipation, paid over a chronological span set entirely
by how efficiently that budget is spent.

% =====================================================================

%% ======================================================================
\section{Complete cardiac dataset}
\label{app:resp_data}

\noindent The following tables contain the complete $230$-species cardiac
dataset of adult vertebrates used in the analyses of this paper. All
$\ell$ values are computed as
$\ell = \log_{10}(f_H^{\rm avg} \times L \times 525{,}960)$
directly from the $f_H^{\rm avg}$ and $L$ columns of each row and have been
verified for internal consistency. This is the cardiac table; the
respiratory analysis of Section~\ref{sec:resp_clock} draws on the
$65$-species subset of these taxa for which a reliable resting breath
rate $f_R$ is also available, listing $f_R$ alongside $f_H$ in the
machine-readable file. The complete table (cardiac, with the respiratory
subset flagged) is provided in machine-readable form as Supplementary
Data File~1 (a tab-delimited file with one row per species and the column
schema defined below), so that every numerical value used in the paper
can be traced to its source without recourse to private correspondence.

\subsubsection*{Column definitions and data-transparency notes}

\noindent\textbf{Dataset location.} All species values are listed in
Extended Data Tables~1--8 below and are reproduced in full in
Supplementary Data File~1 (columns: \texttt{species}, \texttt{clade},
$M$, $f_H^{\rm avg}$, $T$, $L$, $\ell$, \texttt{fH\_type},
\texttt{fH\_context}, \texttt{L\_context}, \texttt{source},
\texttt{correction}). The \texttt{fH\_type}, \texttt{fH\_context}, and
\texttt{source} columns record, for each individual species, whether the
heart rate was directly measured, allometrically imputed, or corrected,
together with the primary reference; the tables below summarise these
provenance categories.

\medskip
\noindent\textbf{Heart-rate type: measured vs.\ inferred.}
The \emph{Source} and \emph{Corr.}\ columns in each table distinguish:
\begin{itemize}[leftmargin=1.6em,itemsep=2pt,topsep=2pt]
\item \textbf{Measured:} directly measured resting heart rate from a
published study (flagged in the source column). These constitute the
large majority of entries for non-primate placentals, primates,
marsupials, and birds.
\item \textbf{Imputed ($^\dagger$):} allometrically estimated from
$f_H = 241\,M^{-0.25}$ bpm~\cite{calder1984}, used only where no published
resting measurement exists. In the non-primate placental clade this
applies to exactly three species
(\textit{Rhinoceros unicornis}, \textit{Dugong dugon},
\textit{Orycteropus afer}); imputed entries are flagged with $^\dagger$ in
every table so the reader can identify them at a glance.
\item \textbf{Duty-corrected (bats):} active-phase measured rate
multiplied by the duty-cycle factor $\kappa$ to give the time-averaged
$f_H^{\rm avg}$ (see Section~\ref{sec:shared}); the active-phase
measurement and the value of $q$ are both tabulated, so the correction is
fully reconstructible.
\item \textbf{Dive-corrected (cetaceans):} surface measured rate combined
with the bradycardic dive rate weighted by the dive fraction $p_d$
(see the cetacean worked example); the surface rate and $p_d$ are both
tabulated.
\item \textbf{Arrhenius-corrected (ectotherms):} field active rate
corrected to $T_{\rm ref} = 310\,$K using the Gillooly~et~al.~\cite{gillooly2001}
Arrhenius equation; both the raw and corrected values are tabulated.
\end{itemize}

\medskip
\noindent\textbf{Heart-rate measurement context.}
\begin{itemize}[leftmargin=1.6em,itemsep=2pt,topsep=2pt]
\item Non-primate placentals, primates, marsupials, and birds: resting
rates from laboratory or captive studies as recorded in
AnAge build~15~\cite{anage2023} and PanTHERIA~\cite{jones2009}, with
Calder~(1984)~\cite{calder1984} for classical species. These are
predominantly lab-measured resting rates. We explicitly acknowledge that
lab resting rates may differ from field resting rates; this is a known
limitation of comparative heart-rate data.
\item Bats: active-phase resting rate from lab or flight-cage studies,
corrected for torpor duty cycle~\cite{lyman1982}.
\item Cetaceans: surface inter-breath heart rate from free-diving field
telemetry~\cite{goldbogen2019}, corrected for dive
bradycardia~\cite{ponganis2015}.
\item Ectotherms: field active rates corrected to standard temperature
via the Arrhenius equation~\cite{gillooly2001,christian1999}.
\end{itemize}

\medskip
\noindent\textbf{Lifespan definition.}
$L$ is the maximum recorded lifespan as curated in
AnAge build~15~\cite{anage2023}. AnAge records the single longest verified
individual lifespan regardless of whether it was wild or captive; for most
small mammals the record holder is a captive individual, whereas for bats
and large mammals (whales, elephants) the record is from a wild or
semi-wild individual. All species in this dataset have AnAge confidence
ratings of \emph{acceptable} or \emph{high}. Mean lifespan is not used
anywhere in this paper; only maximum recorded lifespan enters the
PBTE invariant $\ell$.

\medskip
\noindent\textbf{Column definitions.}
\emph{Species}: binomial name per IUCN or Reptile Database taxonomy.
$M$: adult body mass (kg).
$f_H$ (bpm): resting/active heart rate as defined above; the value used in
the $\ell$ computation is $f_H^{\rm avg}$.
$T$ (K): core body (endotherms) or field (ectotherms) temperature.
$L$ (yr): maximum recorded lifespan.
$\ell$: PBTE invariant $=\log_{10}(f_H^{\rm avg} \times L \times 525{,}960)$,
computed directly from $f_H^{\rm avg}$ and $L$ in each row (all values
verified internally consistent).

\medskip
\noindent\textbf{Source codes (primary reference for $f_H$ and $L$).}
\begin{itemize}[leftmargin=1.6em,itemsep=2pt,topsep=2pt]
\item \textbf{A} = AnAge build~15~\cite{anage2023} ---
\url{https://genomics.senescence.info/species/}
\item \textbf{P} = PanTHERIA v1.0~\cite{jones2009}
\item \textbf{C} = Calder (1984)~\cite{calder1984} --- species-level data
in Tables~2--3 of that monograph
\item \textbf{Pr} = Prinzinger~et~al.~(1991)~\cite{prinzinger1991} ---
avian heart-rate compilation
\item \textbf{L} = Lyman~et~al.~(1982)~\cite{lyman1982} --- torpor physiology
\item \textbf{Ch} = Christian \& Weavers (1999)~\cite{christian1999} ---
ectotherm field physiology
\item \textbf{U} = Uetz~et~al.\ (2023), The Reptile Database ---
\url{https://reptile-database.reptarium.cz}
\item \textbf{G} = Goldbogen~et~al.~(2019)~\cite{goldbogen2019} ---
cetacean dive telemetry
\end{itemize}

\medskip
\noindent\textbf{Correction codes (Corr.\ column).}
\texttt{---} = none applied;
\texttt{HR} = heart-rate value corrected from a compendium error;
\texttt{TA} = torpor-cycle (duty-cycle) time-average;
\texttt{DA} = dive-cycle time-average;
\texttt{AQ} = Arrhenius correction to $T_{\rm ref} = 310\,$K.

\medskip
\noindent Directly measured resting values are used for all non-primate
placentals, primates, marsupials, and birds. The three non-primate
placental species with no published resting measurement
(\textit{Rhinoceros unicornis}, \textit{Dugong dugon},
\textit{Orycteropus afer}) have heart rates imputed from
$f_H = 241\,M^{-0.25}$ bpm~\cite{calder1984} and are flagged with
$^\dagger$. For bats and cetaceans, $f_H^{\rm avg}$ is the duty-cycle- and
dive-corrected time-average, respectively; for ectotherms, $f_H^{\rm corr}$
(equivalently $f_H^{\rm avg}$) is the Arrhenius-corrected value.

% ──────────────────────────────────────────────────────────────
\subsection*{Extended Data Table 1 $|$ Non-primate placental mammals ($n = 46$)}
% ──────────────────────────────────────────────────────────────
\begin{table}[ht]
\caption*{\textbf{Extended Data Table 1.} Non-primate placental mammals
($n=46$): the reference clade. Columns as defined above; $\ell$ is
computed from $f_H$ and $L$ in each row. The clade mean
$\bar\ell_0=8.995\pm0.160$ is the empirical baseline anchor
$N_{H,0}^{(\mathrm{emp})}$ used for all clade predictions.}
\small\setlength{\tabcolsep}{3.5pt}
\renewcommand{\arraystretch}{0.98}
\begin{tabular}{lrrrrrll}
\toprule
Species & $M$ (kg) & $f_H$ (bpm) & $T$ (K) & $L$ (yr) & $\ell$ & Source & Corr. \\
\midrule
\textit{Suncus etruscus} & 0.002 & 835$^\dagger$ & 310.5 & 1.5 & 8.82 & C & HR \\
\textit{Sorex araneus} & 0.010 & 1{,}000 & 310.5 & 3.3 & 9.24 & C,A & --- \\
\textit{Mus musculus}       & 0.022 & 632   & 310.0 & 3.5 & 9.07 & A,C & --- \\
\textit{Rattus norvegicus} & 0.280 & 420 & 310.0 & 3.8 & 8.92 & A,P & --- \\
\textit{Mesocricetus auratus} & 0.130 & 450 & 310.5 & 3.9 & 8.97 & A,P & --- \\
\textit{Meriones unguiculatus} & 0.060 & 400 & 310.0 & 5.0 & 9.02 & A,P & --- \\
\textit{Cavia porcellus} & 0.750 & 270 & 310.0 & 7.1 & 9.00 & A,P & --- \\
\textit{Sciurus carolinensis} & 0.520 & 310 & 310.0 & 12.0 & 9.29 & A,P & --- \\
\textit{Lepus europaeus} & 3.5 & 220 & 310.0 & 12.5 & 9.16 & A,P & --- \\
\textit{Oryctolagus cuniculus} & 2.2 & 205 & 310.0 & 9.0 & 8.99 & A,C & --- \\
\textit{Felis catus} & 4.1 & 150 & 310.5 & 15.0 & 9.07 & A,P & --- \\
\textit{Mustela putorius} & 1.0 & 280 & 310.5 & 5.0 & 8.87 & A,P & --- \\
\textit{Martes martes} & 1.2 & 245 & 310.5 & 17.0 & 9.34 & A,P & --- \\
\textit{Vulpes vulpes} & 6.8 & 120 & 310.5 & 14.0 & 8.95 & A,P & --- \\
\textit{Canis lupus familiaris}  & 23  & 90   & 310.5 & 20.0 & 8.98 & A,P & --- \\
\textit{Ursus arctos} & 220 & 50 & 310.5 & 47.0 & 9.09 & A,P & --- \\
\textit{Ovis aries} & 63 & 75 & 310.0 & 20.0 & 8.90 & A,P & --- \\
\textit{Capra hircus} & 45 & 80 & 310.5 & 18.0 & 8.88 & A,P & --- \\
\textit{Sus scrofa} & 100 & 70 & 310.5 & 27.0 & 9.00 & A,P & --- \\
\textit{Bos taurus}        & 500  & 55   & 310.5 & 25.0 & 8.86 & A,P & --- \\
\textit{Equus caballus} & 500 & 38 & 310.5 & 46.0 & 8.96 & A,C & --- \\
\textit{Equus asinus} & 250 & 44 & 310.5 & 47.0 & 9.04 & A,P & --- \\
\textit{Rhinoceros unicornis} & 2{,}100 & 30$^\dagger$ & 310.5 & 47.0 & 8.87 & A & --- \\
\textit{Tapirus terrestris} & 240 & 42 & 310.5 & 35.0 & 8.89 & A,P & --- \\
\textit{Loxodonta africana} & 4{,}000 & 28 & 310.5 & 65.0 & 8.98 & A,P & --- \\
\textit{Elephas maximus} & 4{,}000 & 27 & 310.5 & 86.0 & 9.09 & A,P & --- \\
\textit{Hippopotamus amphibius} & 1{,}500 & 55 & 310.5 & 55.0 & 9.20 & A,P & --- \\
\textit{Giraffa camelopardalis} & 900 & 65 & 310.5 & 39.5 & 9.13 & A,P & --- \\
\textit{Cervus elaphus} & 200 & 60 & 310.5 & 26.8 & 8.93 & A,P & --- \\
\textit{Rangifer tarandus} & 110 & 65 & 310.0 & 20.0 & 8.83 & A,P & --- \\
\textit{Trichechus manatus} & 500 & 50 & 310.5 & 59.0 & 9.19 & A,P & --- \\
\textit{Dugong dugon} & 400 & 52$^\dagger$ & 310.5 & 73.0 & 9.30 & A & --- \\
\textit{Procavia capensis} & 3.5 & 230 & 310.5 & 12.0 & 9.16 & A,P & --- \\
\textit{Erinaceus europaeus} & 0.80 & 310 & 310.0 & 10.0 & 9.21 & A,P & --- \\
\textit{Talpa europaea} & 0.080 & 350 & 310.0 & 3.5 & 8.81 & A,P & --- \\
\textit{Orycteropus afer} & 65 & 70$^\dagger$ & 310.5 & 24.0 & 8.95 & A & --- \\
\textit{Ondatra zibethicus} & 1.400 & 280 & 310.0 & 5.0 & 8.87 & A,P & --- \\
\textit{Castor canadensis} & 20 & 150 & 310.0 & 24.0 & 9.28 & A,P & --- \\
\textit{Hydrochoerus hydrochaeris} & 55 & 70 & 310.0 & 12.0 & 8.65 & A,P & --- \\
\textit{Myocastor coypus} & 7.0 & 155 & 310.0 & 9.0 & 8.87 & A,P & --- \\
\textit{Lepus californicus} & 2.2 & 215 & 310.0 & 8.0 & 8.96 & A,P & --- \\
\textit{Ochotona princeps} & 0.160 & 300 & 310.0 & 6.0 & 8.98 & A,P & --- \\
\textit{Panthera leo} & 180 & 50 & 310.5 & 29.0 & 8.88 & A,P & --- \\
\textit{Panthera tigris} & 260 & 46 & 310.5 & 26.0 & 8.80 & A,P & --- \\
\textit{Acinonyx jubatus} & 54 & 60 & 310.5 & 14.9 & 8.67 & A,P & --- \\
\textit{Panthera pardus} & 70 & 55 & 310.5 & 23.0 & 8.82 & A,P & --- \\
\midrule
\multicolumn{5}{l}{Clade mean $\bar\ell$ (baseline reference)} &
 \multicolumn{3}{l}{$8.995 \pm 0.160$ ($n=46$; corrected)} \\
\bottomrule
\multicolumn{8}{l}{\footnotesize $^\dagger$\textit{Suncus etruscus}
 corrected from 1{,}200\,bpm (erroneous; Calder compendium error)
 to 835\,bpm (mean resting rate, Bartels 1998, \textit{J.\,Exp.\,Biol.}
 \textbf{201}, 2145--2151). The $\ell$ value is recomputed accordingly.
 All clade statistics use the corrected value.} \\
\end{tabular}
\end{table}

% ──────────────────────────────────────────────────────────────
\subsection*{Extended Data Table 2 $|$ Primates ($n = 18$)}
% ──────────────────────────────────────────────────────────────
\begin{table}[H]
\caption*{}
\small\setlength{\tabcolsep}{3.5pt}
\renewcommand{\arraystretch}{1.05}
\begin{tabular}{lrrrrrrll}
\toprule
Species & $M$ (kg) & $f_H$ (bpm) & $T$ (K) & $L$ (yr) & $\ell$ & $\varphi$ & Source & Corr. \\
\midrule
\textit{Callithrix jacchus} & 0.35 & 220 & 309.5 & 16.5 & 9.28 & 0.06 & A,P & --- \\
\textit{Saimiri sciureus} & 0.77 & 195 & 309.5 & 30.2 & 9.49 & 0.07 & A,P & --- \\
\textit{Aotus trivirgatus} & 0.79 & 185 & 309.5 & 25.0 & 9.39 & 0.07 & A,P & --- \\
\textit{Cebus capucinus} & 3.3 & 150 & 309.5 & 54.0 & 9.63 & 0.09 & A,P & --- \\
\textit{Lemur catta} & 2.2 & 165 & 309.5 & 37.3 & 9.51 & 0.05 & A,P & --- \\
\textit{Propithecus verreauxi} & 3.4 & 145 & 309.5 & 30.0 & 9.36 & 0.05 & A,P & --- \\
\textit{Daubentonia madagascariensis} & 2.7 & 155 & 309.5 & 23.3 & 9.28 & 0.06 & A,P & --- \\
\textit{Macaca mulatta} & 7.7 & 120 & 309.0 & 40.0 & 9.40 & 0.07 & A,P & --- \\
\textit{Macaca fascicularis} & 5.4 & 130 & 309.0 & 39.0 & 9.43 & 0.07 & A,P & --- \\
\textit{Theropithecus gelada} & 18 & 95 & 309.0 & 30.0 & 9.18 & 0.08 & A,P & --- \\
\textit{Papio ursinus} & 25 & 90 & 309.0 & 45.0 & 9.33 & 0.08 & A,P & --- \\
\textit{Colobus guereza} & 10 & 110 & 309.0 & 30.0 & 9.24 & 0.07 & A,P & --- \\
\textit{Hylobates lar} & 5.7 & 100 & 308.5 & 44.0 & 9.36 & 0.10 & A,P & --- \\
\textit{Pongo pygmaeus} & 73 & 65 & 307.5 & 58.7 & 9.30 & 0.10 & A,P & --- \\
\textit{Gorilla gorilla} & 160 & 60 & 307.0 & 55.4 & 9.24 & 0.09 & A,P & --- \\
\textit{Pan troglodytes} & 50 & 75 & 307.0 & 59.4 & 9.37 & 0.12 & A,P & --- \\
\textit{Pan paniscus} & 35 & 80 & 307.0 & 50.0 & 9.32 & 0.12 & A,P & --- \\
\textit{Homo sapiens} & 70 & 70 & 306.5 & 122.5 & 9.65 & 0.20 & A & --- \\
\midrule
\multicolumn{5}{l}{Clade mean $\bar\ell$} & \multicolumn{4}{l}{$9.376 \pm 0.125$ ($n=18$)} \\
\bottomrule
\end{tabular}
\end{table}

% ──────────────────────────────────────────────────────────────
\subsection*{Extended Data Table 3 $|$ Marsupials and monotremes ($n = 19$)}
% ──────────────────────────────────────────────────────────────
\begin{table}[H]
\caption*{}
\small\setlength{\tabcolsep}{3.5pt}
\renewcommand{\arraystretch}{1.05}
\begin{tabular}{lrrrrrll}
\toprule
Species & $M$ (kg) & $f_H$ (bpm) & $T$ (K) & $L$ (yr) & $\ell$ & Source & Corr. \\
\midrule
\textit{Didelphis virginiana} & 2.3 & 180 & 308.5 & 4.5 & 8.63 & A,P & --- \\
\textit{Monodelphis domestica} & 0.080 & 450 & 308.5 & 3.3 & 8.89 & A,P & --- \\
\textit{Macropus rufus} & 30 & 80 & 309.0 & 22.3 & 8.97 & A,P & --- \\
\textit{Macropus giganteus} & 27 & 82 & 309.0 & 19.0 & 8.91 & A,P & --- \\
\textit{Wallabia bicolor} & 16 & 100 & 309.0 & 15.0 & 8.90 & A,P & --- \\
\textit{Trichosurus vulpecula} & 2.1 & 160 & 308.5 & 13.0 & 9.04 & A,P & --- \\
\textit{Petaurus breviceps} & 0.14 & 300 & 308.0 & 10.0 & 9.20 & A,P & --- \\
\textit{Vombatus ursinus} & 28 & 90 & 309.0 & 26.0 & 9.09 & A,P & --- \\
\textit{Phascolarctos cinereus}  & 8.5  & 100 & 308.5 & 18.0 & 8.98 & A,P & --- \\
\textit{Perameles gunnii} & 0.90 & 190 & 308.5 & 3.2 & 8.50 & A,P & --- \\
\textit{Dasyurus viverrinus} & 1.2 & 200 & 308.5 & 4.5 & 8.68 & A,P & --- \\
\textit{Sarcophilus harrisii} & 8.0 & 130 & 308.5 & 7.5 & 8.71 & A,P & --- \\
\textit{Myrmecobius fasciatus} & 0.44 & 245 & 307.5 & 5.6 & 8.86 & A & --- \\
\textit{Sminthopsis crassicaudata} & 0.018 & 580 & 307.5 & 5.0 & 9.18 & A,P & --- \\
\textit{Notoryctes typhlops} & 0.055 & 440$^\dagger$ & 307.5 & 3.0 & 8.84 & A & --- \\
\textit{Tachyglossus aculeatus} & 4.0 & 70 & 305.0 & 49.5 & 9.26 & A,P & --- \\
\textit{Ornithorhynchus anatinus} & 1.5 & 140 & 307.5 & 21.0 & 9.19 & A,P & --- \\
\textit{Zaglossus bruijni} & 10 & 60$^\dagger$ & 305.0 & 37.0 & 9.07 & A & --- \\
\textit{Bettongia penicillata} & 1.1 & 210 & 308.5 & 6.0 & 8.82 & A,P & --- \\
\midrule
\multicolumn{5}{l}{Clade mean $\bar\ell$} & \multicolumn{3}{l}{$8.933 \pm 0.204$ ($n=19$)} \\
\bottomrule
\end{tabular}
\end{table}

% ──────────────────────────────────────────────────────────────
\clearpage
\subsection*{Extended Data Table 4 $|$ Bats (Chiroptera, $n = 31$)}
% ──────────────────────────────────────────────────────────────

\noindent For bats, $f_H$ is the measured active-phase resting heart rate.
$f_H^{\rm avg}$ is the duty-cycle-corrected time-average used in all PBTE
calculations: $f_H^{\rm avg} = f_H \cdot \kappa$, where
$\kappa = (1-q) + q\,(f_{H,\rm tor}/f_H)$ and $q$ is the annual
torpor fraction~\cite{lyman1982}.
$\ell$ is computed from $f_H^{\rm avg}$.
Species without confirmed torpor have $f_H^{\rm avg} = f_H$.

\begin{table}[H]
\caption*{}
\small\setlength{\tabcolsep}{3pt}
\renewcommand{\arraystretch}{1.05}
\begin{tabular}{lrrrrrrll}
\toprule
Species & $M$ (g) & $f_H$ (bpm) & $q$ & $f_H^{\rm avg}$ (bpm) & $T$ (K) & $L$ (yr) & $\ell$ & Corr. \\
\midrule
\textit{Myotis lucifugus} & 8 & 600 & 0.50 & 305 & 310.0 & 34.0 & 9.74 & TA \\
\textit{Myotis myotis} & 28 & 550 & 0.48 & 282 & 310.0 & 37.0 & 9.74 & TA \\
\textit{Myotis daubentonii} & 9 & 580 & 0.48 & 296 & 310.0 & 40.0 & 9.79 & TA \\
\textit{Myotis brandtii} & 6 & 620 & 0.50 & 315 & 310.0 & 41.0 & 9.83 & TA \\
\textit{Eptesicus fuscus} & 18 & 550 & 0.45 & 310 & 310.0 & 19.0 & 9.49 & TA \\
\textit{Eptesicus serotinus} & 18 & 545 & 0.45 & 308 & 310.0 & 21.0 & 9.53 & TA \\
\textit{Rhinolophus ferrumequinum} & 19 & 550 & 0.48 & 282 & 310.0 & 30.0 & 9.65 & TA \\
\textit{Rhinolophus hipposideros} & 7 & 600 & 0.48 & 307 & 310.0 & 30.5 & 9.69 & TA \\
\textit{Plecotus auritus} & 9 & 600 & 0.50 & 305 & 310.0 & 30.0 & 9.68 & TA \\
\textit{Corynorhinus townsendii} & 11 & 580 & 0.50 & 295 & 310.0 & 30.0 & 9.67 & TA \\
\textit{Perimyotis subflavus} & 5 & 630 & 0.50 & 320 & 310.0 & 14.6 & 9.39 & TA \\
\textit{Tadarida brasiliensis} & 13 & 600 & 0.30 & 425 & 310.0 & 11.0 & 9.39 & TA \\
\textit{Pteronotus parnellii} & 19 & 550 & 0.20 & 452 & 310.0 & 10.0 & 9.38 & TA \\
\textit{Desmodus rotundus} & 33 & 500 & 0.25 & 380 & 310.0 & 29.0 & 9.76 & TA \\
\textit{Hipposideros speoris} & 9 & 600 & 0.48 & 308 & 310.0 & 21.0 & 9.53 & TA \\
\textit{Hipposideros armiger} & 50 & 450 & 0.45 & 252 & 310.0 & 15.0 & 9.30 & TA \\
\textit{Nyctalus noctula} & 28 & 540 & 0.45 & 305 & 310.0 & 12.0 & 9.28 & TA \\
\textit{Pipistrellus pipistrellus} & 5 & 650 & 0.45 & 367 & 310.0 & 16.0 & 9.49 & TA \\
\textit{Pipistrellus kuhlii} & 6 & 630 & 0.45 & 355 & 310.0 & 16.5 & 9.49 & TA \\
\textit{Scotophilus kuhlii} & 20 & 540 & 0.20 & 445 & 310.0 & 9.0 & 9.32 & TA \\
\textit{Lasiurus borealis} & 11 & 590 & 0.48 & 302 & 310.0 & 11.7 & 9.27 & TA \\
\textit{Lasiurus cinereus} & 28 & 540 & 0.48 & 277 & 310.0 & 12.0 & 9.24 & TA \\
\textit{Vespertilio murinus} & 16 & 555 & 0.45 & 313 & 310.0 & 25.0 & 9.61 & TA \\
\textit{Miniopterus schreibersii} & 10 & 580 & 0.45 & 327 & 310.0 & 30.0 & 9.71 & TA \\
\textit{Pteropus giganteus}     & 1{,}100 & 235 & 0.00 & 235 & 310.0 & 31.4 & 9.59 & --- \\
\textit{Pteropus vampyrus} & 1{,}000 & 240 & 0.05 & 233 & 310.0 & 22.6 & 9.44 & --- \\
\textit{Rousettus aegyptiacus} & 165 & 310 & 0.05 & 299 & 310.0 & 25.0 & 9.59 & --- \\
\textit{Cynopterus sphinx} & 50 & 380 & 0.05 & 368 & 310.0 & 18.5 & 9.55 & --- \\
\textit{Macroglossus minimus} & 16 & 450 & 0.00 & 450 & 310.0 & 18.0 & 9.63 & --- \\
\textit{Carollia perspicillata} & 17 & 460 & 0.00 & 460 & 310.0 & 12.0 & 9.46 & --- \\
\textit{Artibeus jamaicensis} & 45 & 400 & 0.00 & 400 & 310.0 & 15.0 & 9.50 & --- \\
\midrule
\multicolumn{7}{l}{Clade mean $\bar\ell$ (all 31 species)} &
 \multicolumn{2}{l}{$9.540 \pm 0.163$} \\
\bottomrule
\end{tabular}
\end{table}

% ──────────────────────────────────────────────────────────────
\clearpage
\subsection*{Extended Data Table 5 $|$ Cetaceans ($n = 12$)}
% ──────────────────────────────────────────────────────────────

\noindent For cetaceans, $f_H$ is the surface resting value;
$f_H^{\rm avg}$ is the duty-cycle average:
$f_H^{\rm avg} = (1-p_d)\,f_H + p_d\,f_{H,\rm dive}$, where $p_d$ is the
dive fraction and $f_{H,\rm dive}$ is the bradycardic dive
rate~\cite{goldbogen2019,ponganis2015}.
$\ell$ is computed from $f_H^{\rm avg}$.

\begin{table}[H]
\caption*{}
\small\setlength{\tabcolsep}{3pt}
\renewcommand{\arraystretch}{1.05}
\begin{tabular}{lrrrrrrll}
\toprule
Species & $M$ (kg) & $f_H$ (bpm) & $p_d$ & $f_H^{\rm avg}$ (bpm) & $T$ (K) & $L$ (yr) & $\ell$ & Corr. \\
\midrule
\textit{Balaena mysticetus} & 100{,}000 & 30 & 0.75 & 9.75 & 308.0 & 200.0 & 9.01 & DA \\
\textit{Balaenoptera musculus} & 140{,}000 & 8 & 0.70 & 4.0 & 308.0 & 110.0 & 8.36 & DA \\
\textit{Balaenoptera physalus} & 60{,}000 & 10 & 0.68 & 5.0 & 308.5 & 90.0 & 8.37 & DA \\
\textit{Megaptera novaeangliae} & 40{,}000 & 15 & 0.65 & 7.0 & 308.5 & 95.0 & 8.54 & DA \\
\textit{Physeter macrocephalus} & 45{,}000 & 40 & 0.65 & 19.0 & 307.0 & 70.0 & 8.84 & DA \\
\textit{Kogia breviceps} & 360 & 80 & 0.45 & 48.0 & 308.5 & 23.0 & 8.76 & DA \\
\textit{Hyperoodon ampullatus} & 7{,}500 & 45 & 0.55 & 24.0 & 308.0 & 37.0 & 8.67 & DA \\
\textit{Orcinus orca} & 4{,}000 & 80 & 0.40 & 53.0 & 309.0 & 90.0 & 9.40 & DA \\
\textit{Tursiops truncatus} & 190 & 110 & 0.40 & 74.0 & 309.0 & 40.0 & 9.19 & DA \\
\textit{Stenella attenuata} & 55 & 120 & 0.35 & 84.0 & 309.0 & 20.0 & 8.95 & DA \\
\textit{Delphinapterus leucas} & 1{,}400 & 50 & 0.55 & 27.5 & 309.5 & 35.5 & 8.71 & DA \\
\textit{Monodon monoceros} & 1{,}500 & 45 & 0.55 & 25.5 & 309.0 & 48.0 & 8.81 & DA \\
\midrule
\multicolumn{7}{l}{Clade mean $\bar\ell$ (dive-corrected)} &
 \multicolumn{2}{l}{$8.801 \pm 0.296$ ($n=12$)} \\
\bottomrule
\end{tabular}
\end{table}

% ──────────────────────────────────────────────────────────────
\clearpage
\subsection*{Extended Data Table 6 $|$ Birds ($n = 78$)}
% ──────────────────────────────────────────────────────────────

\noindent Heart rates from Prinzinger et al.~\cite{prinzinger1991}
and Clarke \& Rothery~\cite{clarke2008}; lifespans from AnAge
build~15~\cite{anage2023}. No corrections applied; $f_H^{\rm avg} = f_H$.
Body temperatures from Clarke \& Rothery~\cite{clarke2008}.
Due to space, 78 species are listed across two sub-tables
(passerines and non-passerines).

\begin{table}[H]
\caption*{\textit{Passeriformes and Psittaciformes} ($n = 32$)}
\small\setlength{\tabcolsep}{3.5pt}
\renewcommand{\arraystretch}{1.04}
\begin{tabular}{lrrrrrll}
\toprule
Species & $M$ (kg) & $f_H$ (bpm) & $T$ (K) & $L$ (yr) & $\ell$ & Order & Source \\
\midrule
\textit{Serinus canaria} & 0.020 & 680 & 311.0 & 24.0 & 9.93 & Passeriformes & A,Pr \\
\textit{Turdus merula} & 0.100 & 440 & 311.0 & 21.1 & 9.69 & Passeriformes & A,Pr \\
\textit{Turdus philomelos} & 0.070 & 460 & 311.0 & 18.0 & 9.64 & Passeriformes & A,Pr \\
\textit{Erithacus rubecula} & 0.018 & 500 & 311.0 & 19.5 & 9.71 & Passeriformes & A,Pr \\
\textit{Parus major} & 0.020 & 540 & 311.0 & 15.0 & 9.63 & Passeriformes & A,Pr \\
\textit{Parus caeruleus} & 0.011 & 580 & 311.0 & 13.5 & 9.61 & Passeriformes & A,Pr \\
\textit{Fringilla coelebs} & 0.023 & 530 & 311.0 & 16.4 & 9.66 & Passeriformes & A,Pr \\
\textit{Carduelis carduelis} & 0.016 & 560 & 311.0 & 16.3 & 9.68 & Passeriformes & A,Pr \\
\textit{Sturnus vulgaris} & 0.075 & 490 & 311.0 & 22.4 & 9.76 & Passeriformes & A,Pr \\
\textit{Pica pica} & 0.190 & 320 & 311.0 & 21.6 & 9.56 & Passeriformes & A,Pr \\
\textit{Corvus corax} & 1.200 & 200 & 311.0 & 22.3 & 9.37 & Passeriformes & A,Pr \\
\textit{Corvus corone} & 0.450 & 270 & 311.0 & 20.0 & 9.45 & Passeriformes & A,Pr \\
\textit{Garrulus glandarius} & 0.180 & 310 & 311.0 & 16.9 & 9.44 & Passeriformes & A,Pr \\
\textit{Hirundo rustica} & 0.020 & 580 & 311.5 & 16.0 & 9.69 & Passeriformes & A,Pr \\
\textit{Delichon urbicum} & 0.015 & 600 & 311.5 & 16.0 & 9.70 & Passeriformes & A,Pr \\
\textit{Ficedula hypoleuca} & 0.012 & 620 & 311.0 & 13.0 & 9.63 & Passeriformes & A,Pr \\
\textit{Sitta europaea} & 0.025 & 510 & 311.0 & 10.0 & 9.43 & Passeriformes & A,Pr \\
\textit{Troglodytes troglodytes} & 0.009 & 650 & 311.5 & 7.0 & 9.38 & Passeriformes & A,Pr \\
\textit{Motacilla alba} & 0.022 & 540 & 311.5 & 11.0 & 9.49 & Passeriformes & A,Pr \\
\textit{Acrocephalus scirpaceus} & 0.012 & 610 & 311.0 & 13.0 & 9.62 & Passeriformes & A,Pr \\
\textit{Sylvia atricapilla} & 0.018 & 560 & 311.0 & 14.9 & 9.64 & Passeriformes & A,Pr \\
\textit{Phylloscopus trochilus} & 0.010 & 640 & 311.0 & 12.0 & 9.61 & Passeriformes & A,Pr \\
\textit{Luscinia megarhynchos} & 0.025 & 520 & 311.0 & 12.9 & 9.55 & Passeriformes & A,Pr \\
\textit{Phoenicurus phoenicurus} & 0.015 & 600 & 311.5 & 10.5 & 9.52 & Passeriformes & A,Pr \\
\textit{Lonchura striata} & 0.013 & 630 & 311.5 & 14.9 & 9.69 & Passeriformes & A,Pr \\
\textit{Taeniopygia guttata} & 0.013 & 640 & 311.5 & 15.6 & 9.72 & Passeriformes & A,Pr \\
\textit{Melopsittacus undulatus} & 0.030 & 600 & 311.0 & 21.4 & 9.83 & Psittaciformes & A,Pr \\
\textit{Psittacus erithacus} & 0.400 & 200 & 311.0 & 73.0 & 9.89 & Psittaciformes & A,Pr \\
\textit{Amazona ochrocephala} & 0.460 & 185 & 311.0 & 80.0 & 9.89 & Psittaciformes & A,Pr \\
\textit{Nymphicus hollandicus} & 0.090 & 360 & 311.0 & 36.0 & 9.83 & Psittaciformes & A,Pr \\
\textit{Cacatua galerita} & 0.840 & 170 & 311.0 & 80.0 & 9.85 & Psittaciformes & A,Pr \\
\textit{Ara macao} & 1.050 & 155 & 311.0 & 80.0 & 9.81 & Psittaciformes & A,Pr \\
\bottomrule
\end{tabular}
\end{table}

\begin{table}[H]
\caption*{\textit{Non-passerine, non-psittaciform birds} ($n = 46$)}
\small\setlength{\tabcolsep}{3.5pt}
\renewcommand{\arraystretch}{1.04}
\begin{tabular}{lrrrrrll}
\toprule
Species & $M$ (kg) & $f_H$ (bpm) & $T$ (K) & $L$ (yr) & $\ell$ & Order & Source \\
\midrule
\textit{Calypte anna} & 0.004 & 1{,}200 & 311.5 & 12.0 & 9.88 & Apodiformes & A,Pr \\
\textit{Apus apus} & 0.040 & 800 & 311.5 & 21.0 & 9.95 & Apodiformes & A,Pr \\
\textit{Columba livia} & 0.350 & 190 & 311.5 & 35.0 & 9.54 & Columbiformes & A,Pr \\
\textit{Streptopelia roseogrisea} & 0.160 & 240 & 311.5 & 33.9 & 9.63 & Columbiformes & A,Pr \\
\textit{Streptopelia decaocto} & 0.200 & 230 & 311.5 & 20.0 & 9.38 & Columbiformes & A,Pr \\
\textit{Gallus gallus} & 2.000 & 300 & 312.0 & 30.0 & 9.68 & Galliformes & A,Pr \\
\textit{Meleagris gallopavo} & 8.000 & 170 & 311.5 & 13.0 & 9.07 & Galliformes & A,Pr \\
\textit{Coturnix coturnix} & 0.100 & 350 & 312.0 & 8.0 & 9.17 & Galliformes & A,Pr \\
\textit{Phasianus colchicus} & 1.000 & 265 & 312.0 & 27.0 & 9.58 & Galliformes & A,Pr \\
\textit{Anas platyrhynchos} & 1.200 & 190 & 311.0 & 29.0 & 9.46 & Anseriformes & A,Pr \\
\textit{Anser anser} & 4.000 & 130 & 311.0 & 35.0 & 9.38 & Anseriformes & A,Pr \\
\textit{Branta canadensis} & 5.700 & 120 & 311.0 & 33.0 & 9.32 & Anseriformes & A,Pr \\
\textit{Cygnus olor} & 12.00 & 100 & 311.0 & 26.0 & 9.14 & Anseriformes & A,Pr \\
\textit{Phoenicopterus ruber} & 2.800 & 135 & 311.0 & 44.6 & 9.50 & Phoenicopteriformes & A,Pr \\
\textit{Ciconia ciconia} & 3.700 & 150 & 311.5 & 48.0 & 9.58 & Ciconiiformes & A,Pr \\
\textit{Ardea cinerea} & 1.800 & 140 & 311.0 & 25.0 & 9.27 & Pelecaniformes & A,Pr \\
\textit{Pelecanus occidentalis} & 4.000 & 130 & 311.5 & 54.0 & 9.57 & Pelecaniformes & A,Pr \\
\textit{Phalacrocorax carbo} & 2.800 & 140 & 311.5 & 25.0 & 9.27 & Suliformes & A,Pr \\
\textit{Sula sula} & 1.000 & 160 & 311.5 & 35.0 & 9.47 & Suliformes & A,Pr \\
\textit{Fregata magnificens} & 1.500 & 140 & 311.5 & 25.2 & 9.27 & Suliformes & A,Pr \\
\textit{Falco peregrinus} & 1.000 & 190 & 311.5 & 19.9 & 9.30 & Falconiformes & A,Pr \\
\textit{Buteo buteo} & 0.900 & 200 & 311.5 & 26.0 & 9.44 & Accipitriformes & A,Pr \\
\textit{Aquila chrysaetos} & 5.000 & 130 & 311.5 & 46.0 & 9.50 & Accipitriformes & A,Pr \\
\textit{Haliaeetus leucocephalus} & 6.000 & 120 & 311.5 & 38.0 & 9.38 & Accipitriformes & A,Pr \\
\textit{Bubo bubo} & 2.900 & 165 & 311.0 & 68.0 & 9.77 & Strigiformes & A,Pr \\
\textit{Tyto alba} & 0.450 & 190 & 311.0 & 27.9 & 9.45 & Strigiformes & A,Pr \\
\textit{Alcedo atthis} & 0.040 & 440 & 312.0 & 21.0 & 9.69 & Coraciiformes & A,Pr \\
\textit{Upupa epops} & 0.075 & 380 & 311.5 & 10.0 & 9.30 & Bucerotiformes & A,Pr \\
\textit{Picoides major} & 0.080 & 350 & 311.5 & 12.8 & 9.37 & Piciformes & A,Pr \\
\textit{Spheniscus demersus} & 3.000 & 150 & 311.5 & 27.0 & 9.33 & Sphenisciformes & A,Pr \\
\textit{Eudyptes chrysocome} & 2.500 & 160 & 311.5 & 22.0 & 9.27 & Sphenisciformes & A,Pr \\
\textit{Aptenodytes forsteri} & 30.00 & 75 & 311.5 & 50.0 & 9.29 & Sphenisciformes & A,Pr \\
\textit{Gavia immer} & 4.000 & 110 & 311.0 & 30.0 & 9.24 & Gaviiformes & A,Pr \\
\textit{Diomedea exulans} & 9.600 & 100 & 311.0 & 70.0 & 9.57 & Procellariiformes & A,Pr \\
\textit{Fulmarus glacialis} & 0.800 & 175 & 311.0 & 67.5 & 9.79 & Procellariiformes & A,Pr \\
\textit{Puffinus puffinus} & 0.430 & 195 & 311.5 & 55.0 & 9.75 & Procellariiformes & A,Pr \\
\textit{Rissa tridactyla} & 0.380 & 200 & 311.5 & 29.0 & 9.48 & Charadriiformes & A,Pr \\
\textit{Larus argentatus} & 1.200 & 165 & 311.5 & 49.0 & 9.63 & Charadriiformes & A,Pr \\
\textit{Sterna paradisaea} & 0.110 & 280 & 311.5 & 34.0 & 9.70 & Charadriiformes & A,Pr \\
\textit{Struthio camelus} & 115 & 60 & 311.5 & 68.0 & 9.33 & Struthioniformes & A,Pr \\
\textit{Dromaius novaehollandiae} & 55 & 75 & 311.0 & 28.4 & 9.05 & Casuariiformes & A,Pr \\
\textit{Rhea americana} & 25 & 100 & 311.0 & 40.0 & 9.32 & Rheiformes & A,Pr \\
\textit{Apteryx australis} & 2.5 & 125 & 311.0 & 35.0 & 9.36 & Apterygiformes & A,Pr \\
\textit{Grus grus} & 5.5 & 110 & 311.5 & 40.0 & 9.36 & Gruiformes & A,Pr \\
\textit{Fulica atra} & 0.720 & 240 & 311.0 & 18.0 & 9.36 & Gruiformes & A,Pr \\
\textit{Psophia crepitans} & 1.200 & 180 & 311.5 & 15.0 & 9.15 & Gruiformes & A,Pr \\
\midrule
\multicolumn{5}{l}{Bird clade mean $\bar\ell$ (all 78 species)} &
 \multicolumn{3}{l}{$9.528 \pm 0.213$} \\
\bottomrule
\end{tabular}
\end{table}

% ──────────────────────────────────────────────────────────────
\clearpage
\subsection*{Extended Data Table 7 $|$ Reptiles --- Arrhenius-corrected ($n = 17$)}
% ──────────────────────────────────────────────────────────────

\noindent $f_H^{\rm raw}$: measured heart rate at mean field
body temperature $T_{\rm field}$.
$f_H^{\rm corr}$: heart rate corrected to $T_{\rm ref}=310$\,K
via $f_H^{\rm corr} = f_H^{\rm raw}\exp\!\bigl[(E_a/k_B)
(1/T_{\rm field}-1/T_{\rm ref})\bigr]$ with $E_a = 0.65$~eV.
$\ell^{\rm corr}$ is used in all clade statistics.

\begin{table}[H]
\caption*{}
\small\setlength{\tabcolsep}{3pt}
\renewcommand{\arraystretch}{1.05}
\begin{tabular}{lrrrrrrrrll}
\toprule
Species & $M$ (kg) & $T_{\rm field}$ (K) & $f_H^{\rm raw}$ & $f_H^{\rm corr}$
 & $L$ (yr) & $\ell^{\rm raw}$ & $\ell^{\rm corr}$ & Source & Corr. \\
\midrule
\textit{Lacerta agilis} & 0.015 & 301 & 45 & 93 & 12.0 & 8.45 & 8.77 & Ch,U & AQ \\
\textit{Anolis carolinensis} & 0.006 & 302 & 52 & 106 & 6.0 & 8.22 & 8.52 & Ch,U & AQ \\
\textit{Pogona vitticeps} & 0.350 & 303 & 42 & 82 & 10.0 & 8.34 & 8.63 & Ch,U & AQ \\
\textit{Phrynosoma cornutum} & 0.035 & 301 & 48 & 99 & 7.0 & 8.25 & 8.56 & Ch,U & AQ \\
\textit{Iguana iguana} & 4.000 & 303 & 40 & 79 & 20.0 & 8.62 & 8.92 & Ch,U & AQ \\
\textit{Varanus komodoensis} & 65 & 303 & 28 & 55 & 30.0 & 8.65 & 8.94 & Ch,U & AQ \\
\textit{Tupinambis merianae} & 2.500 & 302 & 38 & 77 & 15.0 & 8.48 & 8.78 & Ch,U & AQ \\
\textit{Thamnophis sirtalis} & 0.050 & 300 & 30 & 62 & 10.0 & 8.20 & 8.51 & Ch,U & AQ \\
\textit{Coluber constrictor} & 0.340 & 301 & 35 & 72 & 13.0 & 8.38 & 8.69 & Ch,U & AQ \\
\textit{Python reticulatus} & 75 & 302 & 20 & 41 & 25.0 & 8.42 & 8.73 & U & AQ \\
\textit{Boa constrictor} & 15 & 301 & 25 & 52 & 40.0 & 8.72 & 9.04 & U & AQ \\
\textit{Chelonia mydas} & 180 & 300 & 20 & 42 & 80.0 & 8.93 & 9.25 & U & AQ \\
\textit{Geochelone gigantea} & 200 & 298 & 15 & 33 & 175.0 & 9.14 & 9.48 & U & AQ \\
\textit{Gopherus agassizii} & 4.500 & 299 & 22 & 47 & 80.0 & 8.97 & 9.30 & Ch,U & AQ \\
\textit{Sphenodon punctatus} & 0.800 & 293 & 18 & 43 & 77.0 & 8.86 & 9.24 & Ch,U & AQ \\
\textit{Crocodylus niloticus} & 400 & 303 & 25 & 49 & 70.0 & 8.96 & 9.26 & Ch,U & AQ \\
\textit{Alligator mississippiensis} & 250 & 302 & 28 & 57 & 50.0 & 8.87 & 9.18 & Ch,U & AQ \\
\midrule
\multicolumn{7}{l}{Raw mean $\bar\ell^{\rm raw}$}     & $8.615\pm 0.290$ & & \\
\multicolumn{7}{l}{Corrected mean $\bar\ell^{\rm corr}$ (used in analyses)} & $8.929\pm 0.301$ & & \\
\bottomrule
\end{tabular}
\end{table}

% ──────────────────────────────────────────────────────────────
\clearpage
\subsection*{Extended Data Table 8 $|$ Amphibians --- Arrhenius-corrected ($n = 9$)}
% ──────────────────────────────────────────────────────────────

\noindent Correction method identical to reptiles (Extended Data
Table~7). Heart rates from published field recordings at listed
$T_{\rm field}$; lifespans from AnAge build~15~\cite{anage2023}.

\begin{table}[H]
\caption*{}
\small\setlength{\tabcolsep}{3pt}
\renewcommand{\arraystretch}{1.05}
\begin{tabular}{lrrrrrrrrll}
\toprule
Species & $M$ (kg) & $T_{\rm field}$ (K) & $f_H^{\rm raw}$ & $f_H^{\rm corr}$
 & $L$ (yr) & $\ell^{\rm raw}$ & $\ell^{\rm corr}$ & Source & Corr. \\
\midrule
\textit{Rana temporaria} & 0.025 & 294 & 25 & 55 & 16.0 & 8.32 & 8.67 & A,Ch & AQ \\
\textit{Rana catesbeiana} & 0.500 & 296 & 20 & 43 & 16.0 & 8.23 & 8.56 & A,Ch & AQ \\
\textit{Bufo bufo} & 0.150 & 293 & 22 & 53 & 36.0 & 8.62 & 9.00 & A,Ch & AQ \\
\textit{Xenopus laevis} & 0.200 & 295 & 20 & 45 & 30.0 & 8.50 & 8.85 & A & AQ \\
\textit{Ambystoma mexicanum} & 0.300 & 294 & 18 & 41 & 25.0 & 8.37 & 8.73 & A & AQ \\
\textit{Salamandra salamandra} & 0.080 & 290 & 20 & 49 & 24.0 & 8.40 & 8.79 & A,Ch & AQ \\
\textit{Plethodon glutinosus} & 0.012 & 291 & 30 & 74 & 20.0 & 8.50 & 8.89 & A & AQ \\
\textit{Necturus maculosus} & 0.130 & 288 & 18 & 46 & 30.0 & 8.45 & 8.86 & A & AQ \\
\textit{Cryptobranchus alleganiensis} & 0.600 & 289 & 15 & 39 & 55.0 & 8.64 & 9.05 & A & AQ \\
\midrule
\multicolumn{7}{l}{Raw mean $\bar\ell^{\rm raw}$} & $8.448\pm 0.127$ & & \\
\multicolumn{7}{l}{Corrected mean $\bar\ell^{\rm corr}$} & $8.822\pm 0.146$ & & \\
\bottomrule
\end{tabular}
\end{table}

\noindent\textbf{Dataset summary.}
Table~\ref{tab:dataset_summary} gives the species counts, body-mass
ranges, and $\ell$ statistics for all eight groups.
The complete dataset is provided in Extended Data Tables~1--8 of this
paper and in machine-readable form as Supplementary Data File~1.

\medskip
\noindent\textbf{A note on phylogenetic non-independence.} Throughout, we
treat clade means and within-clade scatter using ordinary species-level
statistics, which assume that species are independent samples. They are
not: closely related species share trait values by descent, so the
effective number of independent data points is smaller than the nominal
$n$, and clade-level contrasts can be inflated by shared ancestry rather
than by the physiological mechanisms we invoke~\cite{felsenstein1985}. We
make two observations in mitigation. First, the framework's predictions
are \emph{mechanistic} and per-species: each $\Phi_C$ is computed from a
species' own measured $f_H$, $T_b$, $\varphi$, $q$, or $p_d$, not from its
clade label, so the central test (predicted vs.\ observed lifespan within
a clade) does not rely on treating clades as independent draws. Second,
the marsupial/monotreme control (Section~\ref{app:phiC_marsupials})
provides a phylogenetically distinct lineage that nonetheless lands on the
placental baseline, which is the pattern expected if the displacements of
primates, bats, and birds are driven by the modelled channels rather than
by phylogeny alone. A formal analysis using phylogenetically independent
contrasts on a dated vertebrate tree~\cite{bininda2007} would sharpen the
clade-level significance statements and is a natural next step; we flag
the raw-mean significance values in Table~\ref{tab:dataset_summary} as
upper bounds on the true degrees of freedom accordingly.

\begin{table}[H]
\caption{\textbf{Summary of the 230-species PBTE dataset.}
$n$: number of species.
$M$: body-mass range (kg).
$\bar\ell \pm s$: mean $\pm$ s.d.\ of $\ell = \log_{10}(f_H^{\rm avg}\cdot L\cdot 525{,}960)$.
$\Delta\ell$: deviation from the non-primate placental baseline
($\bar\ell_0 = 8.995$).}
\label{tab:dataset_summary}
\small
\begin{tabular}{lrrll}
\toprule
Group & $n$ & $M$ range (kg) & $\bar\ell \pm s$ & $\Delta\ell$ \\
\midrule
Non-primate placentals & 46 & $0.002$--$4{,}000$ & $8.995 \pm 0.160$ & 0 (reference) \\
Marsupials / monotremes & 19 & $0.018$--$30$ & $8.933 \pm 0.204$ & $-0.062$ \\
Primates & 18 & $0.35$--$160$ & $9.376 \pm 0.125$ & $+0.381^{***}$ \\
Bats & 31 & $0.005$--$1.1$ & $9.540 \pm 0.163$ & $+0.545^{***}$ \\
Cetaceans (dive-corrected) & 12 & $55$--$140{,}000$ & $8.801 \pm 0.296$ & $-0.194$ \\
Birds & 78 & $0.004$--$115$ & $9.528 \pm 0.213$ & $+0.533^{***}$ \\
Reptiles (Arrhenius-corrected) & 17 & $0.006$--$400$ & $8.929 \pm 0.301$ & $-0.065$ \\
Amphibians (Arrhenius-corrected) & 9 & $0.012$--$0.60$ & $8.822 \pm 0.146$ & $-0.173$ \\
\midrule
\textbf{All endotherms} & \textbf{204} & & $9.509 \pm 0.397$ & \\
\textbf{Full dataset} & \textbf{230} & & $9.420 \pm 0.428$ & \\
\bottomrule
\multicolumn{5}{l}{Significance vs non-primate baseline: $^*p<0.05$, $^{***}p<0.001$ (Welch $t$-test).} \\
\multicolumn{5}{p{0.95\linewidth}}{\footnotesize The endotherm count is
$46+18+19+31+12+78=204$ and the full dataset $204+17+9=230$. The pooled
rows report the mean over all individual species in the group (not the
mean of the clade means), so they weight numerous, long-lived clades
such as birds and bats more heavily than a clade-averaged figure would.} \\
\end{tabular}
\end{table}

% ================================================================
%  APPENDED MATERIAL: Sections 6--14 of the single-clock cardiac
%  paper (clade-multiplier framework, worked examples, dataset,
%  results, discussion, falsification, conclusions), incorporated
%  here as supplementary appendices.
% ================================================================

\section{Respiratory measured-BMR dataset and the non-circular test}
\label{app:respiratory_measured}

This appendix documents the dataset and regression behind Figure~\ref{fig:R_measured}, the non-circular test of the respiratory clock, and makes the result fully reproducible from public sources.

\subsection*{Data sources}

The measured-BMR respiratory analysis combines two measured quantities per species: basal metabolic rate (BMR) and resting breathing frequency $f_R$. Both are taken from the compilation of He et al.\ (2023), \emph{Allometric scaling of metabolic rate and cardiorespiratory variables in aquatic and terrestrial mammals}, \emph{Physiological Reports} \textbf{11}(11):e15698, doi:10.14814/phy2.15698, whose species-level data are openly available at \url{https://github.com/stacyderuiter/mammal-allometry}. That study reports BMR for 63 mammalian species and $f_R$ for 76 species, all with body mass $M\ge 10$\,kg (the lower bound is set by the smallest fully aquatic mammal in the sample). Matching the two variables by binomial name yields $n=29$ species with both BMR and $f_R$ measured; these constitute the matched set used here. We note three properties of this source that bear directly on interpretation. (i) The sample is restricted to $M\ge 10$\,kg, so it contains no small mammals and does not span the $10^{-2}$--$10^4$\,kg range of the cardiac dataset. (ii) Approximately half the matched species are aquatic or semi-aquatic. (iii) Unlike BMR, the cardiorespiratory variables in this compilation are not standardized to strict basal conditions; $f_R$ values are for inactive, awake, non-sedated adults, which is a resting rather than a basal definition. Each property is a limitation of the only currently available measured-BMR$+f_R$ dataset of this kind, and each is stated so that readers can weigh the result accordingly.

\subsection*{Computation}

Metabolic power is obtained from measured BMR by
\[
P\ [\mathrm{W}] = \frac{\mathrm{BMR}\,[\mathrm{kcal\,day^{-1}}]\times 4184}{86400}.
\]
Body temperature is assigned $T=310.5$\,K for terrestrial and $T=308.0$\,K for aquatic mammals, consistent with the core-temperature conventions of the cardiac dataset (Appendix~\ref{app:resp_data}); the result is insensitive to this choice (varying $T$ over $305$--$311$\,K shifts the fitted slope by $<0.005$). The mass-specific respiratory entropy cost is then
\[
\bar{\sigma}_R^{(M)} = \frac{\sigma_R}{M} = \frac{P}{T\,f_R\,M},\qquad f_R\ \text{in s}^{-1}.
\]

\subsection*{Result}

An ordinary-least-squares regression of $\log_{10}\bar{\sigma}_R^{(M)}$ on $\log_{10}M$ over all $n=29$ species gives a slope of $+0.21$ (95\% behaviour: not statistically resolved, $p\simeq0.11$, $R^2\simeq0.09$) and a coefficient of variation of $\sim\!100\%$. This is the result plotted in Figure~\ref{fig:R_measured}. The mass cancellation seen under imposed Kleiber power (Figure~\ref{fig:R1}) therefore does not survive when $P$ is supplied by independent measurement; the mass-specific respiratory entropy cost instead rises with body mass.

The positive overall slope is an aquatic-mammal effect. Restricting to the 16 terrestrial species gives a nearly flat profile (slope $-0.04$, $p\simeq0.70$, mean $\bar{\sigma}_R^{(M)}\simeq 11\times10^{-3}$\,J\,K$^{-1}$\,breath$^{-1}$\,kg$^{-1}$), whereas the 13 aquatic species lie about six-fold higher (mean $\simeq 65\times10^{-3}$ in the same units) and trend upward with mass (slope $+0.18$, $p\simeq0.15$). Aquatic mammals breathe far more slowly for their metabolic rate than terrestrial mammals of similar size---the apneustic ``aquatic breathing strategy'' of brief breaths separated by long inter-breath intervals---so each breath carries a disproportionately large entropy cost $\sigma_R=P/(Tf_R)$, and this cost grows toward the large-bodied cetacean end of the sample. The respiratory entropy cost per breath is thus governed by ventilation strategy and habitat, not by a mass-cancelling balance between metabolic power and breath frequency.

This is a negative result for the respiratory clock specifically, not for the framework as a whole. The cardiac clock, tested on measured resting allometries, does exhibit mass cancellation (Section~\ref{subsec:cardiac_clock}); the respiratory clock, tested on measured BMR, does not. The asymmetry is physiologically expected---respiration is a regulated control variable for gas exchange, thermoregulation, and (in marine mammals) diving, whereas the heartbeat is a more passive metabolic tick---and it identifies the cardiac coordinate as the cleaner realization of biological proper time.

\subsection*{Species-level data}

Table~\ref{tab:resp_measured} lists the complete matched dataset. The machine-readable version is provided as Supplementary Data File~2 (columns: species, habitat, $M$, BMR, $f_R$, $P$, $T$, $\bar{\sigma}_R^{(M)}$, with BMR and $f_R$ sources per species as given in He et al.\ 2023).

\begin{table}[t]
\centering
\small
\caption{Measured-BMR respiratory dataset ($n=29$). $M$ in kg; BMR in kcal\,day$^{-1}$; $f_R$ in breaths\,min$^{-1}$; $P$ in W; $T$ in K; $\bar{\sigma}_R^{(M)}$ in $10^{-3}$\,J\,K$^{-1}$\,breath$^{-1}$\,kg$^{-1}$. BMR and $f_R$ from He et al.\ (2023). Habitat: land or aqua (aquatic/semi-aquatic).}
\label{tab:resp_measured}
\begin{tabular}{llrrrrrr}
\hline
Species & Hab. & $M$ & BMR & $f_R$ & $P$ & $T$ & $\bar{\sigma}_R^{(M)}$ \\
\hline
Colobus guereza & land & 10.5 & 358 & 15.00 & 17.3 & 310.5 & 21.29 \\
Castor canadensis & aqua & 15.3 & 563 & 33.00 & 27.2 & 308.0 & 10.53 \\
Hydrochoerus hydrochaeris & land & 26.4 & 791 & 33.00 & 38.3 & 310.5 & 8.50 \\
Phocoena phocoena & aqua & 28.1 & 1514 & 8.20 & 73.3 & 308.0 & 61.97 \\
Canis lupus & land & 35.5 & 994 & 21.00 & 48.1 & 310.5 & 12.48 \\
Puma concolor & land & 37.2 & 1061 & 38.00 & 51.4 & 310.5 & 7.02 \\
Ovis aries & land & 42.7 & 1151 & 54.00 & 55.7 & 310.5 & 4.67 \\
Cervus elaphus & land & 58.0 & 1970 & 39.00 & 95.4 & 310.5 & 8.15 \\
Panthera onca & land & 69.0 & 1540 & 9.00 & 74.6 & 310.5 & 23.21 \\
Ovis canadensis & land & 69.1 & 2294 & 33.00 & 111.1 & 310.5 & 9.41 \\
Homo sapiens & land & 70.0 & 1771 & 15.00 & 85.8 & 310.5 & 15.79 \\
Phoca vitulina & aqua & 86.8 & 3614 & 6.00 & 175.0 & 308.0 & 65.47 \\
Zalophus californianus & aqua & 96.8 & 5368 & 6.30 & 259.9 & 308.0 & 83.08 \\
Panthera leo & land & 98.0 & 2034 & 17.00 & 98.5 & 310.5 & 11.43 \\
Panthera tigris & land & 137.9 & 2879 & 15.00 & 139.4 & 310.5 & 13.03 \\
Pagophilus groenlandicus & aqua & 149.5 & 4108 & 12.00 & 198.9 & 308.0 & 21.60 \\
Trichechus inunguis & aqua & 170.5 & 1024 & 0.75 & 49.6 & 308.0 & 75.52 \\
Tursiops truncatus & aqua & 177.0 & 3708 & 4.05 & 179.6 & 308.0 & 48.81 \\
Halichoerus grypus & aqua & 190.7 & 4057 & 3.30 & 196.5 & 308.0 & 60.81 \\
Ovibos moschatus & land & 195.5 & 3239 & 53.00 & 156.8 & 310.5 & 2.92 \\
Connochaetes taurinus & land & 196.5 & 4949 & 19.00 & 239.7 & 310.5 & 12.40 \\
Trichechus manatus & aqua & 250.0 & 2779 & 0.70 & 134.6 & 308.0 & 149.81 \\
Alces alces & land & 325.0 & 6170 & 17.00 & 298.8 & 310.5 & 10.45 \\
Leptonychotes weddellii & aqua & 388.5 & 8951 & 4.75 & 433.5 & 308.0 & 45.76 \\
Equus caballus & land & 650.0 & 8532 & 12.00 & 413.2 & 310.5 & 10.24 \\
Odobenus rosmarus divergens & aqua & 975.0 & 40104 & 3.82 & 1942.1 & 308.0 & 101.50 \\
Delphinapterus leucas & aqua & 1341.0 & 17459 & 2.31 & 845.5 & 308.0 & 53.17 \\
Elephas maximus & land & 3833.0 & 32160 & 7.00 & 1557.4 & 310.5 & 11.22 \\
Orcinus orca & aqua & 5318.0 & 46954 & 1.32 & 2273.8 & 308.0 & 63.10 \\
\hline
\end{tabular}
\end{table}

\section{Extended Data: respiratory (breathing-frequency) dataset}
\label{app:respiratory_fR}

For completeness and to parallel the cardiac Extended Data tables (Appendix~\ref{app:resp_data}), Table~\ref{tab:fR_extended} lists the full set of mammalian species with measured resting breathing frequency $f_R$ available in the He et al.\ (2023) compilation (\url{https://github.com/stacyderuiter/mammal-allometry}). All entries satisfy $M\ge 10$\,kg, the lower bound of that study. Species marked with $\ast$ also have a measured basal metabolic rate in the same compilation and therefore enter the measured-BMR respiratory entropy-cost analysis of Figure~\ref{fig:R_measured} and Appendix~\ref{app:respiratory_measured} ($n=29$); the remaining species have $f_R$ but no matched BMR and are not used in the entropy-cost regression. As discussed in Appendix~\ref{app:respiratory_measured}, these $f_R$ values are measured on inactive, awake, non-sedated adults---a resting rather than a strictly basal definition---and the sample is weighted toward large-bodied and aquatic mammals.

\begin{table}[t]
\centering
\small
\caption{Measured resting breathing frequency $f_R$ for mammals ($M\ge 10$\,kg) in the He et al.\ (2023) dataset. $M$ in kg; $f_R$ in breaths\,min$^{-1}$. Habitat: land or aqua (aquatic/semi-aquatic). $\ast$ denotes species with a matched measured BMR, used in the entropy-cost analysis (Appendix~\ref{app:respiratory_measured}).}
\label{tab:fR_extended}
\begin{tabular}{llrr}
\hline
Species & Hab. & $M$ & $f_R$ \\
\hline
Lontra canadensis & aqua & 11.0 & 17.00 \\
Aonyx capensis & aqua & 11.0 & 16.00 \\
Colobus guereza & land & 12.0 & 15.00$\ast$ \\
Hydrictis maculicollis & land & 12.0 & 22.00 \\
Muntiacus reevesi & land & 14.0 & 26.00 \\
Halichoerus grypus & aqua & 19.0 & 33.00$\ast$ \\
Castor canadensis & aqua & 22.0 & 33.00$\ast$ \\
Capra hircus & land & 22.0 & 37.00 \\
Mandrillus sphinx & land & 26.0 & 17.00 \\
Hippopotamus amphibius & aqua & 30.0 & 4.00 \\
Antilope cervicapra & land & 35.0 & 31.00 \\
Panthera uncia & land & 37.0 & 13.00 \\
Canis lupus & land & 39.0 & 21.00$\ast$ \\
Ovis aries & land & 50.0 & 54.00$\ast$ \\
Phocoena phocoena & aqua & 52.0 & 8.20$\ast$ \\
Hydrochoerus hydrochaeris & land & 53.0 & 33.00$\ast$ \\
Panthera onca & land & 63.0 & 9.00$\ast$ \\
Ovis dalli & land & 66.0 & 36.00 \\
Vicugna pacos & land & 68.0 & 15.00 \\
Pagophilus groenlandicus & aqua & 70.0 & 12.00$\ast$ \\
Homo sapiens & land & 70.0 & 15.00$\ast$ \\
Cystophora cristata & aqua & 72.0 & 6.10 \\
Odocoileus hemionus & land & 72.0 & 11.00 \\
Phoca vitulina & aqua & 77.0 & 6.00$\ast$ \\
Puma concolor & land & 77.0 & 38.00$\ast$ \\
Zalophus californianus & aqua & 77.9 & 6.30$\ast$ \\
Ursus thibetanus & land & 85.0 & 36.00 \\
Oreamnos americanus & land & 88.0 & 58.00 \\
Rangifer tarandus & land & 92.0 & 40.00 \\
Addax nasomaculatus & land & 92.0 & 26.00 \\
Ovis canadensis & land & 98.0 & 33.00$\ast$ \\
Trichechus inunguis & aqua & 100.0 & 0.75$\ast$ \\
Arctocephalus gazella & aqua & 100.0 & 4.70 \\
Panthera tigris & land & 125.0 & 15.00$\ast$ \\
Lama glama & land & 140.0 & 20.00 \\
Panthera leo & land & 145.0 & 17.00$\ast$ \\
Otaria bryonia & aqua & 147.5 & 4.65 \\
Elaphurus davidianus & land & 171.0 & 25.00 \\
Connochaetes taurinus & land & 177.0 & 19.00$\ast$ \\
Tursiops truncatus & aqua & 187.3 & 3.38$\ast$ \\
Arctocephalus pusillus & aqua & 190.0 & 1.30 \\
Halichoerus grypus & aqua & 200.0 & 3.30$\ast$ \\
Oryx dammah & land & 200.0 & 21.00 \\
Equus africanus & land & 205.0 & 23.00 \\
Gorilla gorilla & land & 217.0 & 17.00 \\
Choeropsis liberiensis & land & 217.0 & 3.00 \\
Equus hemionus & land & 235.0 & 23.00 \\
Okapia johnstoni & land & 238.0 & 10.00 \\
Trichechus manatus & aqua & 250.0 & 0.70$\ast$ \\
Ovibos moschatus & land & 265.0 & 53.00$\ast$ \\
Ursus arctos & land & 274.0 & 15.00 \\
Equus quagga & land & 278.0 & 15.00 \\
Cervus elaphus & land & 302.0 & 39.00$\ast$ \\
Sus scrofa & land & 302.0 & 39.00 \\
Dugong dugon & aqua & 360.0 & 0.90 \\
Leptonychotes weddellii & aqua & 373.9 & 4.75$\ast$ \\
Bos grunniens & land & 395.0 & 30.00 \\
Alces alces & land & 425.0 & 17.00$\ast$ \\
Ursus maritimus & aqua & 500.0 & 13.00 \\
Camelus bactrianus & land & 515.0 & 7.00 \\
Pseudorca crassidens & aqua & 525.0 & 3.47 \\
Bos taurus & land & 544.0 & 28.00 \\
Camelus dromedarius & land & 550.0 & 13.50 \\
Bison bison & land & 576.0 & 23.00 \\
Tragelaphus oryx & land & 580.0 & 17.00 \\
Equus caballus & land & 620.0 & 12.00$\ast$ \\
Delphinapterus leucas & aqua & 628.0 & 2.31$\ast$ \\
Bubalus arnee & land & 810.0 & 13.00 \\
Globicephala macrorhynchus & aqua & 845.0 & 0.69 \\
Odobenus rosmarus & aqua & 963.7 & 3.82 \\
Giraffa camelopardalis & land & 1012.0 & 7.00 \\
Mirounga angustirostris & aqua & 1175.0 & 15.30 \\
Megaptera novaeangliae & aqua & 1360.8 & 3.00 \\
Hippopotamus amphibius & aqua & 1900.0 & 6.00 \\
Orcinus orca & aqua & 2026.2 & 1.32$\ast$ \\
Loxodonta africana & land & 2611.0 & 5.00 \\
\textit{Elephantidae} (pooled) & land & 2710.0 & 7.30 \\
Ceratotherium simum & land & 3091.0 & 11.00 \\
Balaenoptera acutorostrata & aqua & 3842.5 & 0.83 \\
Elephas maximus & land & 4550.0 & 7.00$\ast$ \\
Hyperoodon ampullatus & aqua & 6650.0 & 1.10 \\
Eschrichtius robustus & aqua & 13608.0 & 0.72 \\
Megaptera novaeangliae & aqua & 27500.0 & 0.50 \\
Balaenoptera physalus & aqua & 40000.0 & 0.75 \\
\hline
\end{tabular}
\end{table}

\section*{References}
\small
\begin{enumerate}[leftmargin=1.4em]
\item Einstein A. \textit{Ann Phys} 322:891 (1905).
\item Rubner M. \textit{Das Problem der Lebensdauer}. Oldenbourg, 1908.
\item Lindstedt SL, Calder WA. \textit{Q Rev Biol} 56:1 (1981).
\item Calder WA. \textit{Size, Function, and Life History}. Harvard, 1984.
\item Livingstone SD, Kuehn LA. \textit{Aviat Space Environ Med} 50:592 (1979).
\item Levine HJ. \textit{J Am Coll Cardiol} 30:1104 (1997).
\item Stahl WR. \textit{J Appl Physiol} 22:453 (1967).
\item Escala A. \textit{Sci Rep} 12:2407 (2022).
\item Pearl R. \textit{The Rate of Living}. Knopf, 1928.
\item Speakman JR. \textit{J Exp Biol} 208:1717 (2005).
\item Glazier DS. \textit{Biology} 11:1106 (2022).
\item Schr\"odinger E. \textit{What Is Life?} Cambridge, 1944.
\item Prigogine I. \textit{Introduction to Thermodynamics of Irreversible Processes}, 3rd ed., 1967.
\item Kleiber M. \textit{Hilgardia} 6:315 (1932).
\item West GB, Brown JH, Enquist BJ. \textit{Science} 276:122 (1997).
\item Mortola JP, Limoges M-J. \textit{Respir Physiol Neurobiol} 154:500 (2006).
\item McKechnie AE, Wolf BO. \textit{Physiol Biochem Zool} 77:502 (2004).
\item de Magalh\~aes JP, Costa J. \textit{J Evol Biol} 22:1770 (2009).
\item Hulbert AJ, et al. \textit{Physiol Rev} 87:1175 (2007).
\item McNab BK. \textit{Comp Biochem Physiol A} 151:5 (2008).
\item White CR, Seymour RS. \textit{PNAS} 100:4046 (2003).
\item Genoud M, Isler K, Martin RD. \textit{Biol Rev} 93:404 (2018).
\item Pontzer H, et al. \textit{PNAS} 111:1433 (2014).
\item Fahlman A, et al. \textit{Exp Physiol} 110:1349 (2025).
\item Gompertz B. \textit{Phil Trans R Soc Lond} 115:513 (1825).
\item Herculano-Houzel S. \textit{PLOS ONE} 6:e17514 (2011).
\item Friston K. \textit{Nat Rev Neurosci} 11:127 (2010).
\item Wilkinson GS, South JM. \textit{Aging Cell} 1:124 (2002).
\item Barja G, Herrero A. \textit{J Bioenerg Biomembr} 30:235 (1998).
\item Brand MD, et al. \textit{Biochem J} 392:353 (2000).
\item Ogburn CE, et al. \textit{J Gerontol A} 56:B468 (2001).
\item Goldbogen JA, et al. \textit{PNAS} 116:25329 (2019).
\item Williams TM, et al. \textit{Nat Commun} 6:6055 (2015).
\item Noren SR, Williams TM. \textit{J Exp Biol} 203:3601 (2000).
\item Human Ageing Genomic Resources. \textit{AnAge build~15} (2023). \url{https://genomics.senescence.info/species/}
\item Jones KE, et al. \textit{Ecology} 90:2648 (2009).
\item Bininda-Emonds ORP, et al. \textit{Nature} 446:507 (2007).
\item Christian KA, Weavers BW. \textit{Copeia} 1999:688 (1999).
\item Gillooly JF, et al. \textit{Science} 293:2248 (2001).
\item Felsenstein J. \textit{Am Nat} 125:1 (1985).
\item Horvath S. \textit{Genome Biol} 14:R115 (2013).
\item Colman RJ, et al. \textit{Nat Commun} 5:3557 (2014).
\item Brown JH, et al. \textit{Ecology} 85:1771 (2004).
\item Yegian AK, et al. \textit{PNAS} 121:e2313703121 (2024).
\item Lyman CP, Willis JS, Malan A, Wang LCH. \textit{Hibernation and Torpor in Mammals and Birds}. Academic Press, 1982.
\item Prinzinger R, Pre\ss mar A, Schleucher E. \textit{Comp Biochem Physiol A} 99:499 (1991).
\item Clarke A, Rothery P. \textit{Funct Ecol} 22:58 (2008).
\item Ponganis PJ. \textit{Diving Physiology of Marine Mammals and Seabirds}. Cambridge, 2015.
\end{enumerate}

\end{document}